\documentclass[useAMS,usenatbib,usegraphicx]{mn2e}
\usepackage{times}
\usepackage{aas_macros} 
\usepackage{amssymb}

\usepackage{subfigure}
\usepackage{lscape}
\usepackage{natbib}
\usepackage{color}
\usepackage[utf8]{inputenc}

\usepackage{ulem}

\newcommand{\red}{\textcolor{black} } 

\title%
[Internal rotation of low-mass
main-sequence stars]%
{Nearly-uniform internal rotation of  solar-like main-sequence stars revealed by space-based asteroseismology and spectroscopic measurements} 

\author[O. Benomar \it et al.]{
O. \textsc{Benomar}$^{1}$\thanks{E-mail:othman.benomar@astron.s.u-tokyo.ac.jp},
M. \textsc{Takata}$^{1}$,
H. \textsc{Shibahashi}$^{1}$,
T. \textsc{Ceillier}$^{2}$,
R. A. \textsc{Garc{\'{\i}}a}$^{2}$ \\
{$^{1}$Department of Astronomy,
School of Science, The University of Tokyo, Bunkyo-ku, Tokyo 113-0033, Japan}\\
{$^{2}$Laboratoire AIM, CEA/DSM - CNRS - Univ. Paris Diderot - IRFU/SAp, Centre de Saclay, 91191, Gif-sur-Yvette Cedex, France}\\
}

\begin{document}
\date{Accepted 2015 3 July. Received 2015 May 22; in original form 2015 February 2015}

\maketitle

\begin{abstract}
 The rotation rates in the deep interior and at the surface of 22 main-sequence stars with masses between $1.0$ and $1.6\,{\rm M}_{\sun}$ are constrained by combining asteroseismological analysis with spectroscopic measurements. The asteroseismic data of each star are taken by the {\it Kepler} or CoRoT space mission. It is found that the difference  between the surface rotation rate and the average rotation rate (excluding the convective core) of most of stars is small enough to suggest that an efficient process of angular momentum transport operates during and/or before the main-sequence stage of stars.
If each of the surface convective zone and the underlying radiative zone, for individual stars, is assumed to rotate uniformly, the difference in the rotation rate between the two zones turns out to be no more than a factor of two in most of the stars independently of their ages.
\end{abstract}

\section{Introduction and summary
\label{sec:intro}}
Rotation is one of the fundamental issues that affect stellar structure and evolution. It induces material mixing processes which
may have a significant influence on nucleosynthesis in stars and affects their evolution tracks. 
In addition, the stellar rotation plays a crucial role in the dynamo process, which is
essential for the generation and maintenance of the magnetic field in stars. 

A key process to understand the stellar rotation is angular momentum transport.
In almost all phases of evolution, the core of stars  gradually contracts, while the envelope expands, so that the rotation rate of the core is larger than that of the envelope, as long as the angular momentum is locally conserved. This general trend is compensated by the transport of the angular momentum from the core to the envelope.
While many uncertainties still remain in the theoretical treatment, some clear evidences to support the presence of efficient angular momentum transport have been found in observations.
Helioseismology has demonstrated nearly-uniform rotation of the Sun in the radial direction from 30 per cent of the total radius (at least) to the surface
{\sloppy \citep[e.g.][]{Schou1998, Thompson2003, Darwich2013}}.
The surface rotation rates of white dwarfs, which are essentially cores of red giants, are much smaller than the value obtained by shrinking the Sun to the typical size of those stars assuming local conservation of angular momentum.
Asteroseismology has also recently shown that the core-to-envelope ratio of the rotation rate of several subgiants and young red giants is about 60
at most \citep{Beck2012Nature,Deheuvels2012,Deheuvels2014}, and that almost uniform rotation is detected in a main-sequence A star \citep{Kurtz2014} and in a main-sequence F star \citep{Saio2015}.
All of these results indicate that the angular momentum in the core is efficiently transported to the envelope. On the other hand, we remark that some model-dependent studies about B stars
based on ground-based observation
have presented only weak arguments of non-rigid rotation
\citep{Aerts2003sci,Pamyatnykh2004}.

In spite of these results, the number of main-sequence stars in which the structure of the internal rotation has been measured is still very limited. In this paper, we present the first statistical study of the radial rotation structure for main-sequence solar-like stars.
A method is proposed to constrain the ratio between the internal and surface rotation rates of a star that shows solar-like oscillations. On one hand, we measure the average interior rotation rate, $f_{\mathrm{seis}}$, by asteroseismology. On the other hand, the surface rotation rate, $f_{\mathrm{surf}}$, is determined either by spectroscopy or by measuring periodic luminosity variation due to stellar spots. 
An agreement between $f_{\mathrm{seis}}$ and $f_{\mathrm{surf}}$ implies uniform rotation, whereas a disagreement suggests a differential rotation.
The determination of the core rotation rate based on high-order and low-degree p modes has been tried in helioseismology
\citep[e.g.][]{1995Natur.376..669E,1996SoPh..166....1L, Garcia2004}, though these have been found to be only weakly sensitive to the solar rotation bellow $0.2R_{\sun}$ \citep{1999MNRAS.308..405C,Garcia2008}. Comparison of the core rotation rate
with the surface rotation rate derived from luminosity variation due to spots has been suggested in the case of solar-like pulsators by, e.g., \cite{2010aste.book.....A}.
In this paper, we consider not only spots but also spectroscopic measurements
to constrain the surface rotation rate,
and find that the latter is more reliable.

%

The method has been applied to a sample of main-sequence solar-like pulsators, 20 of which have been observed by the {\it Kepler} mission \citep{Borucki2010}, and two by the CoRoT spacecraft \citep{Baglin2006a, Baglin2006b}.
These are $1.0$--$1.6\,{\rm M}_{\sun}$ stars with the
solar composition, while their evolutionary stages cover
the entire main-sequence phase.
None of these stars are known to be members of a close binary system.
In addition, they rotate faster than the Sun by (typically) an order of magnitude.

The comparison of the surface rotation rate
with the average internal rotation rate
shows that
the two rates are generally close to each other. For 10 stars, their difference
is too small to be explained by
simple evolutionary models that assume
local conservation of
angular momentum in the radiative layers,
and constant rotation rates in the convection zones.
This clearly demonstrates that angular momentum is transported
from the inner layers to the outer layers
(in the radiative layers and/or
at the interfaces between the radiative and convective zones)
of these stars.
The result would generally be valid for
main-sequence stars with similar masses to the Sun.
Under the assumption that the surface convective zone and the internal radiative zone rotate rigidly (with different rates
from each other in general), we found that \red{21} out of 22 stars show rotation rates
in the radiative interior that are
consistent with those in the convective envelope within a
factor of two.

The paper is structured as follows. section \ref{sec:Kernels} presents how the radial structure of the rotation profile can be measured in solar-like stars. Then section \ref{sec:methods}
describes the observational constraints and the method, followed by
section \ref{sec:results} that shows our results. Finally,
section \ref{sec:discussion} is devoted to discussion.
%
\section{Asteroseismic measurement of internal rotation}
\label{sec:Kernels}
\subsection{Mode properties} \label{sec:Kernels:general}
Asteroseismology is a powerful method to probe physical properties of invisible stellar interiors, using pulsations detected at the surface. Recent space-borne observations made by the CoRoT and {\it Kepler} satellites have
detected oscillations of a large number of solar-like stars
with an unprecedented precision \citep[e.g][]{Appourchaux2012,Mathur2012},
opening the possibility of scrutinising the internal structure of the stars similar to the Sun statistically \citep{Chaplin2014}. 
More importantly, the
rotation rates both at the surface and in the deep layers have been measured for a significant number of stars at various evolutionary stages \citep[e.g.][]{Benomar2009, Barban2009, Deheuvels2012, Deheuvels2014, Davies2015}.
The measure of the internal rotation is made possible because rotation induces
a splitting of oscillation eigenfrequencies,
which could be compared to the Zeeman effect observed in spectral lines in presence of a magnetic field.
More precisely, for a non-rotating star, a pulsation mode is described by two
integers
$n$ and $l$, which correspond to the number of nodes along the radius of a star (radial order of the mode\footnote{
This is only an approximate
explanation about the radial order.
More detailed discussion is found in \cite{takata2012}.})
and the number of nodal lines on the surface (degree of the mode), respectively.
There exist $2l+1$ modes with the same $n$ and $l$ that
have a common eigenfrequency (degenerated solutions),
but different dependence on the azimuthal angle,
if the star is spherically symmetric.
However, the rotation lifts
the degeneracy among non-radial pulsation modes ($l>0$), revealing
a discrete structure of modes identified by their azimuthal order $m$ such as $-l \leq m \leq +l$, with $|m|$ equal to the number of nodal lines on the surface
that coincide with meridians.
Slowly rotating stars like the Sun almost
preserve their spherical symmetry,\footnote{%
Strictly speaking,
the effect of the centrifugal force
that causes distortion of the stellar configuration
might not always be negligible
in the stars dealt in this paper,
because they
rotate significantly faster than the Sun.
This problem is discussed in
section \ref{subsec:centrifugal_effects}.
On the other hand,
we assume for simplicity that the effect of
the magnetic field is much smaller
than that of rotation.
}
so that each mode frequency $\nu_{n,l,m}$ is given by
\begin{equation} \label{eq:splitting}
	\nu_{n,l,m} =  \nu_{n,l} + m \delta\nu_{n,l}.
\end{equation}
Here, $\nu_{n,l}$ is the frequency for a spherically symmetric star,
while $\delta\nu_{n,l}$ is called
the {\it rotational splitting}, which represents
a small perturbation induced by the rotation. 
The rotational splitting is the observable
parameter that enables us to measure the rotation of the stellar interior.

\subsection{Rotational kernels} \label{sec:Kernels:kernels}
We explain how the rotational splittings
are sensitive to the internal rotation profile of the star,
assuming for simplicity that the rotation rate is a function of only the distance from the centre of the star.
The sensitivity depends on the properties of observed oscillation modes,
which are high-order (large $n$) and low-degree (small $l$) p modes
consisting of acoustic waves, in the case of solar-like pulsators.
The constituent waves can travel from the surface to deep layers before
reaching their turning point, where they are refracted back toward the surface.
The sensitivity of such modes to the rotation profile within the visited layers is represented by the
so-called {\it  rotational kernels}, hereafter simply referred as {\it kernels}, $K_{n,l}(r)$. The kernel of each mode mostly depends on the eigenfunction and is related to the rotational splitting,
  $\delta\nu_{n,l}$,  and to the cyclic frequency of rotation, $f(r)$, by
\begin{eqnarray}
		\delta\nu_{n,l} &=& (1 - C_{n,l}) \int_0^R K_{n,l}(r) f(r) \,{\rm d}r  \nonumber\\
 &=& (1 - C_{n,l}) f_{n,l}.
 \label{eq:kernel}    
\end{eqnarray}%
Thus the splitting is proportional to
the average cyclic rotation frequency $f_{n,l}$, across the region visited by a mode of degree $l$ and radial order $n$. Here, $C_{n,l}$ is the {\it Ledoux} constant \citep{cowling1949, Ledoux1951}. This constant measures the effect of the Coriolis force and tends toward $0$ for high-order p modes. For example, $C_{n,l} \simeq 10^{-2}$ in the range of observed modes in the Sun.  This
implies that the splitting may differ from the rotation frequency ${f_{n,l}}$ by about $1$ per cent at most.
Because the typical uncertainty in the rotational splittings of solar-like stars is as large as 5 per cent,
the Coriolis term $C_{n,l}$ is in practice negligible in equation (\ref{eq:kernel}).

Figs \ref{fig:krn-l1}(a) and \ref{fig:krn-l2}(a) show examples of kernels for $n=10$ and $n=25$
for
the models of the
three different stars; the
Sun \citep{JCD1996Science}, {\it Kepler}-25 \citep[$M=1.26 \pm 0.03\,{\rm M}_{\sun}$, see][]{Benomar2014b} and HAT-P-7 ($M=1.59 \pm 0.03\,{\rm M}_{\sun}$, \citealt[][]{Benomar2014b, Lund2014}). Only modes of degrees $l=1$ and $l=2$ are shown, as it is rare to observe $l=3$ (and higher-degree)
modes in {\it Kepler} or CoRoT data. The radius is normalised by the stellar radius $R$ in each case. The kernels are
denser and of greater amplitude near the stellar surface, indicating that the sensitivity to the rotation is the highest close to the surface.
This reflects the fact that acoustic waves, the main constituents of high-order p modes, slow down as they come close to the surface, so that the mode periods, which are essentially determined by the residence time of the waves at each layer, are most influenced by the near-surface structure. \red{Note that at the outer edge of the convective core of {\it Kepler} 25 and HAT-P-7, a small kernel discontinuity appears. This is caused by those of the chemical composition at the top of the convective core and is due to the fact that diffusion is neglected in our models.}

Figs \ref{fig:krn-l1}(b) and \ref{fig:krn-l2}(b) show the cumulative integrals $\int_0^{r_0} K_{n,l}(r) \,{\rm d}r$ for degrees $l=1$ and $l=2$, respectively. Note that within the observable range of $n$, which is approximately between $10$ and $25$, all of the kernels with
$l=1$ and $l=2$ are nearly identical to each other.
Still, there is clear dependence of the base of the convective envelope on the mass, such that more massive stars have shallower convective envelopes. This implies that the contribution of the radiative zone to the rotational splittings is greater in more massive stars than in less massive ones.
Furthermore,
the profiles in the radiative zone are almost linear functions of radius,
between $\sim0.15 R$ and the outer edge of the zone.
Since the gradient of the profiles indicates the sensitivity to each layer,
each observed eigenmode probes the radiative zone nearly uniformly
from the very deep inner layers to the outer edge.

\subsection{Rotation rates in the envelope and in the radiative zone} \label{sec:kernels:Orad}
\begin{figure}
  \begin{center}
   \includegraphics[angle=0,width=8cm]{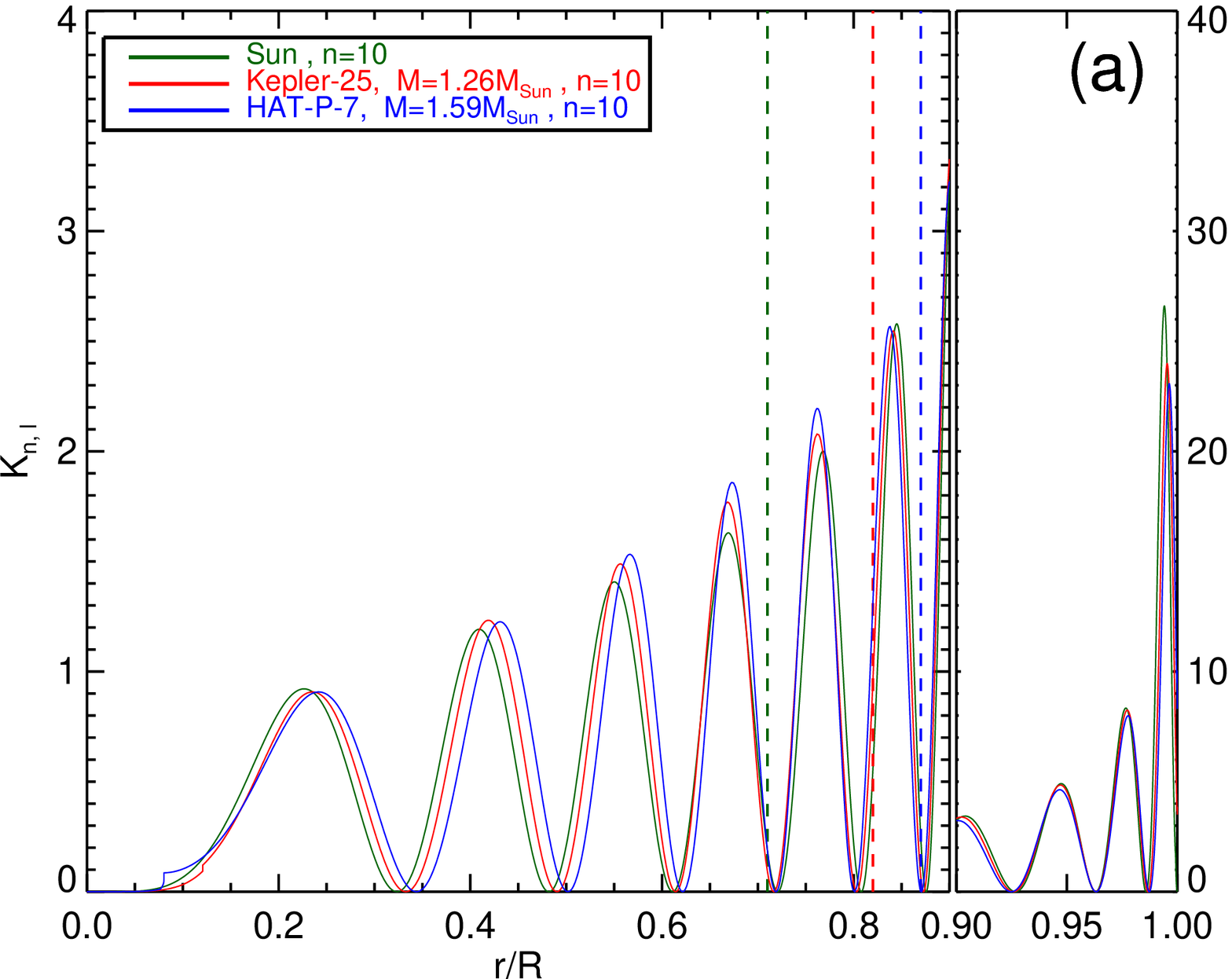}\\
    \includegraphics[angle=0,width=8cm]{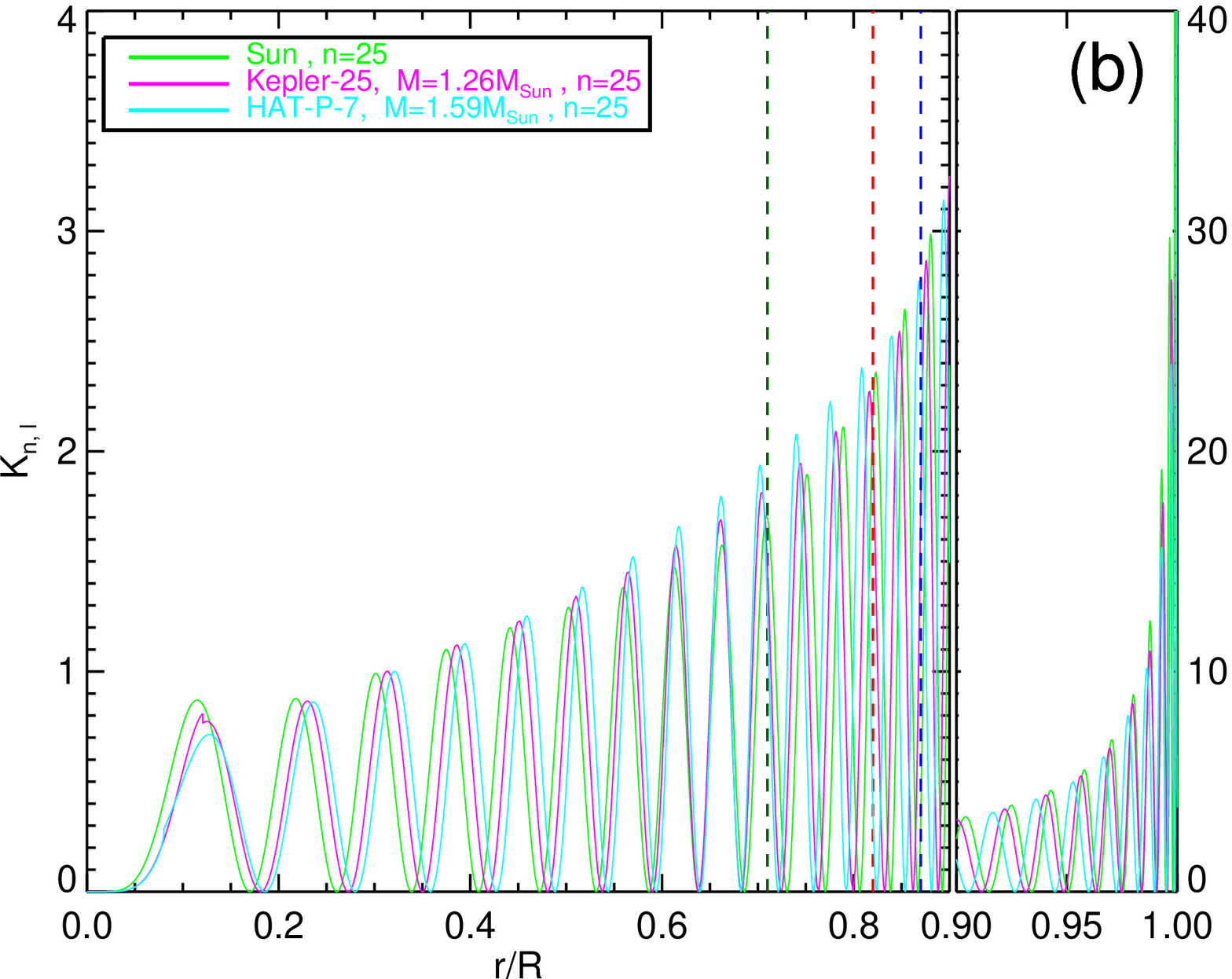}\\  
   \includegraphics[angle=0,width=8.2cm]{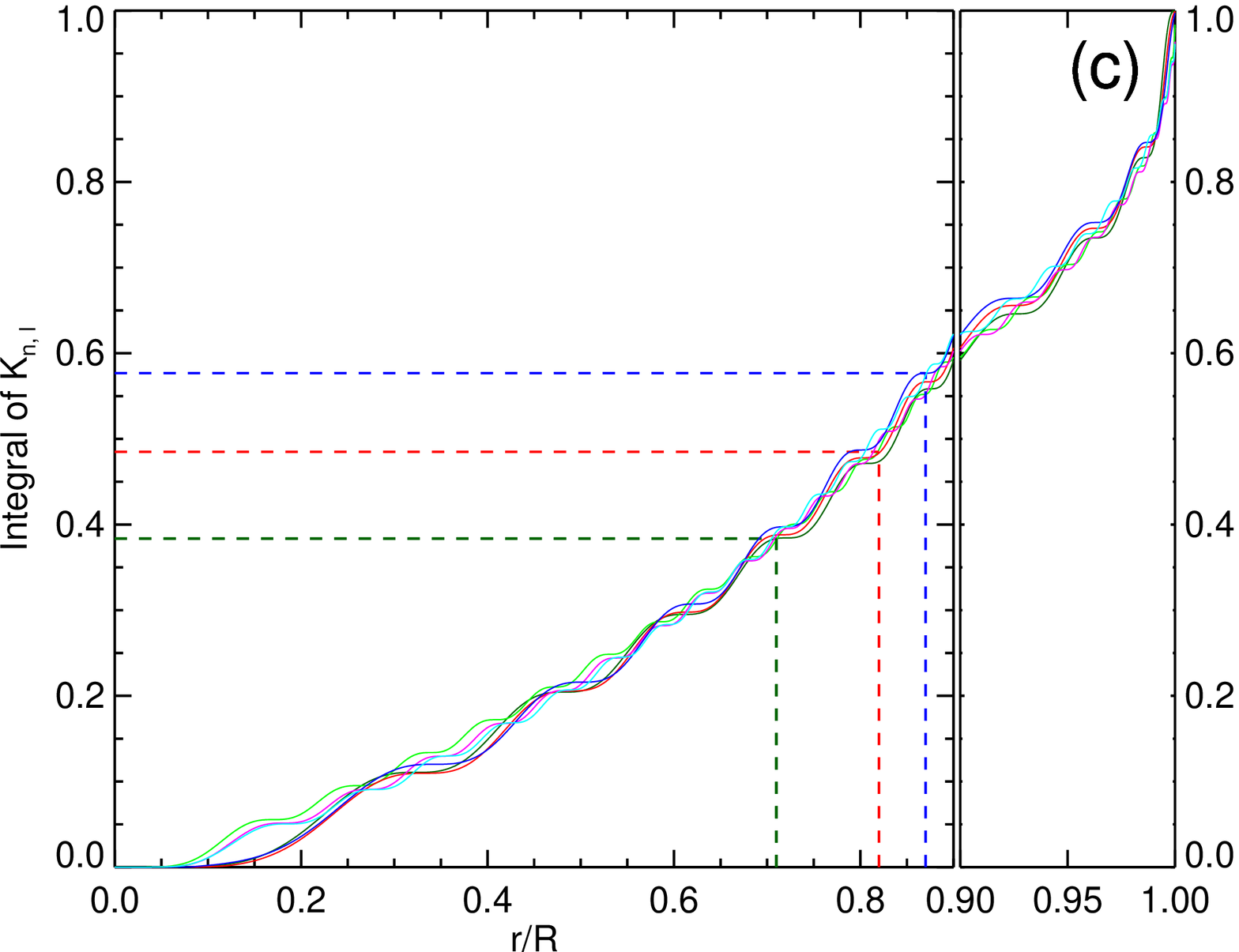}
  \end{center}
 \caption{The figures show the rotational kernels \red{(a,b)} and their integral over the radius \red{(c)} for the Sun, {\it Kepler}-25 and HAT-P-7 for $l=1$ modes. \red{Plots have two scales to magnify the Kernel of near-surface layers.} Only
 modes with radial orders
 $n=10$ \red{(a)} and $n=25$ \red{(b)} are shown. Integrals of kernels are almost identical to each other
 independently of the star. Vertical dotted lines indicate bases
 of the outer convective zones. \red{Small discontinuities around 0.1 in (a) and (b) are caused by those of the chemical composition at the top of the convective core.}}  
\label{fig:krn-l1}
\end{figure}

\begin{figure}
 \begin{center}
  \includegraphics[angle=0,width=8cm]{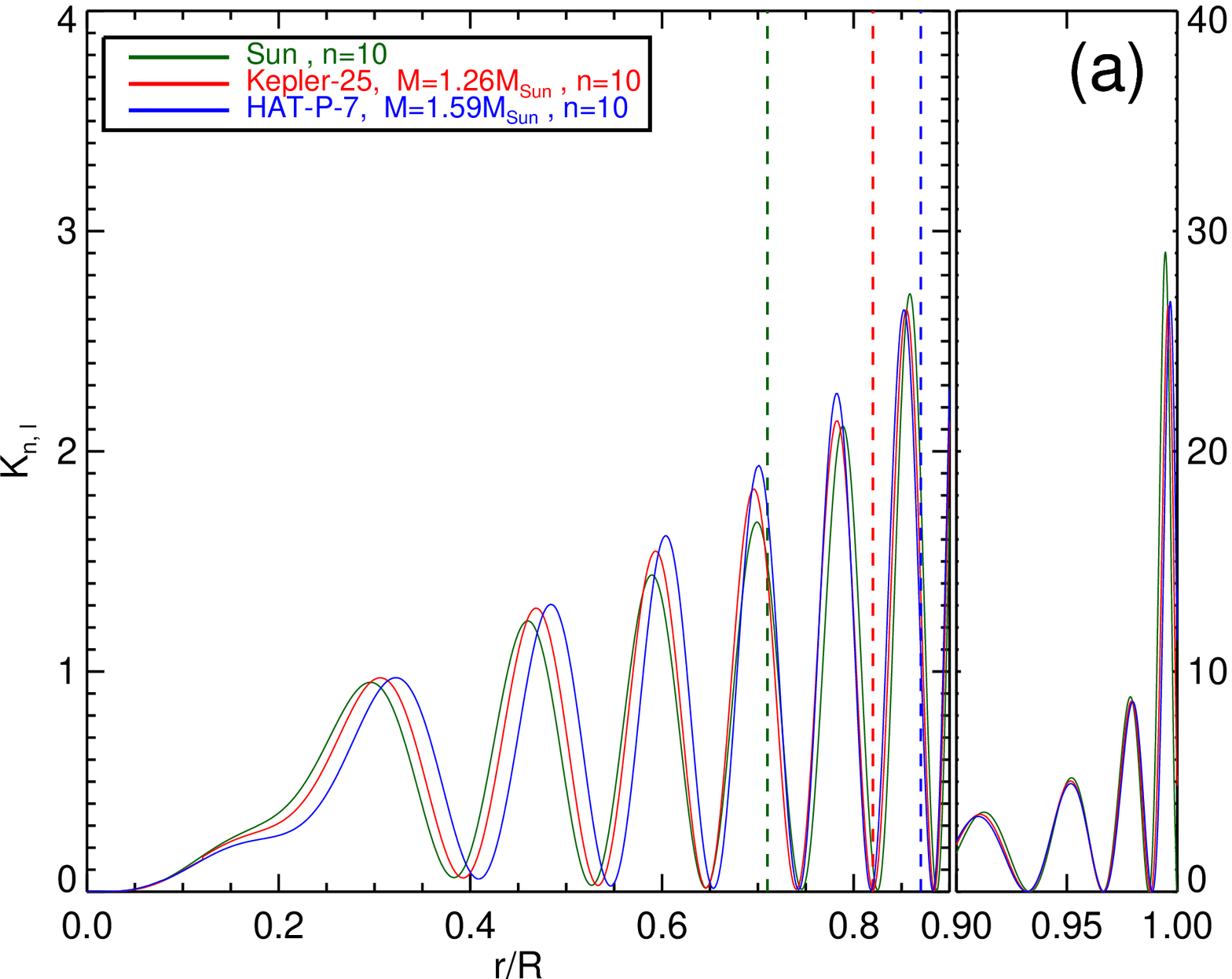}\\
    \includegraphics[angle=0,width=8cm]{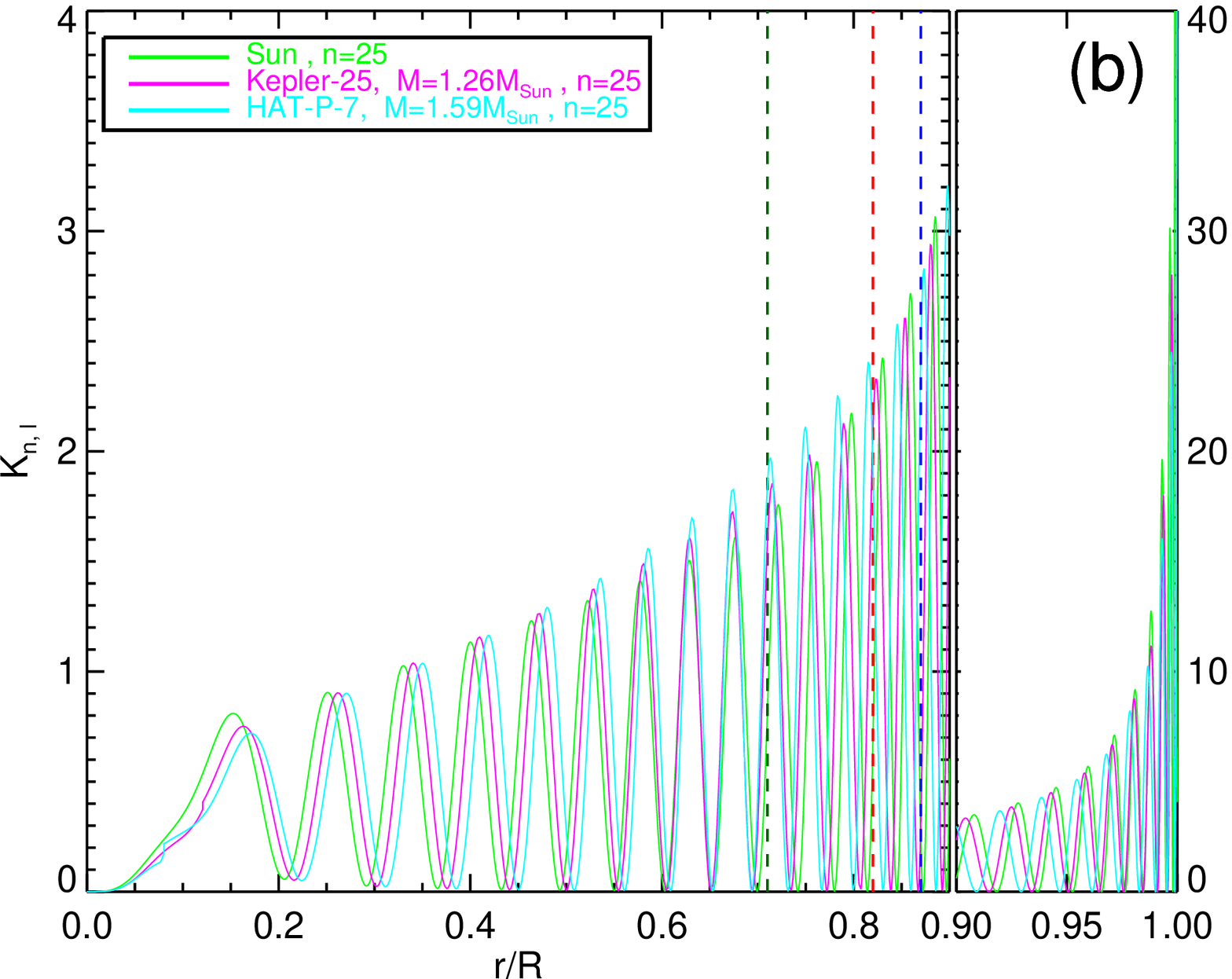}\\
  \includegraphics[angle=0,width=8.2cm]{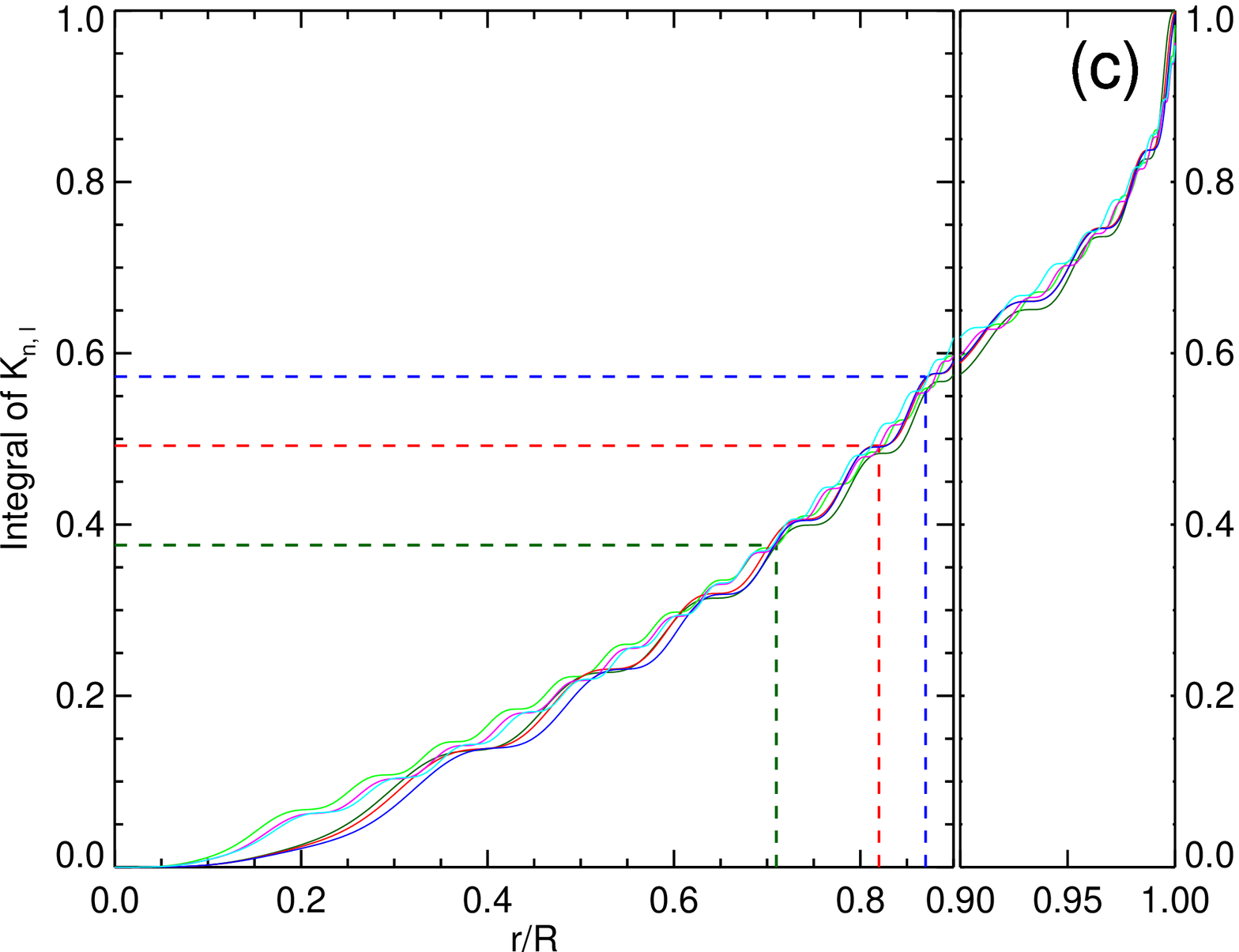}
 \end{center}
 \caption{Same as Fig.\,\ref{fig:krn-l1}, but for $l=2$ modes.}  
\label{fig:krn-l2}
\end{figure}

A solar-like star (slightly more massive than the Sun)
consists of
an outer convective zone, an intermediate
radiative zone and a core where nuclear fusion occurs.
Figs \ref{fig:krn-l1} and \ref{fig:krn-l2} demonstrate that low-degree modes are insensitive to the stellar core.
In order to probe the rotation rate within the (radiative) core, it is required
to measure g modes \citep[e.g.][]{Garcia2008b, Mathur2008,Appourchaux2010},
which are still
elusive in solar-like stars, or mixed modes \citep[e.g.][]{Deheuvels2014}, which are seen in evolved solar-like stars only.  However,  we stress that the rotational splittings of low-degree p modes in main-sequence solar-like stars are sensitive to
the rest of the structure, both of the radiative zone and the convective envelope.

In order to separate the contributions from the radiative and (outer) convective zones, it is assumed that the two zones rotate uniformly with different rates along the same axis.
Then,
it is possible to express the rotational splitting, $\delta\nu_{n,l}$, by a weighted average of the rotation rate of the radiative zone, ${f_{\rm rad}}$, and that of the convective zone, ${f_{\rm conv}}$, as
\begin{equation}
 \delta\nu_{n,l} \simeq I_{\rm rad} \, {f_{\rm rad}}  + I_{\rm conv} \, {f_{\rm conv}}, 
 \label{eq:kernel:rot_gen_b}
\end{equation}
where $I_{\rm rad}=\int_0^{r_{\rm c}} K_{n,l}(r)\,{\rm d}r$ and $I_{\rm conv}=\int_{r_{\rm c}}^R K_{n,l}(r)\,{\rm d}r$ denote the integrals of the kernel in the radiative and convective zones,
respectively, with $r_{\rm c}$ being the radius at the base of the convective zone. Note that
\begin{equation}
 I_{\rm conv} + I_{\rm rad} = 1,
  \label{eq:kernel:rot_gen_a}  
\end{equation}
because each kernel is normalised to 1.

It is required to make two additional assumptions in order to make use of equation (\ref{eq:kernel:rot_gen_b}). 
Firstly, the rotation rate in the convective envelope, ${f_{\rm conv}}$, is assumed to be equal to that at the surface, ${f_{\mathrm{surf}}}$. This is supported by the solar case, which shows 30 per cent difference at most in the rotation rate in the latitudinal direction and much smaller in the radial direction in the convection zone
{\sloppy\citep[e.g.][]{Schou1998, Thompson2003, Darwich2013}}.
Secondly, it is assumed that the rotational splittings remain nearly constant over the observed ranges of $n$ and $l$, so that the average rotational splitting, $\langle\delta\nu_{n,l}\rangle$, can be used as a representative value of the seismically measured internal rotation rate, $f_{\rm seis}$.  This point has been justified in section
\ref{sec:Kernels:kernels} by demonstrating that the integrals of the kernels of the observed p modes differ little from each other.
We can now use equation (\ref{eq:kernel:rot_gen_b})
to express the relation linking
the surface rotation rate, $f_{\mathrm{surf}}$, the average rotation rate in the radiative zone, $\langle f_{\rm rad} \rangle$,
and the average rotation rate of the stellar interior, $f_{\rm seis}=\langle\delta\nu_{n,l}\rangle$, as
\begin{equation} \label{eq:kernel:rot_gen_final}
	\langle f_{\rm rad}\rangle = f_{\mathrm{surf}} + \langle I_{\rm rad}\rangle^{-1} (f_{\rm seis}- f_{\mathrm{surf}}).
\end{equation}
Note that $\langle\, \rangle$ denotes the average values over the range of observed $n$ and $l$, with $10 \la n \la 25$ and $1 \le l \le 2$.

If we go back to equation (\ref{eq:kernel:rot_gen_b}),
it tells that $f_{\rm seis}$ is reduced to a simple average of $f_{\rm rad}$ and $f_{\rm conv}$, if $\langle I_{\rm rad} \rangle = \langle I_{\rm conv}\rangle$.
However, in stars with a thin convective envelope ({\it i.e.} the most massive solar-like stars), the sensitivity of the splitting to the rotation within the radiative zone may dominate,
which means $\langle I_{\rm rad}\rangle \gg \langle I_{\rm conv}\rangle$, so
 that these stars are ideal candidates to probe a differential rate between outer and inner layers.
 In fact, equation (\ref{eq:kernel:rot_gen_final}) is equivalent to
 $f_{\rm seis} - f_{\rm surf} =
 \langle I_{\rm rad}\rangle
 \left(\langle f_{\rm rad}\rangle - \langle f_{\rm conv}\rangle\right)$,
 whose left-hand side is an observable.  This equation shows that,
 for a given difference in the rotation rate between the radiative zone
 and the convective envelope, the observed difference between $f_{\rm seis}$ and  $f_{\rm surf}$ is larger (in magnitude)
 in stars with larger $\langle I_{\rm rad}\rangle$. 
 
 \red{Applying the error propagation law to equation (\ref{eq:kernel:rot_gen_final}), we estimate that an interior with faster-than-surface rotation can be detected at a $1\sigma$ detection level if $(f_{\rm seis} - f_{\rm surf})/f_{\rm surf} \gtrsim 25$ per cent, which corresponds to $\langle f_{\rm rad}\rangle/f_{\rm surf} \gtrsim 1.5$. Similarly, detecting a slower-than-surface rotation requires $(f_{\rm seis} - f_{\rm surf})/f_{\rm surf} \lesssim -45$ per cent, or $\langle f_{\rm rad}\rangle/f_{\rm surf} \lesssim 0.2$. This is obtained by assuming that the relative uncertainty on $f_{\rm surf}$ is of 30 per cent, based on the Sun. Here, the uncertainty account for the fact that $f_{\rm conv}$ and $f_{\rm surf}$ may not be identical, depending on the latitude at which the star is observed (latitudinal differential rotation)\footnote{Actually, this is a very conservative value as our formal uncertainties on $f_{\rm surf}$ are in fact of $\approx 10$ per cent.}. We also set the uncertainties on  $f_{\rm seis}$ to 5 per cent and to 4 per cent on $\langle I_{\rm rad} \rangle$ (justified in section \ref{sec:results}).}
 
In the following, equation (\ref{eq:kernel:rot_gen_final})
is used to measure the rotation rate in the radiative zone.


\section{Deriving the stellar rotation} \label{sec:methods}

This section describes how the surface and internal rotation is derived from the observables.


\subsection{Surface rotation rate
\label{subsec:spectro-spots}}
 The surface rotation rate $f_{\mathrm{surf}}$ can be estimated using either (1) rotational broadening of spectroscopic absorption lines, which gives $v\,\sin i=2 \upi R \,f^{(1)}_{\rm surf} \sin i$, with the radius, $R$, and the inclination angle, $i$, determined by theoretical models and asteroseismic analysis, respectively,
 or (2) the brightness modulation of period $P_{\rm rot}=1/f^{(2)}_{\rm surf}$ due to structures at the surface of the star (e.g. magnetic spots).   Note that the rotational broadening is 
determined by contributions from a wide range of the area on the visible hemisphere,
 while the brightness modulation measures the rotation period at the latitude where spots evolve. For the Sun, it is well known that the spots migrate from $\approx 30{\degr}$ \cite[][]{Maunder1904} toward the equator during their 11-year cycle, so that $v\,\sin i$ and $P_{\rm rot}$ may not be sensitive to rotation of the same regions. 
 This means that the observed rotation frequency $f_{\mathrm{surf}}^{(1)}$ and $f_{\mathrm{surf}}^{(2)}$ may intrinsically differ from each other in case of significant
 latitudinal differential rotation. This point is further discussed in section \ref{sec:results}.
 
 The spectroscopic $v\,\sin i$ of the analysed CoRoT stars has been measured by \cite{Bruntt2009}, while, except
 for HAT-P-7 and {\it Kepler}-25, the studied {\it Kepler} stars have been observed spectroscopically by \cite{Bruntt2012}. Since
 \cite{Bruntt2012} do not provide uncertainties, we assumed here
 a typical precision of $10$ per cent.  Regarding HAT-P-7 and {\it Kepler}-25,
 we use
 values determined by \cite{Pal2008} and \cite{Marcy2014}, respectively.

 The surface rotation frequencies $f_{\mathrm{surf}}^{(2)}$ of the two CoRoT stars, HD 49933 and HD 181420 are from \cite{Benomar2009b} and \cite{Barban2009}, respectively, who have detected the low-frequency periodicity in the power spectrum. Note that alternative values were derived by using a stellar spot modelling technique in \cite{Mosser2009}. These are in agreement with \red{the} surface rate used in this study (see Table \ref{tab:rot}). For {\it Kepler} stars, they are determined
 from rotation periods $P_{\rm rot}$, which have been
 measured by \cite{Garcia2014} based on
 the wavelet transform and on the analysis of the autocorrelation function of the light curve\footnote{Using KADACS and PDC-MAP data, see \cite{Garcia2011, Thompson2013}.}. When the comparison is possible, their derived surface rotation is consistent with the rotation from \cite{Karoff2013}. Note that $f_{\mathrm{surf}}^{(2)}$ cannot be evaluated for HAT-P-7 and {\it Kepler}-25 because they do not show detectable surface activity.
 
\subsection{Internal rotation and stellar inclination
\label{subsec:seismo}}

Asteroseismology of solar-like stars, in the present case, is based on the photometry of the luminosity variation. Because of its stochastic nature, each solar-like mode has a Lorentzian profile in the power spectrum \citep{harvey1985}. In that case, the stellar oscillations can be expressed as a sum of Lorentzians over $n$, $l$ and $m$,
\begin{equation}
\label{eq:power}
P(\nu)= \sum_{n,l}\sum_{m=-l}^l 
\frac{ H_{n,l, m}}
{1+4(\nu-\nu_{n,l,m})^2/\Gamma^2_{n,l,m}} .
\end{equation}
Here, $\nu_{n,l,m}$ is given by equation (\ref{eq:splitting})
, while $H_{n,l,m}$ and $\Gamma_{n,l,m}$ correspond to the mode height and width at half maximum, respectively.

The turbulent convection, which is the
origin of acoustic modes, does not prefer any particular direction to others,
so that the rotationally split modes with the same $l$ and $n$ are expected to have
almost the same amplitudes.
Due to a geometrical projection effect \citep{Ballot2006,Gizon2003}, in disk-integrated photometry, the height of the split modes depend on
the stellar inclination angle $i$ as
$H_{n,l,m}=\mathcal{E}_{l,m}(i)
H_{n,l}$. Here $H_{n,l}$ is the intrinsic height for the mode, while
 $\mathcal{E}_{l,m}(i)$ is the visibility of the $m$-components
in the power spectrum,
\begin{equation}
\label{eq:legendre}
\mathcal{E}_{l,m}(i)=\frac{(l-|m|)!}{(l+|m|)!} \left[P^{|m|}_l (\cos i)\right]^2 ,
\end{equation}
with $P^{|m|}_l$ being the associated Legendre function.
It is therefore possible to determine $i$ from $\mathcal{E}_{l,m}(i)$.
Several authors have already
exploited that approach
to measure the stellar inclination \cite[e.g.][]{Ballot2008, Benomar2009b, Appourchaux2012, Chaplin2013, Benomar2014, Lund2014}. 

In practice, the rotational splittings and the inclination angle are determined, together with the Lorentzian parameters, by fitting the power spectrum of the light curve with the formula given by equation (\ref{eq:power}). Therefore, the precision on the rotation is limited by the correlation between
the rotational splitting and of the stellar inclination.
Note that the splitting is assumed constant over the observed range of modes in this analysis.
As shown by \cite{Benomar2014}, the correlation problem is particularly important in slow rotators, because the $2l+1$
$m$-components of the same order $n$ and degree $l$
are difficult to disentangle from each other.  
Contrary to heat driven pulsators, stochastically driven pulsators have an important intrinsic noise of stellar origin that renders mode pulsations hard to analyse and requires
 sophisticated statistical methods. Here, we use a Markov Chain Monte Carlo method \cite[see e.g.][for applications in asteroseismology]{Benomar2009, Handberg2011}. 
This allows us to evaluate the probability density function for the rotational splitting and stellar inclination, as well as their correlations. Note that part of studied stars have been
analysed by \cite{Appourchaux2012} and \cite{Appourchaux2014}, whose work focuses
on mode frequencies, widths and heights.  Stars of the current work in common with these studies are revisited here using the same data, but focusing on the extraction of the rotation and the stellar inclination angle. 

\subsection{Centrifugal distortion}
\label{subsec:centrifugal_effects}

\red{In section \ref{sec:Kernels:general}, we have assumed that the internal rotation produces equal spacings in each multiplet of the power spectrum, as the first-order perturbation analysis due to the Coriolis force 
predicts. However, this cannot be fully justified if the second-order effect caused by the centrifugal force affects the pulsation modes significantly. Assuming uniform rotation, we
can estimate the effects of the Coriolis and centrifugal forces by the two dimensionless parameters,
$\epsilon_1=f/\nu$ and $\epsilon_2=\left(f/\nu_{\rm K}\right)^2$, respectively,
in which
$f$ and $\nu$ are the rotational and pulsational frequencies, respectively,
and
$\nu_{\rm K} = (2\upi)^{-1} (GM R^{-3})^{1/2}$
\citep[e.g.][]{Saio:1981aa}.}

\red{These two parameters are related to the asymmetry
of each multiplet by
\begin{equation}
\left|
 \left[
  \nu_{0} - \frac{\nu_{m} + \nu_{-m}}{2}
 \right]
 \left[
  \frac{\nu_{m} - \nu_{-m}}{2 m}
 \right]^{-1}
\right|
\sim
\frac{\epsilon_2}{\epsilon_1}
\end{equation}
for $m \ge 1$,
where $\nu_{m}$ represents the frequency
of the component specified by
the azimuthal order $m$.
It is possible that the Coriolis and the centrifugal force are of the same order in some of the solar-like stars, which would induce an asymmetry of the multiplets in the power spectrum.
Still, fitting such asymmetric structure with the symmetric formula should mainly
introduce additional systematic errors in our estimates of the rotational splittings (and the stellar inclination),
rather than lead to completely wrong values of them.
In fact, the frequency difference,
$(\nu_m-\nu_{-m})/(2m)$,
which is equal to the rotational splitting,
is not modified by the centrifugal distortion,
but only changed by the third- and higher-order effects, which are negligibly small
in all the stars. Moreover, if the asymmetry is significant, our analysis should suggest multiple
different values of the rotational splittings.
In the case of dipole modes, for example,
we should find not only the correct value, $(\nu_{m=1}-\nu_{m=-1})/2$,
but also two other different spacings (at most),
one between $m=1$ and $m=0$ components,
and the other between $m=0$ and $m=-1$ components.
Since most of the probability density functions, which are shown in
Fig. \ref{fig:jointpdf}, do not have such multiple-peak structure,
we presume that the asymmetry is generally smaller than
(or comparable to) the width of the main peak in the probability density functions.}
%
%

\subsection{Stellar models
\label{subsec:seism:compare}}

In order to evaluate $I_{\rm conv}$ and $I_{\rm rad}$ in equation
(\ref{eq:kernel:rot_gen_b}), it is necessary
to determine the thickness of the convective and radiative zones. In addition, an estimate of the radius of each star is needed to evaluate the surface rotation rate based on the $v \sin i$ measurements. All of these quantities are obtained by computing stellar models that simultaneously match non-seismic observables ($T_{\rm eff}$, $\log g$ and $[\rm Fe/\rm H]$)  and seismic observables (mode frequencies). The used constraints, their source and our results are summarised in Table \ref{tab:1}.

The best fitting models have been
found using the `astero' module of
the Modules for Experiments in Stellar Astrophysics (MESA) evolutionary
code \citep{paxton2011,paxton2013}.
Stellar models have been
calculated assuming a fixed mixing-length parameter $\alpha_{\rm MLT}=2.0$.
We have used
OPAL opacities from \cite{Iglesias1996} and
the solar composition (for the mixture of heavy elements)
from \cite{Asplund2009}. The initial hydrogen abundance has been
fixed to $X_{0}=0.7$. No diffusion has been taken into account.
Nuclear reactions have been
set to include standard hydrogen and
helium burning; the pp-chain and the CNO cycle in addition to the triple alpha reaction, which is the default setup of the MESA `astero' module. We have utilised
the NACRE compilation of nuclear reaction rates
\citep{Angulo1999} with the updates for
$^{14}$N(p, $\gamma$)$^{15}$O and $^{12}$C($\alpha$, $\gamma$)$^{16}$O reactions \citep{Kunz2002,Formicola2004}.

We have adopted
the M. Schwarzschild treatment to define the boundary between the
convective and radiative zones. Note that we do not take account of the convective overshooting,
which we understand modifies
the age of the stars, the position of the Base of the Convective Zone (BCZ) and the stellar radius to some extent. In order to assess the importance of this effect, we have computed models
of HD 49933 (index 21),
 HD 181420 (index 22),
 {\it Kepler}-25 (index 19), and
 HAT-P-7 (index 20)
 including the overshoot. These models have shown
that by neglecting the overshoot, the BCZ and the radius are changed by only a few per cent, at a level that does not significantly influence the inferred internal rotation rate $\langle f_{\rm rad}\rangle$. However, it can
affect the age (and hence the central hydrogen abundance)
by $25$ per cent typically.

\begin{figure}
  \begin{center}
   \includegraphics[width=9.0cm, angle=0]{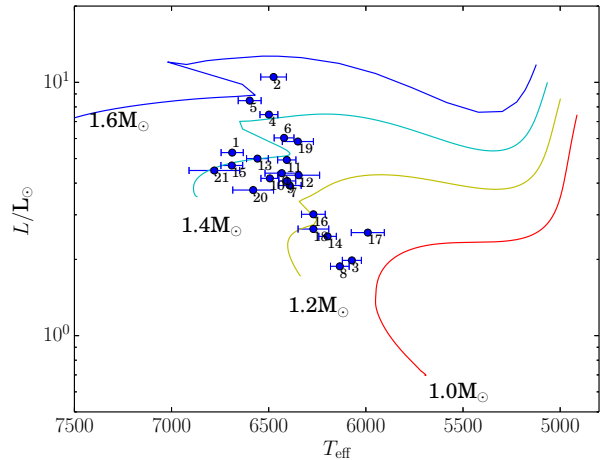}
  \end{center}
 \caption{ Best fitting models of the 22 analysed stars in the HR diagram. The numbers identify each of stars and correspond to those in
the index column of Tables A1 and A2. These models have been
 computed with the MESA code so that they match both the non-seismic observables ($T_{\rm eff}$, $\log g$ and $[\rm Fe/\rm H]$)  and the seismic observables (mode frequencies) simultaneously. Evolutionary tracks are for the solar metallicity.}  
\label{fig:HRdiag}
\end{figure} 

Eigenfrequencies have been
calculated using {\tt
ADIPLS} \citep{JCD2008b}, which assumes adiabaticity.
We have applied
surface effect corrections to the
frequencies, following the method of \cite{Kjeldsen2008}.  The search
for the best fitting model has involved
a simplex minimisation approach
\citep{Simplex} using the $\chi^2$ criteria, where the mass $M$, metallicity [Fe/H] and age are the free parameters of the fit. 
%
\section{Results} \label{sec:results}
  The sensitivity of the average rotational splitting to the rotation rate within the radiative zone is estimated to be $\langle I_{\rm rad} \rangle \,= 32$--$64$ per cent
  (the full list is given in Table \ref{tab:1} in appendix \ref{sec:appendix}). This range is consistent with what is reported in Figs \ref{fig:krn-l1}(b) and \ref{fig:krn-l2}(b). Using the uncertainty in
  the radius, we estimate the error in $\langle I_{\rm rad}\rangle$ to be
  approximately $4$ per cent. This is used to calculate the total uncertainty in
  the average rotation rate of the radiative zone.
  
  Fig.\ \ref{fig:HRdiag} shows the HR diagram of our samples (blue dots). The numbers identify the stars and correspond to
  those in the index column of Tables \ref{tab:1} and \ref{tab:rot}.
  Luminosities of these models are determined so that the models fit both the non-seismic and seismic observables,
  while the effective temperatures $T_{\rm eff}$ are those from spectroscopy.  The coloured lines
  are evolutionary tracks with the adopted physics and for
  the
  solar metallicity. Our sample contains predominantly F stars --- some of which have been extensively analysed for activity variations and magnetic cycles by \cite{Mathur2014} ---, with masses spanning from $\simeq1.0\,{\rm M}_{\sun}$ to
  $\simeq1.6\,{\rm M}_{\sun}$, an effective temperature $5990\,{\rm K}<T_{\rm eff}<6690$\,K and a metallicity similar to the Sun.
  Table \ref{tab:1} 
  provides more detailed information about the best fitting models of each star.
  
\subsection{Mode identification} \label{modeid}
A fundamental step in asteroseismology
is to fix the indices $n$ (radial order), $l$ (degree) and $m$ (azimuthal order) of each observed mode.
This is generally called mode identification.
Eigenmodes of cool G type stars among our sample can easily be
identified (at least for $l$)
by visual inspection of the so-called \'echelle diagram \citep{Grec1983}. Furthermore,
the relative height of each low-degree mode in the power spectrum
(mode visibility $V_l$ of degrees $l$, defined by $H_{n,l}=V^2_l \, H_{n, l=0}$)
 is mostly due to a geometrical
 effect,
 and therefore is nearly independent of
 the fundamental properties of the stars \citep{Ballot2011}. This property is also used to identify modes in the star's power spectrum.

 However, in some cases, particularly for relatively hot
 stars, eigenmodes of the same parity ({\it e.g.} $l=0$ and $l=2$) are hard to disentangle from each other
 by visual inspection, mostly because the mode width at half maximum [see equation (\ref{eq:power}) for its definition] is of the same order as
the frequency spacing between the modes. In such a case, the most likely mode identification relies on a statistical criteria such as the Bayes factor, between competitive solutions\footnote{There are actually only two possible solutions in the case of main sequence solar-like stars. Readers are recommended to refer to the papers cited above for additional explanations.} \citep{Benomar2009, Benomar2009b, Appourchaux2012, Corsaro2014}. This problem of mode
identification is actually recurrent in F stars and was first encountered in the CoRoT star HD 49933 \citep{Appourchaux2008}. In this paper, the mode identification of {\it Kepler} stars is based on \cite{Appourchaux2012},
which used
the methods discussed above.
As for HD 49933 (index 21)
and HD 181420 (index 22),
the identification relied on the Bayes factor.
It is the same as in \cite{Benomar2009b}
for HD 49933, and corresponds to Scenario 1 of \cite{Barban2009}, for HD 181420. These identifications are also found as the most likely by \cite{White2012}, using an independent approach.

Interestingly, an incorrect mode identification of $l$
will often lead
to a disagreement between the spectroscopic $v\,\sin i$ and the seismic $v_{\rm seis}\,\sin i=2\upi R\, f_{\rm seis} \sin i$ (see Table \ref{tab:rot}). Therefore, stars with indications of important disagreement should be carefully discussed. Here, we check that the mode identification is correct for
KIC 9139163 (index 10) \red{which is} 
also discussed in section \ref{difrotcandidates}.
We note that the modes of the
same parity are disentangled and visible in the \'echelle diagram, with the expected mode visibilities. 
We therefore conclude that our current mode identification of $l$
is correct for \red{this star.} 

As for KIC 3424541 (index 3), the mode identification
reported in \cite{Appourchaux2012} leads to the seismic $v_{\rm seis}\,\sin i=12.1 \pm 2.3\, {\rm km}\,{\rm s}^{-1}$, which is incompatible with the spectroscopic $v\,\sin i=29.8 \,{\rm km}\,{\rm s}^{-1}$ from \cite{Bruntt2012}. In order to check the mode identification, we have used the latest {\it Kepler} data (Q5 to Q17), computed the Bayes factor of the two possible solutions and found that it is not in favour of the identification of  \cite{Appourchaux2012}.
Note that \red{due to its relatively fast rotation} this star \red{has a rotational splitting of the same order as the frequency spacing between modes of degree $l=2$ and $l=0$, while the stellar inclination is of $\approx 90\degr$. In such a configuration, the mode identification is not obvious because the azimuthal order $m=-1$ of the degree $l=1$ can easily be falsely identified as an $l=2$ multiplet, while the $m=+1$ could be falsely identified as an $l=0$.}  \red{Note also that \cite{Appourchaux2012}} used a Bayesian framework, but with a shorter dataset.
The new identification, which has permuted $l=1$ and the pairs of $l=0, 2$,
leads to $v_{\rm seis}\,\sin i=26.89^{+0.80}_{-0.84} \, {\rm km}\,{\rm s}^{-1}$. \red{This new result is in agreement} with the spectroscopic $v\,\sin i$.
\red{Additional information regarding the analysis of this star is available in appendix \ref{sec:appendix-342}.}

\subsection{Probability density functions}
\begin{figure}
 	\subfigure[KIC 7206837]{
 \includegraphics[angle=0,width=7.5cm]{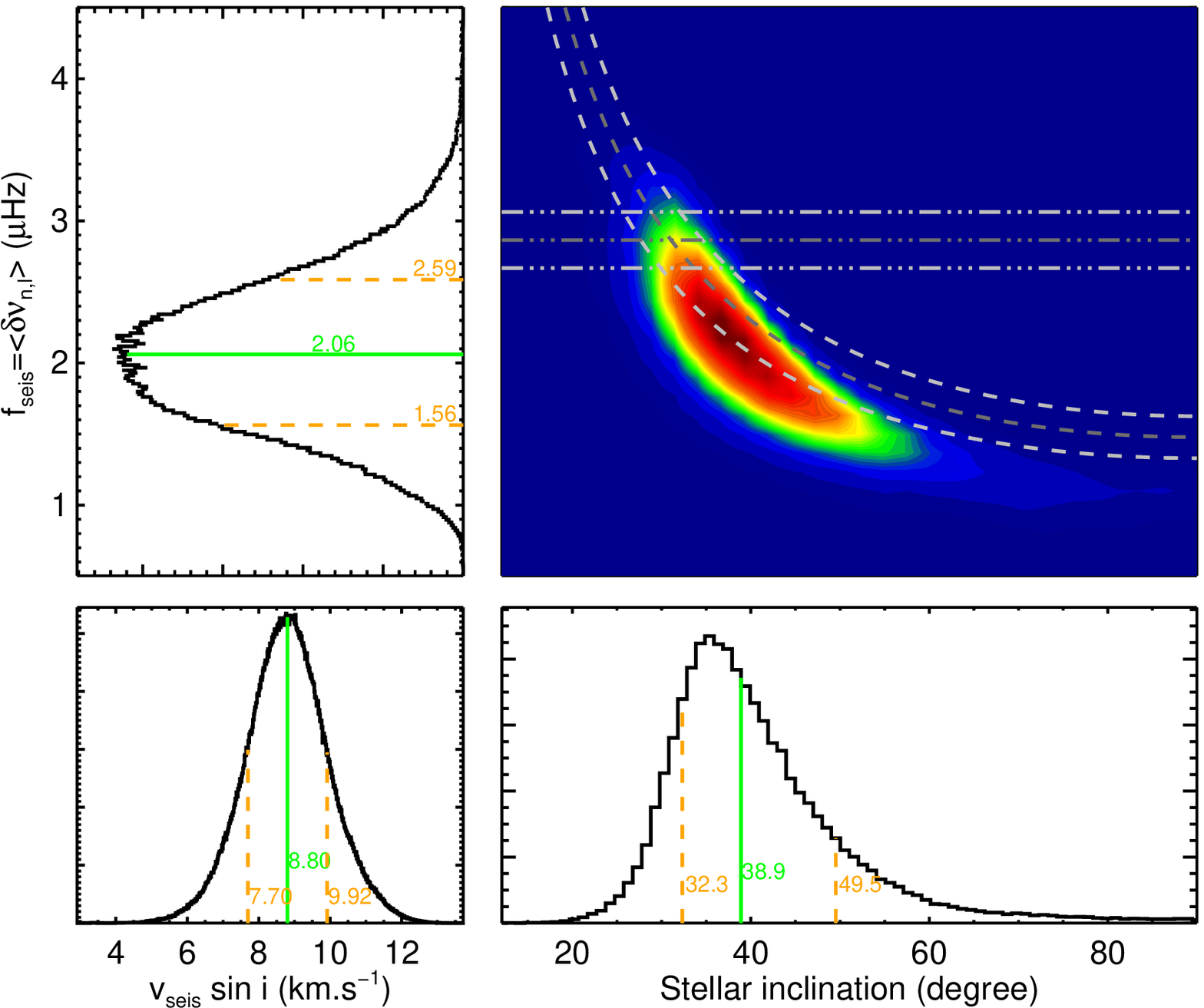}%
		}
	\subfigure[KIC 9139163]{
 \includegraphics[angle=0,width=7.5cm]{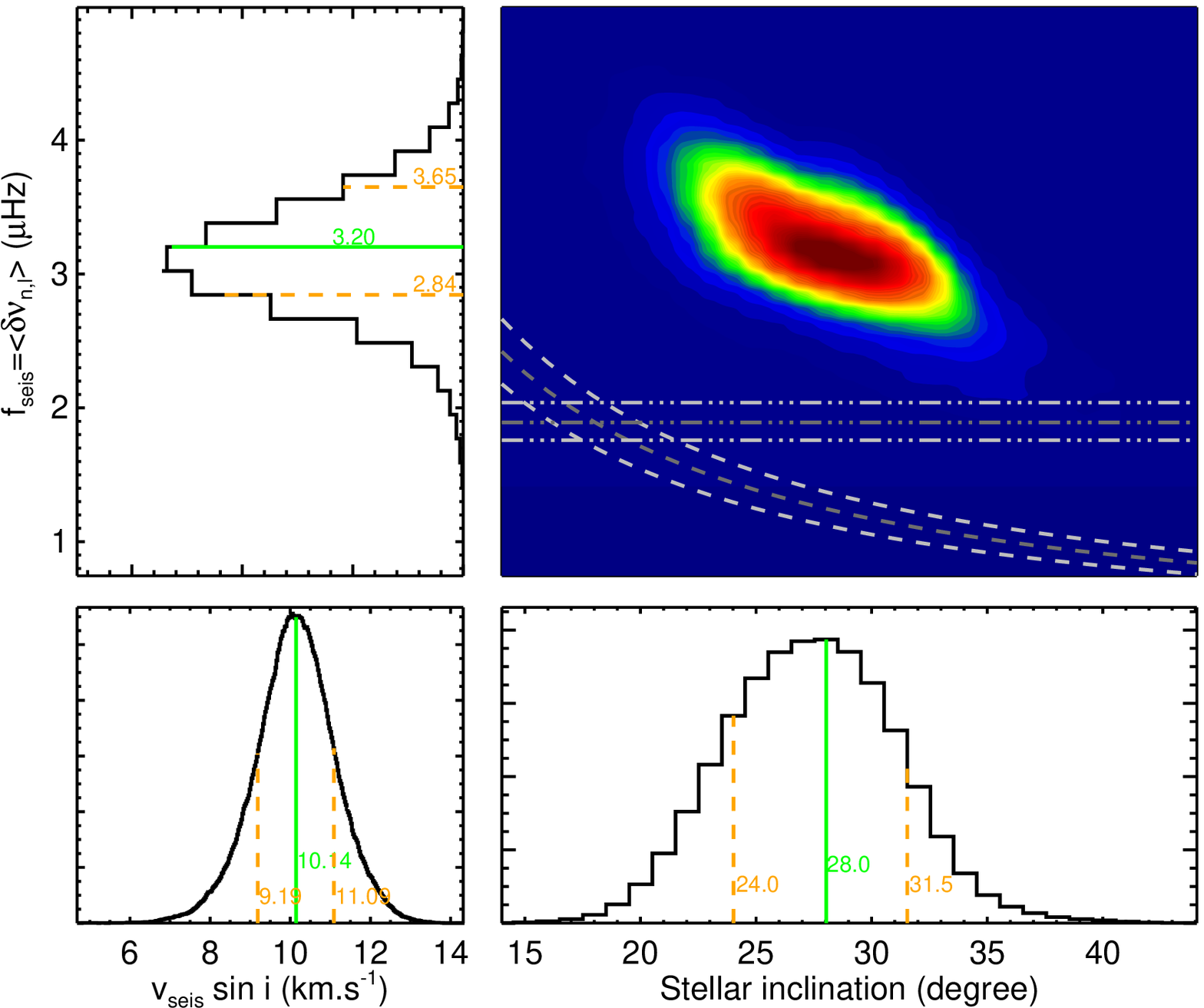}%
		}
 \caption{The colour maps show the correlation between the stellar internal rotation rates measured by the rotational splitting and the stellar inclination angles of three stars, on which 
 superimposed are
 the surface rotation rates derived from the stellar spot (horizontal black dash-dotted lines) and the spectroscopic $v\sin i$ (curved dotted lines). The confidence intervals are shown in grey. In case (a), the surface and internal
 rotation rates
 as well as the stellar inclination angle
 are all consistent with each other, which
 indicates a near solid-body rotation.
 In case (b), although the \red{indicator} of the surface rotation \red{is} consistent with the seismically derived stellar inclination,  \red{it is} in disagreement with the internal rotation rate.}
\label{fig:corrmap}
\end{figure}

Fig.\ \ref{fig:corrmap} shows the joint-probability density functions of the stellar inclination and the rotational splitting, derived from asteroseismology for
KIC 7206837 (index 8),
KIC 9139163 (index 10) and
KIC 9206432 (index 11).
We also estimate the projected equatorial radial velocity from this diagram by assuming uniform rotation; 
$v_{\rm seis}\,\sin i=2\upi R\, f_{\rm seis} \sin i$.
It also shows the surface rotation rates derived from two different kinds of observations as a
function of the stellar inclination $i$;  one from the lightcurve modulation due to spots, and the other from the rotational broadening of spectroscopic absorption lines.
KIC 7206837 is an example of good match between spectroscopic $v\,\sin i$, $P_{\rm rot}$, $f_{\rm seis}$ and $i$, while KIC 9139163 and KIC 9206432
are examples of mismatch among
 the observables (candidates for significant radial differential rotation).
 The joint-probability density functions for all of
 the 22 stars are shown
 in Fig. \ref{fig:jointpdf} in appendix \ref{sec:appendix}. 

\subsection{Surface rotation} \label{describe:surfrot}
As the surface rotation rate $f^{(1)}_{\rm surf}$ depends on $v\,\sin i$, the radius $R$ and the stellar inclination $i$, it is useful to assess their respective contribution to the total uncertainty of $f^{(1)}_{\rm surf}$.
We estimate
the uncertainties of the radius, $v \sin i$ and the inclination $i$
to be $\delta R/R \simeq 1$ per cent, $\delta v\,\sin i/v\,\sin i  \la 10$ per cent and $\delta \sin i/\sin i \simeq 16$ per cent, respectively (see Table \ref{tab:rot} in appendix \ref{sec:appendix}). 
 Thus, the uncertainty of the stellar inclination is the dominant error term in $f^{(1)}_{\rm surf}$, while the uncertainty in $v\,\sin i$ is the second most important term. 
 Meanwhile, as for our estimates of stellar radii listed in Table\,\ref{tab:1}, the results of $91$ per cent of the stars are compatible within $2\sigma$ with the estimates made by \cite{Metcalfe2014}\footnote{They modelled all {\it Kepler} stars of our sample except {\it Kepler}-25 and HAT-P-7, by taking account of more physical effects than our models, such as the overshooting and the diffusion.}.
This fraction drops to $72$ per cent when comparing with stellar radii derived from the scaling relation that is calibrated by the Sun \citep{Kjeldsen1995}.
Still, compatibility at $3\sigma$ is kept%
--- see further discussions in \cite{Mosser2013} ---. Even if
we consider a $3\sigma$ uncertainty in the radius, the main error sources remain
  the stellar inclination and $v\,\sin i$.

\begin{figure}
  \begin{center}
   \includegraphics[angle=0,width=8cm]{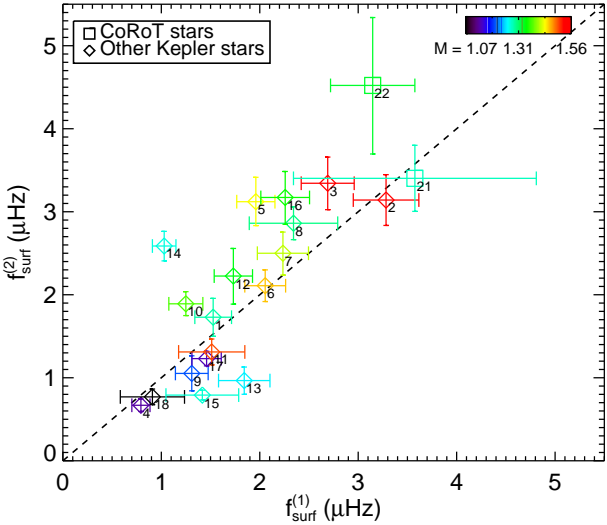}
  \end{center}
 \caption{ Comparison of the surface rotation rates estimated from (1) the rotational broadening of spectroscopic absorption lines, $f_{\rm surf}^{(1)}$, and (2) the lightcurve modulation due to spots, $f_{\rm surf}^{(2)}$. Colours show the mass of stars, while the black dotted line shows a one-to-one relation.
 The indices for the stars correspond to those in
 the first columns of Tables \ref{tab:1} and \ref{tab:rot}.
 }
\label{fig:Os1vsOs2}
\end{figure} 

\begin{figure}
  \begin{center}
   \includegraphics[angle=0,width=8cm]{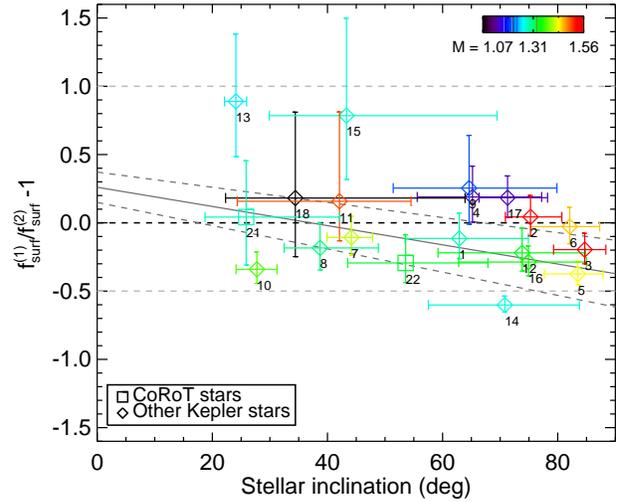}
  \end{center}
 \caption{ Ratio of the surface rotation rates deduced from two different kinds of observational data (one from the rotational broadening of spectral lines and the other from the lightcurve modulation due to spots) as a function of the stellar inclination.
The indices for the stars correspond to those in the first columns of Tables \ref{tab:1} and \ref{tab:rot}. \red{The linear fit that includes all stars --case (a)-- is shown in dark gray. The best fit is the solid line, uncertainties at $1\sigma$ are the dotted lines.}
 }
\label{fig:Is-vsini-Prot}
\end{figure} 

Fig.\ \ref{fig:Os1vsOs2}
compares the two independent estimates of the surface rotation rates,
$f^{(1)}_{\rm surf}$ and $f^{(2)}_{\rm surf}$,
which have been
derived from the spectroscopic $v\,\sin i$ and  
  the lightcurve modulation due to surface structures, respectively.
  Although they are overall
  consistent
  with each other, 
  some stars deviate from a one-to-one relation by approximately a factor of two. This could be (1) due to
systematic errors in
the spectroscopic $v\,\sin i$, (2) because multiple spot
clusters appear at the same time at the surface of the stars,
whereas the spot method is accurate only if one main structure exists, or
 (3) due to
 an intrinsic difference between the two estimates that originates from
 a latitudinal differential rotation.
 When small, the spectroscopic $v\,\sin i$ can be poorly estimated because of the growing contribution of the microturbulence and macroturbulence relative to the actual rotational broadening of spectral lines. Microturbulence is of the order of one kilometre
per second \cite[e.g.][]{Bruntt2009, Doyle2014}, {\it i.e.} significantly smaller than our $v\,\sin i$ and is not expected to be a major source of bias. However, macroturbulence increases with the effective temperature \citep{Valenti2005}, and reaches $4$--$8 \,{\rm km}\,{\rm s}^{-1}$  in the case of F stars. As macroturbulence broadens the spectral lines, neglecting it overestimates $v\,\sin i$. This could explain that Fig.\ \ref{fig:Os1vsOs2} shows a group of stars, such as KIC 10162436 (index 13) and KIC 10454113 (index 15), that lie below a one-to-one relation, but cannot explain the population of stars with $f^{(2)}_{\rm surf} \gg f^{(1)}_{\rm surf}$ such as KIC 6508366 (index 5) and KIC 10355856 (index 14). However, we remark that these stars are seen from latitude near the equator ($60{\degr}<i<90{\degr}$), where most stellar spots are expected to emerge. Therefore, the important discrepancy between $f^{(1)}_{\rm surf}$ and $f^{(2)}_{\rm surf}$ could be due to the presence of multiple spots.  

A latitudinal differential rotation may also lead to discrepancies between $f^{(1)}_{\rm surf}$ and $f^{(2)}_{\rm surf}$. To estimate this, we
looked at the ratio  $f^{(1)}_{\rm surf}/ f^{(2)}_{\rm surf}$ as a function of the stellar inclination $i$ (Fig.\ \ref{fig:Is-vsini-Prot}), \red{which exhibits a trend. To assess its significance, we fitted a linear function $f_{\rm surf}^{(1)}/f_{\rm surf}^{(2)}= a + b i$, in three different situations,}
\begin{enumerate}
	\item[(a)] \red{with all observed stars. We obtain $a=0.26 \pm 0.12$ and $b=0.0015\pm 0.008 $,}
	\item[(b)] \red{excluding the star which shows the most extreme difference (star 14). We obtain $a=-0.085 \pm 0.15$ and $b=0.001\pm 0.003 $,}
	\item[(c)] \red{excluding the three most extreme stars (star 13,14,15). We obtain $a=-0.18 \pm 0.15$ and $b= 0.000 \pm 0.003 $.}
\end{enumerate}
\red{While case (a) indicates a significant slope (see also Fig.\ \ref{fig:Is-vsini-Prot}), case (b) and (c) show that the fit result strongly dependents on a few extreme observations. We conclude that our measures do not strongly indicate evidences of a trend with $i$ that could explain discrepancies described above. Further study on a larger sample,} that better accounts
for the macroturbulence is required before drawing \red{firm conclusions} on the latitudinal differential rotation.
In any case, possible biases in
$v\,\sin i$ are expected to be less important than those in $P_{\rm rot}$, and in the following, we choose to $f^{(1)}_{\rm surf}$ for
the indicator of the surface rotation.

\subsection{Comparison between internal and surface rotation} \label{compare:rot}

\begin{figure}
  \begin{center}
   \includegraphics[width=8cm]{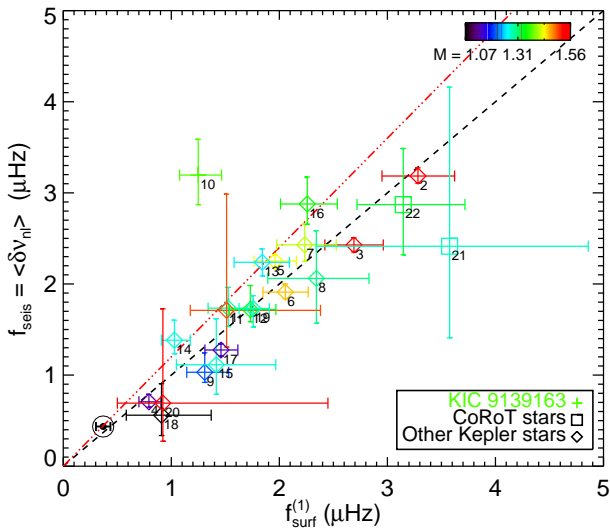}

  \end{center}
 \caption{Rotational splittings versus surface rotation rates measured using the spectroscopic $v\sin i$ for sampled stars.
 The solar symbol near the lower left corner indicates the Sun.
 The numbers attached to the stars correspond to those in the first columns of Tables \ref{tab:1} and \ref{tab:rot}.
 The ratios of these two quantities are mostly close to unity, 
 implying that there is no significant radial differential rotation in most cases.
 The dashed triple-dotted line represents the expected positions of ZAMS models
 that assume local conservation of angular momentum in the radiative layers
 and constant rotation rates in the convection zone.
}
\label{fig:dnuvsosurf}
\end{figure} 

Fig.\,\ref{fig:dnuvsosurf} compares $f^{(1)}_{\rm surf}$ with the seismically obtained rotation rates $f_{\rm seis}$. For reference, the Sun is also shown. The solar
values
of the rotational splitting for $l=1$ and $l=2$ with $14<n<23$ were
taken from \cite{Toutain2001}.
We obtained their average of
$\langle\delta\nu_{nl}\rangle_{\sun}=0.370 \pm 0.007$ $u\mu$Hz. The surface velocity is calculated for the Sun-as-a-star, using the solar $v_{\sun}\,\sin\,i_{\sun} = 1.6 \pm 0.3$ km\,s$^{-1}$ from \cite{Pavlenko2012} and adopting
${\rm R_{\sun}}=6.96 \times 10^5$ km. The ecliptic plane
is inclined from the solar equatorial plane by approximately $7 \degr$,
which means that $\sin i_{\sun}$ can be different from $1$ by
$0.8$ per cent at most. In Fig.\,\ref{fig:dnuvsosurf}, $f^{(1)}_{\rm surf}$ and $f_{\rm seis}$ are in good agreement, which already
suggests that there is not a significant radial differential rotation in most cases.
We thus confirm that
nearly uniform rotation,
which has been well established in the case of the Sun,
extends to the faster rotating main-sequence stars with similar masses.
However, we note that KIC 9139163 (index 10)
depart significantly from the one-to-one relation, indicating that \red{it seems} to have faster-than-surface interior rotation.

In order to discuss the implication of this result,
evolution of the rotation profiles
of $1.2$ and $1.5$ ${\rm M}_{\sun}$ stellar models
with the solar composition
has been calculated
(Saio, private communication) 
from the (pre-main-sequence) Hayashi phase to the
Zero-Age Main-Sequence (ZAMS) stage, with the following assumptions:
(1)
rotation rates are slow enough
to have negligible effects on
the structure, which keeps spherical symmetry,
(2)
total angular momentum is conserved
(no addition to and no removal from the surface layers
are considered),
(3)
the rotation profiles are functions of only radius,
(4)
angular momentum is locally conserved in the radiative zone
and (5)
it is redistributed instantaneously
in the convection zone
to establish
a constant rotation rate
(a limiting case of the efficient transport)
with the total angular momentum of the zone fixed.
Note that point (5) leads to uniform initial rotation profiles,
because the models are fully convective in the Hayashi phase.

As the models evolve towards the ZAMS,
the central mass concentration gets higher, which
results in spin-up of the central part, whereas
the rotation rate in the envelope augments more weakly.
Consequently, the ratio between the central and surface rotation rates
reaches about $2.5$ ($2.7$) for the $1.2$ ($1.5$) ${\rm M}_{\sun}$ model.
Note that this ratio is independent of the initial rotation rate itself.
By substituting the calculated rotation profiles of the ZAMS models
into equation (\ref{eq:kernel}),
we have computed the average rotational splittings
normalised by the surface rotation rate
over all modes with $l=1$, $2$ and $n=10$--$25$.
The resultant ratios are  equal to
$1.22$ and $1.24$ for the $1.2$ and $1.5$ ${\rm M}_{\sun}$ models,
respectively, which suggest that
the ratio increases as a function of mass very mildly
(at least in the range that we discuss in the present analysis).
To take account of the smallest mass in our sample ($\sim 1.0\,{\rm M}_{\sun}$),
a common ratio of $1.2$ is adopted for the ZAMS models
between $1.0$ and $1.6\,{\rm M}_{\sun}$.
Fig.\,\ref{fig:dnuvsosurf} shows
a dashed triple-dotted line with this value of slope.
During the subsequent evolution of the models after ZAMS, the central mass condensation continues, leading to an increase of the ratio as the
models age.
The line thus indicates
the lower limit of the ratio
that main-sequence stars
in the mass range under consideration
would have,
if angular momentum were locally conserved
in the radiative layers
and
fully redistributed
in the convection zone to realise a constant rotation rate.
Therefore, the stars that are found below the line in the figure
provide clear evidence that
angular momentum has been transported
inside the radiative zone
and/or
between the radiative and convective zones
of these stars. With a $1\sigma$ confidence interval,
the number of such stars is equal to
13 (those with indices 2, 3, 4, 6, 8, 9, 12, 15, 17, 18, 19, 21 and 22).
At the $2\sigma$ level\footnote{The calculation has been done using the probability density functions and thus account for the skewness of the distributions.}, it still remains 10 stars (2, 3, 4, 6, 9, 15, 17, 18, 19, 22).\footnote{However, the values of $v\sin i$ of stars with indices 4 and 18 are below $5\,{\rm km}\,{\rm s}^{-1}$, so that the contribution of macroturbulence could possibly be more significant than the estimated uncertainties.}
Because
the same physical process should operate in the same physical conditions,
we presume that
angular momentum is efficiently transported
inside solar-like stars in general.
Note that
the magnetic braking of the surface layers,
which is supposed to be effective
for low-mass stars in our sample
\citep{1967ApJ...150..551K},  is expected to increase the ratio of the rotation rates between the interior and the surface.
The stars that have
$\sim 1.3\,{\rm M}_{\sun}$ or below,
and
found on or above the line on the figure
(those with indices 10, 13, 14 and 16) could reflect a significant effect of this mechanism.

\subsection{Interpretation based on the simple two-layer model}

\begin{figure}
  \begin{center}
   \includegraphics[angle=0,width=8cm]{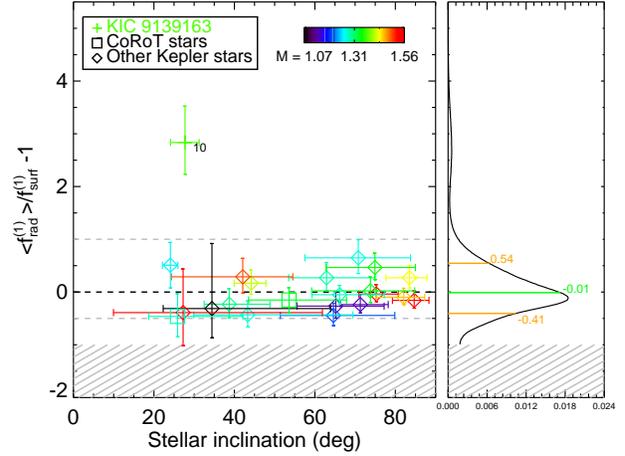}
  \end{center}
\caption{\textbf{Left.} Average radial differential rotation between the radiative zone and the surface as a function of the stellar inclination. Grey lines indicate a differential rotation of a factor of two. The dashed area indicate negative rotation rates. \textbf{Right.} Probability density function for $\langle f^{(1)}_{\rm rad}\rangle/f^{(1)}_{\rm surf} - 1$ of the sampled stars. The median is shown in  green, while orange lines indicate the confidence interval at $1\sigma$.}   
\label{fig:Is-Orad}
\end{figure}

To quantify the level of radial differential rotation in a different way, we evaluated the ratio $\langle f^{(1)}_{\rm rad}\rangle/f^{(1)}_{\rm surf}$
based on equation (\ref{eq:kernel:rot_gen_final}).
Fig.\ \ref{fig:Is-Orad} shows that if we pickup randomly a solar-like star of our sample,  there is $68$ per cent probability that it has a rotation rate in the radiative zone not greater by  \red{54} per cent and not slower by 41 per cent
than the surface. Individually, the radial differential rotation does not exceed a factor of two, except for
KIC 9139163 (index 10).
Although it is in principle
possible to obtain negative rotation rates, the observables do
not provide
us with enough information
to distinguish the direction of rotation.
Therefore, we exclude the possibility of opposite rotation rates (dashed area in Fig.\ \ref{fig:Is-Orad})

\subsection{Particular cases of KIC 9139163 and KIC 9206432} \label{difrotcandidates}
\red{In this section, we further discuss the case KIC 9139163 and KIC 9206432, which were more thoroughly analysed than other stars.}

\subsubsection{KIC 9139163}
\red{Our analysis of KIC 9139163 leads to values of spectroscopic $v\sin i$ which are less than half as small as 
the seismic $v_{\rm seis}\sin i$. The spectroscopy gives $v\,\sin i=4.00 \, {\rm km}\,{\rm s}^{-1}$, while the seismically derived value is $v_{\rm seis}\sin i=10.15 \pm 0.95 \,{\rm km}\,{\rm s}^{-1}$. It does not seem that these meaningful differences can be fully explained by only the known biases in
the spectroscopic $v\,\sin i$
(see section \ref{describe:surfrot}). The surface rotation derived from spots still appears to
indicate an important radial differential rotation, as $f^{(2)}_{\rm rad}/f^{(2)}_{\rm surf} \simeq 2.3$ (instead of $f^{(1)}_{\rm rad}/f^{(1)}_{\rm surf} \simeq 3.8$). In general,  the mode identification of oscillations in F stars is harder than in G stars, and one might think that this could also be a source of discrepancy. However, that latter possibility is ruled out for this star (see section \ref{modeid} for further discussion).}

\red{A careful review of the assumptions in our analysis revealed that the centrifugal distortion may explain the apparent discrepancy. In our analysis and as discussed in section\ \ref{subsec:centrifugal_effects}, this was neglected so that the internal rotation is assumed to produce
equal spacings in each multiplet of the power spectrum.}
\red{However, by comparing the two parameters for a representative frequency of the star, it is found that they can be of the same order. We thus performed a new analysis that includes the centrifugal distortion, in similar fashion as it has been done for {\it Kepler}-410A by \citet{2014ApJ...782...14V}. We found that the centrifugal force effect is negligible and did not modify neither the splitting nor the stellar inclination\footnote{\red{We found $\delta\nu_{nl}=3.16_{-0.51}^{+0.56}$ and $i=23.6_{-2.8}^{+3.1}$.}}. We conclude that second order effects on the splitting are not the cause of the discrepancy between the surface rotation and the internal rotation.}

\red{Mode visibilities $V^2_{l=1}$ and $V^2_{l=2}$ which were considered as free parameters may also modify the results as an important correlation with the splitting may exist. This is because in hot stars, modes of degree $l=0$ and $l=2$ (see section \ref{modeid}), potentially overlap. However, a new analysis that fixes them to the expected visualities $V^2_{l=1}=1.5$ and $V^2_{l=2}=0.5$ did not modify neither the splitting nor the inclination.}



\subsubsection{KIC 9206432}

\red{For this star, the mode identification is not evident from the echelle diagram and requires to rely on the Bayes factor, which is in favour of the mode identification from \cite{Appourchaux2012}. Furthermore, the same identification is obtained using the empirical approach from \cite{White2012}. Note \red{also that the less likely} mode identification gives $i \simeq55{\degr}$ and $f_{\rm seis} \simeq 3\, \umu$Hz, which \red{leads to a great} discrepancy between spectroscopic and seismic $v\,\sin i$.} \red{We thus assume in the following, that correct mode identification is the one already presented in \cite{Appourchaux2012}}. 

\red{The inferred rotation for the most likely mode identification is very sensitive to the fit assumptions. Our analyses revealed that the stellar inclination is relatively low, so that the $m \ne 0$ components of the degree $l=1$ do not have a significant power and do not provide an accurate constraint on the rotational splitting. Furthermore, the modes of degree $l=0$ and $l=2$ overlap significantly, so that these can hardly be disentangled. Without tight constraints on the mode visibilities (i.e. when mode visibilities are free parameters) we noticed that the fit is unable to distinguish the small separation $d_{02}$ from the rotational splitting $\langle\delta\nu_{nl}\rangle$. This leads to visibilities far greater than expected. We found $V^2_{l=1}=2.2 \pm 0.3$ and $V^2_{l=2}=1.3 \pm 0.2$ instead of $V^2_{l=1}=1.5$ and $V^2_{l=2}=0.5$.  It also gives a biased rotational splitting which leads to an apparent disagreement between spectroscopic and seismic, as $v\sin i=6.70 \, {\rm km}\,{\rm s}^{-1}$ while $v_{\rm seis}\sin i=13.66^{+0.93}_{-1.64}\, {\rm km}\,{\rm s}^{-1}$.}

\red{Fixing $V^2_{l=1}=1.5$ and $V^2_{l=2}=0.5$, reduce the degeneracy of solutions and gives a better agreement ($v_{\rm seis}\sin i=8.21^{+1.30}_{-1.17}\, {\rm km}\,{\rm s}^{-1}$).
We also evaluated the influence of $l=3$ modes in our analysis (with all visibilities fixed and $V^2_{l=3}=0.07$) and found an even better agreement ($v_{\rm seis}\sin i=7.77^{+0.99}_{-0.98}\, {\rm km}\,{\rm s}^{-1}$).}

\red{Thus in that particular case, it is necessary to provide strong constraints on the mode visibilities in order to disentangle the power associated to the $l=0$ from its neighbour $l=2$ and consequently, to accurately measure the rotational splitting. Note that in the following, we provide the results of the fit with fixed visibilities and that includes $l=3$ because the calculation of the Bayes ratio indicates that these are statistically significant\footnote{\red{By comparing the model that includes $l=3$ with the one that does not, we found a Bayes ratio of $\approx 70$, which corresponds to a probability of 98\% in favour of the model with $l=3$.}.}}

\section{Discussion}  \label{sec:discussion}
In the following, we further discuss the implications of our results. 

 \subsection{Sample selection}
\begin{figure}
  \begin{center}
   \includegraphics[angle=0,width=7.5cm]{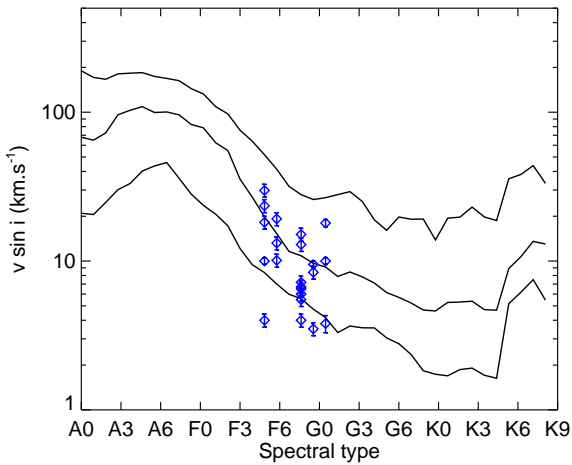}
  \end{center}
\caption{Comparison of the $v\,\sin\,i$  in our sample (blue diamonds) with $v\,\sin\,i$ from the catalog of \protect\cite{Glebocki2005Cat}. Upper and lower lines indicate the $1\sigma$ interval of catalog population, The middle line is the median.}   
\label{fig:vsini}
\end{figure}

The ensemble of 22 stars analysed in this paper are selected
by the following criteria: (1)
they are not known to
be members of close binary systems;
(2) they must show solar-like oscillations; (3) they
must have a surface rotation velocity
of at least a few kilometres per second to have accurate $v\sin i$;
(4) they are preferably hot solar-like stars (F stars), so that they have a thin convective envelope, a condition that improves the accuracy on the measured rotation rate of the radiative zone. While (1) is an important criteria for having reliable spectroscopy, our sample representativity for the ensemble of solar-like population has to be discussed for (2, 3, 4). Among the initial 2000 stars selected for asteroseismic observation (in short-cadence mode) in the {\it Kepler} field, only 500 showed sufficiently low activity to enable the detection of solar-like oscillations \citep{Chaplin2011Sci}. This may indicate that we are biased towards low-activity stars. Furthermore, the least massive and cooler solar-like pulsators are not represented in our sample mostly because of their weak, hard to detect, pulsations. Yet, we hypothesise that because the same physical processes should operate in all solar-like pulsators, our results might be extrapolated to cooler or more active stars.
One may also argue that our sample is not representative in terms of $v\,\sin\,i$. However no evident bias can be seen in Fig. \ref{fig:vsini}, which compares the spectroscopic $v\,\sin\,i$ of our sample (blue symbols) with a catalog of $v\,\sin\,i$ for 28179 stars \citep{Glebocki2005Cat}. In this catalog, we only selected stars of K9 to A0 type for which uncertainties were provided (over 9000 stars).
Thus we think that our conclusion that most stars slightly more massive than the Sun, rotate nearly uniformly is valid for most of solar-like stars. Further analysis, based on a larger and homogeneous sample of stars would help to confirm this.

\subsection{Main results and their implications}
\label{subsec:main_results}
We have
demonstrated that,
for most of the 22 main-sequence stars in our sample, which have
masses between $1.0$ and $1.6\,{\rm M}_{\sun}$,
the surface rotation rate and the average rotation rate
of the nearly whole star (except for the central 15 per cent region in radius)
are almost consistent with each other.
The difference between the two rates
is so small
for 10 stars
that the simple evolutionary models without
transport of angular momentum
inside the radiative zone
and between the radiative and convective zones
are clearly rejected
(see Fig.\,\ref{fig:dnuvsosurf}).
Expecting that the physical mechanism is universal,
it is inferred that
efficient transport of angular momentum
generally occurs in solar-like stars.
Assuming uniform rotation
in each of the convective envelope and the radiative zone,
it has been found for \red{21} out of the 22 stars
that the rotation rates of the two zones
do not differ by more than a factor of two. 

\begin{figure}
  \begin{center}
   \includegraphics[angle=0,width=8cm]{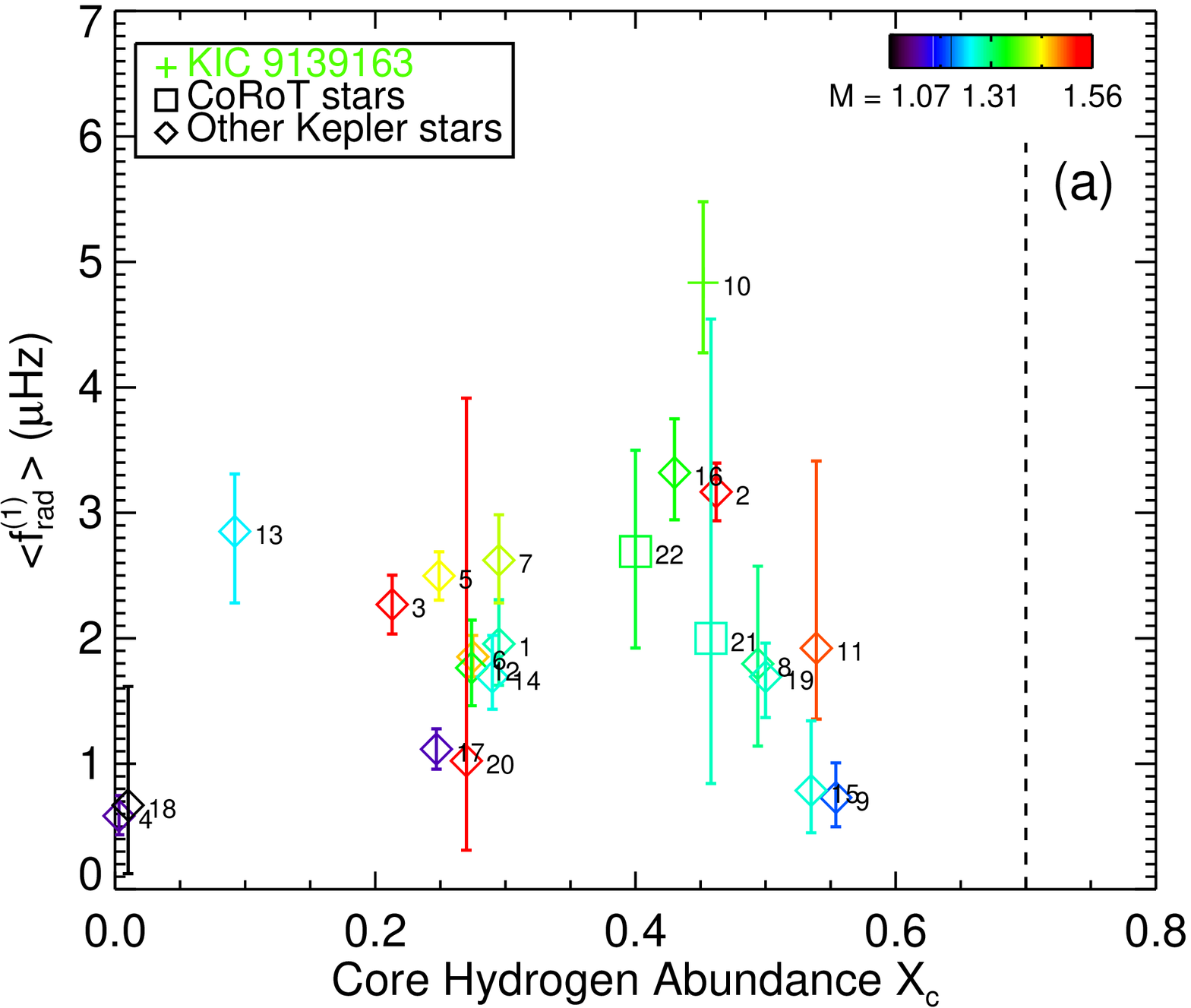}\\
   \includegraphics[angle=0,width=8cm]{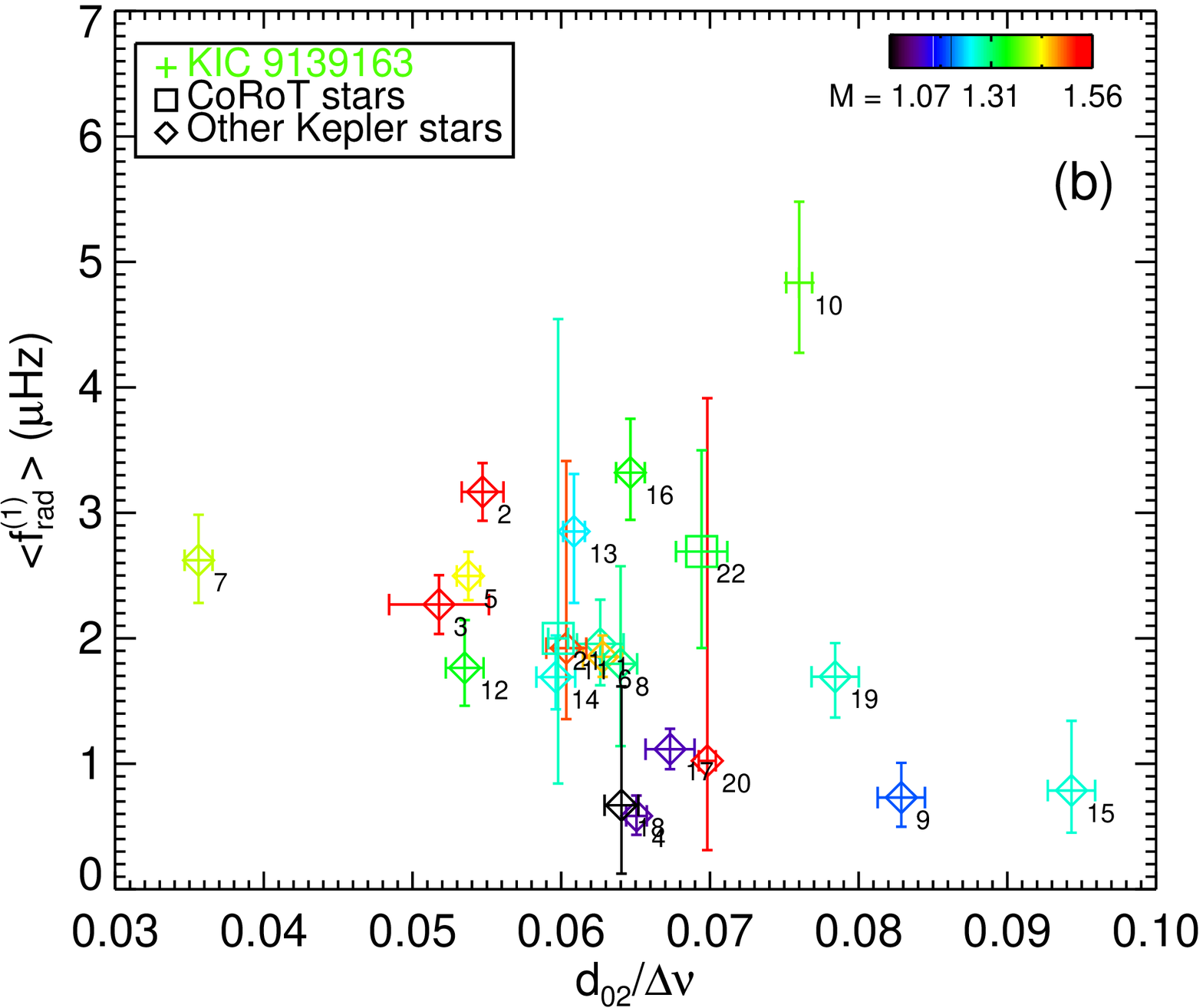}
  \end{center}
 \caption{Internal rotation as a
 function \red{two age indicators: (a) the core hydrogen abundance $X_\mathrm{c}$ derived from stellar modelling and (b) the reduced small separation $d_{02}/\Delta\nu$. In (a),} the vertical dot line indicate the initial hydrogen abundance. Models do not account for overshoot which can modify $X_\mathrm{c}$ by 25 per cent.  Selected stars cover a large part of the main sequence\red{, but no evident slow down of the interior is observed along evolution. The trend seen in (b) indicate a stellar mass-rotation relation. }.  
 }
\label{fig:rot-xc}
\end{figure}

\red{As gyrochronology suggests that younger stars rotate faster than older stars, we attempted to identify signs of decrease of rotation rate in the interior $\langle f^{(1)}_{\rm rad}\rangle$ by using two age indicators, }
\begin{enumerate}
	\item[(a)] \red{the hydrogen central abundance $X_{\rm c}$ from the models.}
	\item[(b)] \red{the seismic observed ratio between the small separation $d_{02}=\nu_{n-1,l=2} - \nu_{n, l=0}$ and the frequency spacing between consecutive modes of same degree $\Delta\nu=\nu_{n,l} - \nu_{n-1,l}$.}
\end{enumerate}
As shown by Fig.\,\ref{fig:rot-xc}\red{(a)}, the
\red{models of the stars have various ages since
the core-hydrogen abundance} $X_{\rm c}$ spans between $\sim0.01$ and $\sim0.54$, leading to $X_{\rm c}/X_{0} \sim 1$ per cent at the lower limit and $X_{\rm c}/X_{0} \sim 78$ per cent at the upper limit.\footnote{$X_{0}=0.7$ is the initial hydrogen abundance (see section \ref{subsec:seism:compare}).} \red{No trend suggesting a change of rotation rate along evolution is visible. 
This indicates that an angular momentum transport mechanism is
efficient from the early main-sequence stage, and possibly
even since the pre-main-sequence phase.}

\red{Along evolution, models indicate that $d_{02}/\Delta\nu$ decreases, which might be associated to a decrease of rotation rate in low-mass stars with a significant magnetic braking. Instead, in Fig. \ref{fig:rot-xc}(b), the rotation appear to increase with $d_{02}/\Delta\nu$. Actually, because this indicator also decreases as the mass increase, the seen trend is mostly due to a mass-rotation relation.}

Interestingly, only \red{one star},
KIC 9139163 (index 10)
 seem\red{s} to have a significantly faster rotation rate
in the interior than the surface. We note that \red{this} star \red{is} among the youngest and massive stars in our sample, which may indicate that
the near solid body rotation has not been established in these stars yet.
The study of a larger sample, with possibly younger stars, may help to better evaluate when the angular momentum transfer mechanism
becomes effective and to derive timescales of such a process.
On the other hand, additional analysis of the current candidate for a large radial differential rotation is also required. For example, new high resolution spectroscopic observations would be useful to confirm measures from \cite{Bruntt2012} and to assess the uncertainties in $v\sin i$.
KIC 9139163 could be an interesting case for the study of the differential rotation because it shows \red{a large} contrast between interior and surface rotation.
\red{However we stress that} due to the low stellar inclination ($\simeq 20 \degr$, which implies that the power of $|m| > 0$ relative to the $m=0$ is only of $\simeq 15 \%$ for $l=1$ modes and $\simeq 35 \%$ for $l=2$ modes), our result could be \red{very} sensitive to the noise realisation and therefore, on the noise background model. This issue has also been discussed by \cite{Corsaro2014} for that star. 
Thus, further studies are certainly required to confirm the unusually \red{fast interior rotation} of KIC 9139163.

Because the mass range of our sample
between $1.0$ and $1.6\,{\rm M}_{\sun}$ is about the same as that of
six subgiants and young red giants analysed by \cite{Deheuvels2014},
the comparison of the two studies allows us to examine the
evolution of internal rotation of stars in this mass range
during the post-main-sequence stage.
Since our analysis
has demonstrated
that these stars
rotate almost uniformly in the main-sequence stage,
the contrast (of about factor 60 at most)
in the rotation rate
between the core and the envelope
detected in the subgiants
and young red giants
should be understood as
the outcome of the competition between the two effects,
the efficient angular momentum transport
and
the expansion of the envelope after the main-sequence phase.
This picture was already suggested by \cite{Deheuvels2012},
with the assumption of uniform rotation at the end of the main-sequence stage.
The present analysis strongly supports their assumption.

\subsection{Implication in the problem of misaligned exoplanet systems}
Some of exoplanet systems show
significant misalignment between the rotation axis of the host star
and the normal of the orbital plane.
It is even claimed that some exoplanets orbit in the opposite direction
to the rotation of the host star \cite[e.g.][]{Xue2014}.
In order to explain this phenomenon,
\cite{Rogers2012} proposed a hypothesis that
the near-surface layers of the host star
rotate in the opposite direction to the rest of the star, being caused by
the angular momentum transport by gravity waves that are generated
at the outer boundary of the convective core.
Since our analysis simply assumes that 
all layers of the stars rotate in the same
direction about a common axis,
the sign change of the rotation rate cannot be measured directly by our analysis. However, if such reversal occurs, it should be still detectable by
the analysis in this paper as a large difference between
the surface rotation rate and the average rotation rate over the large range of the star
(the rotational splitting).
The fact that
most of the stars in our sample,
including those with known exoplanets (HAT-P-7 and {\it Kepler}-25),
rotate almost uniformly
(see Fig.\,\ref{fig:dnuvsosurf})
clearly contradicts such reversal.
Thus, the present analysis
does not support the hypothesis proposed by
\cite{Rogers2012} as
a general picture of F type main-sequence stars.

\subsection{Characteristics of the method}
Our method of measuring the rotation rates of solar-like pulsators
is based on
(1) usage of asteroseismic analysis
and
(2) combination with spectroscopic observations. In the following, these two points are explained within the context of the internal rotation measurements
in asteroseismology. 

The first point is justified by the fact that
the rotational splittings detected in the main-sequence solar-like pulsators
are sensitive to the rotation not only in the convective envelope
but also in the radiative zone.
\cite{Lund2014b}
examined the rotation kernels of high-order and low-degree modes of a solar-mass main-sequence stellar model,
taking account of dependence on not only radius but also co-latitude.
They wrote in section 2.1,
``It is clear for all these modes that the kernels are mainly
sampling the outer parts of the star''.
But this statement should be accepted with great care for general solar-like stars.
While the contribution from
the radiative zone
amounts to about 37 per cent in their model \citep{Lund2014Erratum},
the value can rise to more than 50 per cent
for more massive solar-like stars,
as demonstrated by Figs \ref{fig:krn-l1}(b) and \ref{fig:krn-l2}(b) of this paper.
We understand that this is mainly because the convection envelope is thinner
(in fractional radius)
for more massive stars.
The non-negligible contribution of the radiative interior
means that the rotational splittings observed in solar-like stars also contain a considerable amount of information about the rotation in the radiative zone.

The second point reflects
that it is difficult to distinguish
the rotation rates between the radiative and convective zones
based only on asteroseismic analysis.
This is because the observed eigenmodes of each star
have very similar rotation kernels to each other,
as shown in Figs \ref{fig:krn-l1} and \ref{fig:krn-l2},
and therefore
all of the rotational splittings provide almost the same
weighted average of the internal rotation rates.
This
has been
known for a long time \citep[e.g.][]{2010aste.book.....A} and
confirmed by recent
space-based observations, for which only the average rotational splitting within the range of observed modes is robustly measured \cite[e.g.][]{Appourchaux2008, Ballot2011, Chaplin2013}. Recently, \cite{Nielsen2014} were unsuccessful in detecting variations of the rotational splittings in {\it Kepler} stars. Therefore, an independent source of information is required to
examine radial dependence of the rotation rate.
For this purpose, we have
exploited the surface rotation rate measured by the spectroscopic $v\,\sin i$ combined with the average rotational splitting, allowing us to evaluate
the degree of radial differential rotation.

\section*{ACKNOWLEDGEMENTS}
We thank NASA and CNES, as well as the {\it  Kepler} and CoRoT teams for their revolutionary data.
We are grateful to Enrico Corsaro, Benoit Mosser and Thierry Appourchaux for their useful comments.
We express our gratitude to Hideyuki Saio for calculating
the rotation profiles of evolutionary stellar models.
O.B. is supported by Japan Society for Promotion of Science (JSPS)
Fellowship for Research (No. 25-13316).
M.T. thanks the support
by JSPS KAKENHI Grant Number 26400219.
R.A.G. and T.C. thank the support of the CNES.
\red{We thank the referee for his/her comments that greatly improved the quality of the manuscript.}

\bibliography{mybib} 


\clearpage
\appendix
%
\section{Supplementary contents \red{for the main results}}
\label{sec:appendix}

Provided in this appendix are the following pieces of supplementary information about
the analysis in the main part:
\begin{enumerate}
 \item 
       summary of the best fitting models
       in Table \ref{tab:1}
 \item 
       summary of the rotation rates at the surface and the radiative interior
       in Table \ref{tab:rot}
 \item 
       joint-probability density functions
       in Fig. \ref{fig:jointpdf}
\end{enumerate}

\onecolumn
\begin{landscape}
 \begin{table*}%
  \caption{Parameters used to constrain models (left), those derived from models (middle) and mass/radius derived from the scaling relations (right). Effective temperatures are from \protect\cite{Pinsonneault2012} and the other non-seismic constraints ([Fe/H] and $\log g$)
are from (i) \protect\cite{Bruntt2012}, (ii) \protect\cite{Marcy2014}, (iii) \protect\cite{Pal2008} or (iv) \protect\cite{Bruntt2009}. The uncertainty in $\left\langle I_{\rm rad}\right\rangle$ is about 4 per cent.
  }
\begin{center}
    \begin{tabular}{cc|cccc|cccccc|cc}
\hline
         &      &   \multicolumn{4}{c|}{Constraints}  & \multicolumn{6}{|c|}{Results from models} & \multicolumn{2}{|c}{Results from scal. rel.}\\ 
ind. & Name                 &    Spec. type  & $T_{\rm eff}$             & [Fe/H]  &  $\log g$   &   $L/{\rm L}_{\sun}$   & $M/{\rm M}_{\sun}$           &  $R/{\rm R}_{\sun}$      & $X_{\rm c}$                &  BCZ                        &   $\langle I_{\rm rad}\rangle$                 & $M/{\rm M}_{\sun}$    & $R/{\rm R}_{\sun}$ \\
          &                            &         (MK)                &            (K)                &              &      (cgs)    &                                  &                                            &                                     &                               &  (frac. rad.)           &               $(\%)$                      &                              &               \\
\hline
1 & KIC 1435467${}^{(\rm i)}$ & F8         &  $6433 \pm 86$  & $-0.11 \pm 0.06$  &   $4.09 \pm 0.08$   &  $4.38 \pm 0.08$     & $ 1.294   \pm 0.039   $  & $1.686      \pm 0.020         $  & 0.295    &  $ 0.825                $ &  $    49.6$                              &  $1.47 \pm 0.04$     & $1.75    \pm  0.02 $ \\
2 & KIC 2837475${}^{(\rm i)}$ & F5         &  $6688 \pm 57$  & $-0.02 \pm 0.06$  &   $4.16 \pm 0.08$   &  $5.28 \pm 0.03$     & $ 1.553   \pm 0.028     $  & $1.691      \pm 0.020         $  & 0.462   &  $ 0.902                $ &  $    62.8$                             &  $1.51 \pm 0.03$     & $1.69     \pm  0.01 $\\
3 & KIC 3424541${}^{(\rm i)}$ & G0         &  $6475 \pm 66$  & $0.01 \pm 0.06$    &  $3.82 \pm 0.08$   &  $10.50 \pm 0.20$   & $ 1.556   \pm 0.018     $  & $2.551      \pm 0.026         $    & 0.213   &  $ 0.875                  $ &  $    58.1$                          &  $1.80 \pm 0.05$     & $2.67     \pm  0.02 $ \\
4 & KIC 6116048${}^{(\rm i)}$ & F9        &  $6072 \pm 49$    & $-0.24 \pm 0.06$   & $ 4.28 \pm 0.08$ &   $1.98 \pm 0.02$     & $ 1.134   \pm 0.027     $  & $1.274      \pm 0.020         $    & 0.003   &  $ 0.730                  $ &  $    38.8$                        &  $1.02 \pm 0.02$ & $1.23     \pm  0.01 $ \\
5 & KIC 6508366${}^{(\rm i)}$ & F6         &  $6499 \pm 46 $   & $-0.08 \pm 0.06$   & $3.94 \pm 0.08$  &  $7.46 \pm 0.08$      & $ 1.436   \pm 0.021     $  & $2.142      \pm 0.020         $    & 0.249   &  $ 0.864                  $ &  $    55.9$                         & $1.61 \pm 0.03$ & $2.22     \pm  0.02 $ \\
6 & KIC 6679371${}^{(\rm i)}$ & F5         &  $6598 \pm 59$    & $-0.13 \pm 0.06$   & $3.91 \pm 0.08$  &  $8.47 \pm 0.07$      & $ 1.457   \pm 0.021     $  & $2.167      \pm 0.024         $   & 0.275   &  $ 0.909                 $ &  $    64.2$                        &  $1.70 \pm 0.04$ & $2.30    \pm   0.02 $ \\
7 & KIC 7103006${}^{(\rm i)}$ & F8         &  $6421 \pm 51$    & $0.05  \pm 0.06$   & $4.01 \pm 0.08$  &  $6.04 \pm 0.08$       & $ 1.417   \pm 0.025     $  & $1.941      \pm 0.020         $    & 0.295   &  $ 0.853                  $ &  $    53.8$                        &  $1.57 \pm 0.03$ & $2.00    \pm   0.01 $ \\
8 & KIC 7206837${}^{(\rm i)}$ & F8    &  $6392 \pm 59$   & $0.14 \pm 0.06$    & $4.17 \pm 0.08$  &   $3.92 \pm 0.06$       & $1.307   \pm 0.027     $  & $ 1.562      \pm 0.018         $ & 0.494   &  $ 0.842                  $  &  $ 52.0$                        &  $1.38 \pm 0.03$   & $1.59 \pm   0.01 $ \\
9 & KIC 9139151${}^{(\rm i)}$ & G0.5${}^{(\rm a)}$         &  $6134 \pm 48$     & $0.11 \pm 0.06$    & $4.38 \pm 0.08$  & $1.88 \pm 0.03$       & $ 1.206   \pm 0.077     $  & $1.170      \pm 0.008         $   & 0.554   &  $ 0.780                  $ &  $    44.3$                        &  $1.18 \pm 0.02$ & $1.16    \pm  0.01 $ \\
10 & KIC 9139163${}^{(\rm i)}$ & F8         &  $6405 \pm 44$     & $0.15 \pm 0.06$    & $4.18 \pm 0.08$  & $4.08 \pm 0.03$       & $ 1.378   \pm 0.028     $  & $1.561      \pm 0.018         $   & 0.452   &  $ 0.860                  $ &  $    55.0$                        &  $1.49 \pm 0.03$ & $1.60   \pm   0.01 $ \\
11 & KIC 9206432${}^{(\rm i)}$ & F8         &  $6494 \pm 46$     & $0.23  \pm 0.06$   & $4.23 \pm 0.08$  & $4.18 \pm 0.05$       & $ 1.500   \pm 0.050     $  & $1.551      \pm 0.010         $   & 0.539   &  $ 0.878                  $ &  $    58.3$                        &  $1.57 \pm 0.03$ & $1.59    \pm   0.01 $ \\
12 & KIC 9812850${}^{(\rm i)}$ & F8        &  $6407 \pm 47$     & $-0.16 \pm 0.06$   & $4.05 \pm 0.08$  & $4.94 \pm 0.07$       & $ 1.343   \pm 0.013     $  & $1.809      \pm 0.010         $   & 0.274   &  $ 0.822                  $ &  $    49.1$                           & $1.40 \pm 0.03$ & $1.83   \pm   0.01  $\\
13 & KIC 10162436${}^{(\rm i)}$ & F8         &  $6346 \pm 108$  & $-0.08  \pm 0.06$ & $3.95 \pm 0.08$  &  $4.30 \pm 0.03$       & $ 1.259   \pm 0.015     $  & $1.971      \pm 0.018         $   & 0.092   &  $ 0.735                  $ &  $    38.6$                            &  $1.45 \pm 0.05$ &$ 2.05   \pm    0.02 $ \\
14 & KIC 10355856${}^{(\rm i)}$  & F5        &  $6558 \pm 56$    & $-0.19 \pm 0.06$  & $4.08 \pm 0.08$  &  $5.00 \pm 0.05$       & $ 1.273   \pm 0.016     $  & $1.721      \pm 0.018         $   & 0.290   &  $ 0.862                  $ &  $    55.3$                              &  $1.43 \pm 0.03$ &  $1.79  \pm   0.01 $ \\
15 & KIC 10454113${}^{(\rm i)}$  & F9        & $6197 \pm 47$     & $-0.06 \pm 0.06$  & $4.31 \pm 0.08$  &  $2.47 \pm 0.02$       & $ 1.279   \pm 0.024     $  & $1.285      \pm 0.018         $  & 0.535   &  $ 0.811                  $ &  $    47.9$                             &   $1.22 \pm 0.02$ & $1.27 \pm      0.01 $ \\
16 & KIC 11253226${}^{(\rm i)}$  & F5       &  $6690 \pm 56$   & $-0.08  \pm 0.06$ &  $4.16 \pm 0.08$  & $4.70 \pm 0.07$       & $1.355   \pm 0.016     $  & $1.603      \pm 0.023         $   & 0.430   &  $ 0.893                  $ &  $    60.7$                                    &  $1.69 \pm 0.06$ & $1.73 \pm     0.02 $ \\
17 & KIC 12009504${}^{(\rm i)}$  & F9       &  $6230 \pm 51$    & $-0.09  \pm 0.06$    &  $4.21 \pm 0.08$ & $3.01 \pm 0.03$     & $ 1.140   \pm 0.074     $  & $1.386      \pm 0.008      $  & 0.247  &  $ 0.851              $  & $    52.5$                                           &  $1.28 \pm 0.03$ & $1.45   \pm    0.01 $ \\
18 & KIC 12258514${}^{(\rm i)}$  & G0.5${}^{(\rm a)}$       &  $5990 \pm 85$    & $0.04  \pm 0.06$   & $4.11 \pm 0.08$   &  $2.55 \pm 0.01$      & $ 1.068   \pm 0.034     $  & $1.543      \pm 0.035         $   & 0.010  &  $ 0.661                  $ &  $    32.0$                                          & $1.47 \pm 0.04$ & $1.64  \pm    0.01 $ \\
19 & {\it Kepler}-25${}^{(\rm ii)}$  &  F${}^{(\rm b)}$       &  $6270 \pm 79$     & $-0.04 \pm 0.03$ &  $4.278 \pm 0.03$     & $2.64 \pm 0.06$         & $ 1.285   \pm 0.036 $  & $1.348\pm 0.011 $ & 0.500 &  $ 0.820               $   &  $ 49.0$        &  $1.32 \pm 0.03$  & $1.36   \pm   0.01 $ \\
20 & HAT-P-7${}^{(\rm iii)}$   & F6    &  $6350 \pm 80$        & $0.26 \pm 0.08$   &  $4.070 \pm 0.08$      &  $5.84 \pm 0.05$         & $ 1.540   \pm 0.030 $  & $ 2.000 \pm 0.009 $  &  0.270 & $ 0.870                 $   &  $ 57.4$        & $1.48 \pm 0.03$ & $1.97   \pm   0.01$ \\
21 & HD 49933${}^{(\rm iv)}$ &  F8        &  $6570 \pm 60$  & $-0.44 \pm 0.03$   & $4.28 \pm 0.06$        & $3.76 \pm 0.08$          & $ 1.287   \pm 0.009 $  & $1.463 \pm 0.006 $  &   0.458 & $ 0.880                  $   &  $ 58.3$        &  $1.34 \pm 0.02$  & $1.49 \pm 0.01$\\
22 & HD 181420${}^{(\rm iv)}$ & F6         &  $6580 \pm 105$ & $0.00 \pm 0.06$   & $4.26 \pm 0.06$      & $4.49 \pm 0.13$          & $ 1.335   \pm 0.010 $  & $1.622 \pm 0.007$ &  0.400 & $ 0.877              $   &  $    57.8$ &  $1.61 \pm 0.05$  & $1.73 \pm 0.02$\\ 
\hline
\multicolumn{14}{c}{$(\rm a)$ Assumed as G0 in Fig. \ref{fig:vsini}. $(\rm b)$ Assumed as F5  in Fig. \ref{fig:vsini}.} \\  
 \end{tabular}
  \end{center}
  \label{tab:1}
 \end{table*}
 \end{landscape}

\onecolumn
\begin{landscape}
\begin{table}
 \caption{Rotation rates at the surface of the stars and in the radiative zone. The penultimate column indicates the length of the data set while the last column gives the first and last used {\it Kepler} data quarters.}
\begin{center}
\begin{tabular}{cc|ccc|cc|cc|cc|cc} 
\hline 
index & Name                  & $f_{\rm seis}$  & $i$  &  $v_{\rm seis}\,\sin i$                  & $v\,\sin i$                      &     $P_{\rm rot}$ &    $f^{(1)}_{\rm surf}$          &  $f^{(2)}_{\rm surf}$      &   $\langle f^{(1)}_{\rm rad}\rangle$      &   $\langle f^{(2)}_{\rm rad}\rangle$  & Obs. duration     & Quarters \\
         &                             &($\umu$Hz)      & (${\degr}$) & $({\rm km}\,{\rm s}^{-1})$ & $({\rm km}\,{\rm s}^{-1})$  &      (d)            &     ($\umu$Hz)                     &   ($\umu$Hz)                &             ($\umu$Hz)                             &               ($\umu$Hz)                        &        (d)               &                \\
\hline 
1 & KIC 1435467    &      $  1.73^{ +0.23}_{ -0.20} $ & $    62.90^{+   12.60}_{ - 9.50} $ & $   11.29^{+  0.67}_{ -0.66} $    & $10.00 \pm 1.00$ &  $6.68 \pm 0.89$ & $    1.53_{-   0.18}^{+   0.22} $  &  $    1.73_{-   0.23}^{+   0.23} $  &   $    1.96_{-   0.33}^{+   0.35} $  &  $    1.76_{-  0.47}^{+   0.50} $  & 738   &  Q5-Q12 \\
2 & KIC 2837475    &  $  3.19^{ +   0.08}_{ - 0.09} $ & $    75.40^{+   5.60}_{ - 4.50} $ & $   22.79^{+  0.59}_{ -0.61} $   & $23.50 \pm 2.35$ &  $3.68 \pm 0.36$ & $    3.28_{-   0.33}^{+   0.34} $  &  $    3.14_{-   0.31}^{+   0.31} $  &   $    3.17_{-   0.23}^{+   0.23} $  &  $    3.25_{-   0.23}^{+   0.23} $  & 738   & Q5-Q12\\
3 & KIC 3424541    & $  2.43^{  +  0.08}_{ - 0.08} $ & $    84.90^{+   4.10}_{ - 5.50} $ & $   26.88^{+  0.84}_{ -  0.86} $   & $29.80 \pm 2.98$  & $3.46 \pm 0.33$ & $    2.69_{-   0.27}^{+   0.27} $  &  $    3.34_{-   0.32}^{+   0.32} $  &   $    2.27_{-   0.23}^{+   0.23} $  &  $    1.80_{-   0.27}^{+   0.27} $  & 1148   & Q5-Q17 \\
4 & KIC 6116048  & $ 0.71^{ +   0.08}_{ - 0.06} $ & $    65.30^{+   12.10}_{ -   9.50} $ & $   3.57^{+  0.15}_{ -  0.14} $         & $4.00 \pm 0.40$  &  $17.26 \pm 1.96$ & $    0.79_{-   0.09}^{+   0.11} $  &  $    0.67_{-   0.08}^{+   0.08} $  &  $    0.59_{-   0.15}^{+   0.16} $  &  $    0.78_{-   0.20}^{+   0.24} $  & 738   & Q5-Q12\\
5 & KIC 6508366  & $  2.25^{  +  0.07}_{ - 0.07} $ & $    83.90^{+   4.70}_{ -   6.00} $ & $   20.87^{+  0.59}_{ -  0.61} $        & $18.20 \pm 1.82$ &  $3.70 \pm 0.35$ & $    1.96_{-   0.20}^{+   0.20} $  &  $    3.13_{-   0.29}^{+   0.29} $  &   $    2.50_{-   0.19}^{+   0.19} $  &  $    1.58_{-   0.26}^{+   0.26} $  & 738   & Q5-Q12\\
6 & KIC 6679371  & $  1.91^{+ 0.09}_{ - 0.07} $ & $    82.00^{+   5.50}_{ -   6.50} $ & $   17.88^{+  0.67}_{ -  0.64} $       & $19.20 \pm 1.92$ & $5.48 \pm 0.50$ & $    2.06_{-   0.21}^{+   0.21} $  &  $    2.11_{-   0.19}^{+   0.19} $  &  $    1.85_{-   0.16}^{+   0.17} $  &  $    1.82_{-   0.16}^{+   0.18} $  & 738   & Q5-Q12\\
7 & KIC 7103006   & $  2.43^{+  0.23}_{ -  0.18} $ & $    44.30^{+   4.00}_{ -   4.50} $ & $   14.37^{+  0.73}_{ -  0.79} $       & $13.20 \pm 1.32$ & $4.62 \pm 0.48$ & $    2.23_{-   0.27}^{+   0.29} $  &  $    2.50_{-   0.26}^{+   0.26} $  &   $    2.62_{-   0.34}^{+   0.36} $  &  $    2.39_{-   0.42}^{+   0.48} $  & 738   & Q5-Q12\\
8 & KIC 7206837	& $  2.06^{+  0.53}_{ -  0.50} $ & $    38.90^{+   10.60}_{ -   6.60} $ & $   8.80^{+   1.12}_{ -   1.10} $      & $10.10 \pm 1.02$ & $4.04 \pm 0.28$ & $    2.34_{-   0.45}^{+   0.48} $  &  $    2.86_{-   0.20}^{+   0.20} $ &   $    1.80_{-   0.66}^{+   0.78} $  &  $    1.51_{-   0.89}^{+   1.10} $  & 738   & Q5-Q12\\
9 & KIC 9139151   & $  1.03^{+  0.21}_{ -  0.11} $ & $    65.00^{+   15.00}_{ -   13.60} $ & $   4.79^{+  0.29}_{ -  0.28} $       & $6.00 \pm 0.60$ &  $10.96 \pm 2.22$ & $    1.31_{-   0.17}^{+   0.22} $  &  $    1.05_{-   0.21}^{+   0.21} $  &   $    0.73_{-   0.23}^{+   0.28} $  &  $    1.06_{-   0.40}^{+   0.50} $  & 738   & Q5-Q12\\
10 & KIC 9139163    & $  3.20^{+  0.39}_{ -  0.34} $ & $    28.00^{+   3.50}_{ -   4.00} $ & $   10.15^{+  0.94}_{ -  0.97} $    & $4.00 \pm 0.40$ & $6.10 \pm 0.47$ & $    1.26_{-   0.18}^{+   0.21} $  &  $    1.89_{-   0.19}^{+   0.19} $  &   $    4.83_{-   0.56}^{+   0.65} $  &  $    4.32_{-   0.63}^{+   0.74} $  & 738   & Q5-Q12\\
11 & KIC 9206432   & $  1.71^{+  1.27}_{ -  0.43} $ & $    42.50^{+   12.2}_{ -   18.1} $ & $   7.77^{+  0.99}_{ -   0.98} $       & $6.70 \pm 0.67$ & $8.80 \pm 1.06$      & $    1.51_{-   0.34}^{+   0.87} $  &  $    1.31_{-   0.16}^{+   0.16} $  &    $    1.92_{-   0.57}^{+  1.50} $  &  $    2.02_{-   0.72}^{+   2.20} $  & 738   & Q5-Q12\\
12 & KIC 9812850    & $  1.72^{+  0.26}_{ -  0.14} $ & $    74.70^{+   11.70}_{ -   15.20} $ & $   12.98^{+  0.81}_{ -  0.74} $ & $12.90 \pm 1.29$ &  $5.19 \pm 0.79$   &  $    1.73_{-   0.19}^{+   0.24} $  &   $2.22_{-   0.34}^{+   0.34} $      &    $    1.77_{-   0.30}^{+   0.38} $  &    $1.26_{-   0.48}^{+   0.60} $      & 738   & Q5-Q12\\
13 & KIC 10162436   & $  2.24^{+  0.15}_{ -  0.15} $ & $    24.20^{+   2.10}_{ -   2.50} $ & $   7.81^{+  0.55}_{ -  0.77} $     & $6.50 \pm 0.65$ & $11.96 \pm 2.05$ & $    1.84_{-   0.26}^{+   0.25} $  &  $    0.96_{-   0.17}^{+   0.17} $  &   $    2.85_{-   0.57}^{+   0.46} $  &  $    4.29_{-   0.55}^{+   0.50} $ & 738    & Q5-Q12\\
14 & KIC 10355856    & $  1.38^{+  0.22}_{ -  0.15} $ & $    71.40^{+   13.10}_{ -   13.50} $ & $   9.74^{+  0.82}_{ -  0.80} $    & $7.20 \pm 0.72$ & $4.47 \pm 0.31$   & $    1.03_{-   0.12}^{+   0.15} $  &  $    2.59_{-   0.18}^{+   0.18} $  &  $    1.69_{-   0.26}^{+   0.33} $  &  $    0.49_{-   0.29}^{+   0.48} $  & 462  & Q5-Q9\\
15 & KIC 10454113     & $  1.11^{+  0.51}_{ -  0.32} $ & $    43.50^{+   26.00}_{ -   13.50} $ & $   4.33^{+  0.46}_{ -  0.46} $  & $5.50 \pm 0.55$ & $14.61 \pm 1.09$ & $    1.41_{-   0.37}^{+   0.56} $  &  $    0.79_{-   0.06}^{+   0.06} $  &  $    0.79_{-   0.34}^{+   0.55} $  &  $    1.48_{-   0.68}^{+   1.07} $  & 738   & Q5-Q12\\
16 & KIC 11253226   & $  2.88^{+  0.29}_{ -  0.22} $ & $    75.00^{+   10.50}_{ -   12.00} $ & $   19.28^{+   1.22}_{ -   1.26} $    & $15.1 \pm 1.51$ &  $3.64 \pm 0.37$      & $    2.26_{-   0.25}^{+   0.28} $  &  $3.17_{-   0.32}^{+   0.32} $   &   $    3.32_{-   0.38}^{+   0.43} $  &   $2.72_{-   0.44}^{+   0.53} $    & 276 &  Q5-Q7 \\
17 & KIC 12009504   & $  1.28^{+ 0.06}_{ - 0.07} $ & $    72.10^{+   7.20}_{ -   5.60} $ & $   7.32^{+  0.23}_{ -  0.23} $    & $8.40 \pm 0.84$ & $9.39 \pm 0.68$   & $    1.46_{-   0.15}^{+   0.15} $  &  $    1.23_{-   0.09}^{+   0.09} $  &   $    1.12_{-   0.16}^{+   0.16} $  &  $    1.33_{-   0.15}^{+   0.15} $  & 738   & Q5-Q12\\
18 & KIC 12258514   & $ 0.56^{+  0.35}_{ -  0.22} $ & $    34.50^{+   30.30}_{ -   12.10} $ & $   2.16^{+  0.27}_{ -  0.26} $  & $3.50 \pm 0.35$ & $15.00 \pm 1.84$ & $    0.91_{-   0.33}^{+   0.46} $  &  $    0.77_{-   0.09}^{+   0.09} $  &   $    0.63_{-   0.51}^{+   0.88} $  &  $    1.11_{-   0.78}^{+   1.22} $   & 738   & Q5-Q12\\
19 & {\it Kepler}-25   & $  1.73^{+  0.14}_{ -  0.20} $ & $    67.10^{+   11.20}_{ -   7.60} $ & $   9.18^{+  0.61}_{ -  0.70} $&  $9.50 \pm 0.50$ &            /                 & $    1.76_{-   0.13}^{+   0.15} $  &     /                                            &   $    1.69_{-   0.33}^{+   0.27} $  &      /                                         & 1114   & Q5-Q16 \\
20 & HAT-P-7   & $0.68^{+   1.04}_{ -  0.41} $ & $    27.80^{+   35.40}_{ -   18.20} $ & $   2.95^{+   1.68}_{ -   1.97} $  &    $3.80 \pm 0.50$   &            /                 & $    0.92_{-   0.42}^{+   1.53} $      &        /                                             &   $    1.02_{-   0.71}^{+   2.90} $  &     /                                          & 1437   & Q0-Q16 \\ 
21 & HD 49933    & $  2.42^{+   1.70}_{ -   1.09} $ & $    25.70^{+   16.20}_{ -   7.20} $ & $   7.31^{+   2.45}_{ -   2.21} $&    $ 10.00 \pm 0.50$  &  $3.40 \pm 0.40$${}^{(\rm a)}$     & $    3.58_{-   1.23}^{+   1.28} $  &  $    3.40_{-   0.40}^{+   0.40} $  &   $    2.00_{-   1.16}^{+   2.55} $  &  $    2.54_{-   1.82}^{+   3.02} $  & 180   &  /\\
22 & HD 181420  & $  2.86^{+  0.62}_{ -  0.53} $ & $    53.70^{+   14.60}_{ -   10.10} $ & $   16.31^{+   2.12}_{ -   2.32} $ &    $ 18.00 \pm 1.00$  &  $2.56 \pm 0.57$${}^{(\rm a)}$   & $    3.15_{-   0.43}^{+   0.57} $  &  $    4.50_{-   0.85}^{+   0.85} $ &   $    2.69_{-   0.77}^{+   0.81} $  &  $    1.87_{-   1.05}^{+   1.29} $  & 156   &  /\\
\hline 
\multicolumn{12}{c}{${(\rm a)}$ Using spot modelling, \cite{Mosser2009} provide alternative surface rotation rate for the CoRoT stars. For HD 49933, they found $3.45 \pm 0.05$ days and $2.25 \pm 0.03$ days for HD 181420.}  \\
\end{tabular}
\end{center}
\label{tab:rot}
\end{table}
\end{landscape}

\renewcommand*\thesubfigure{(\arabic{subfigure})}
\begin{figure*}
   \begin{center}
    \hspace*{\fill}%
    \subfigure[KIC 1435467]{%
   \includegraphics[angle=0,width=8cm,height=5.95cm]{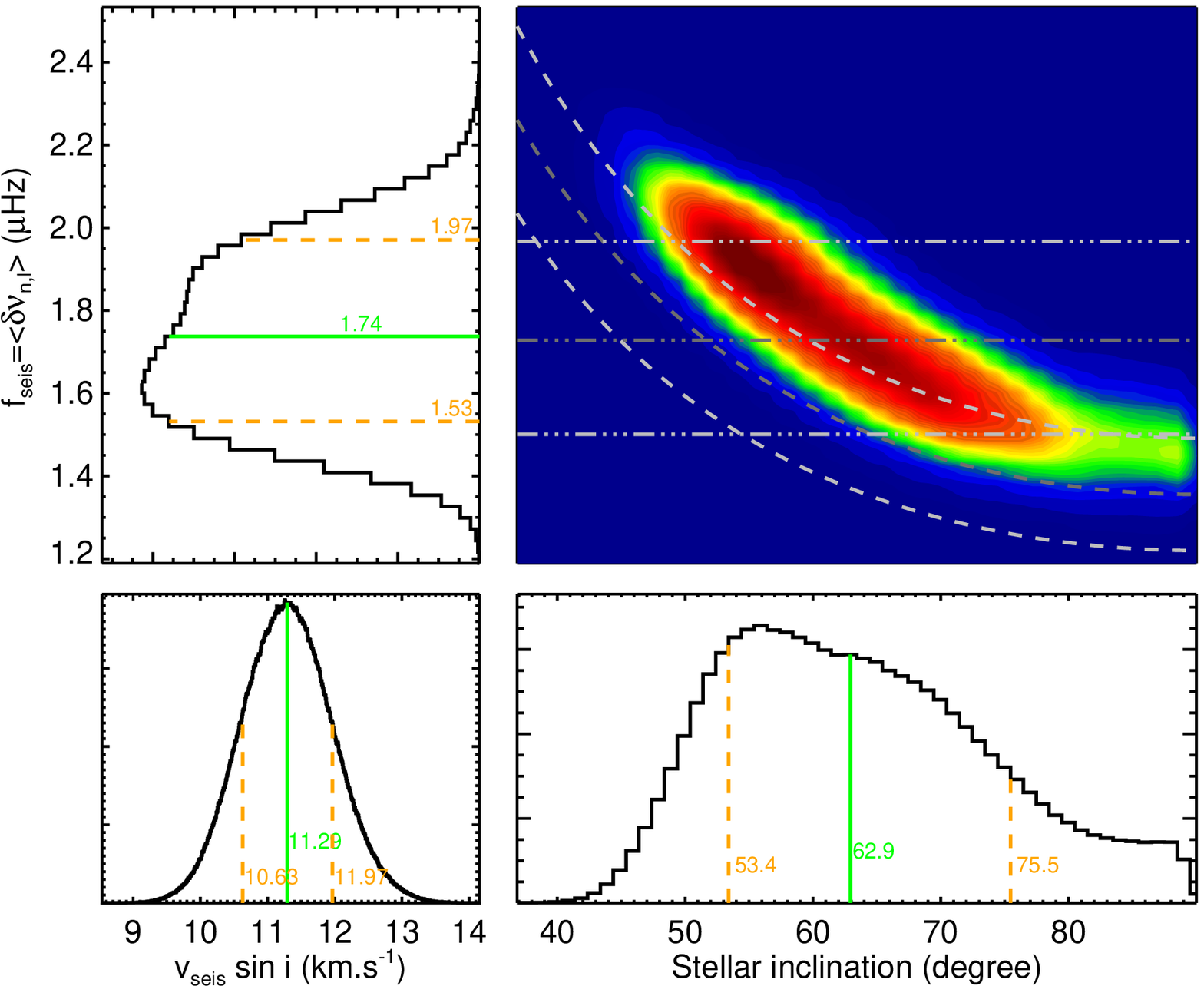} %
	}\hfill%
	\subfigure[KIC 2837475]{%
   \includegraphics[angle=0,width=8cm,height=5.95cm]{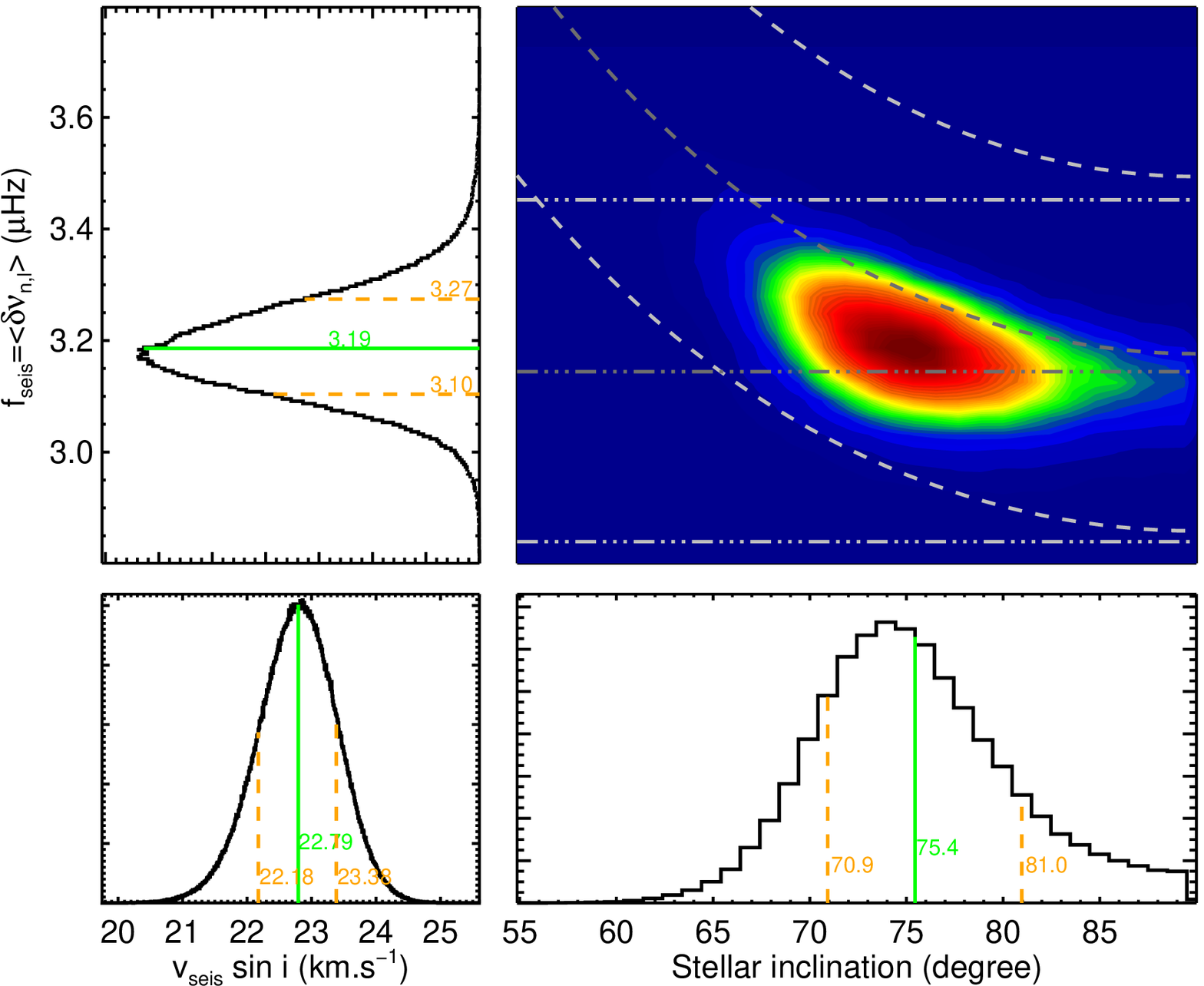}%
    }\hspace*{\fill}
    
    \hspace*{\fill}%
    \subfigure[KIC 3424541]{%
   \includegraphics[angle=0,width=8cm,height=5.95cm]{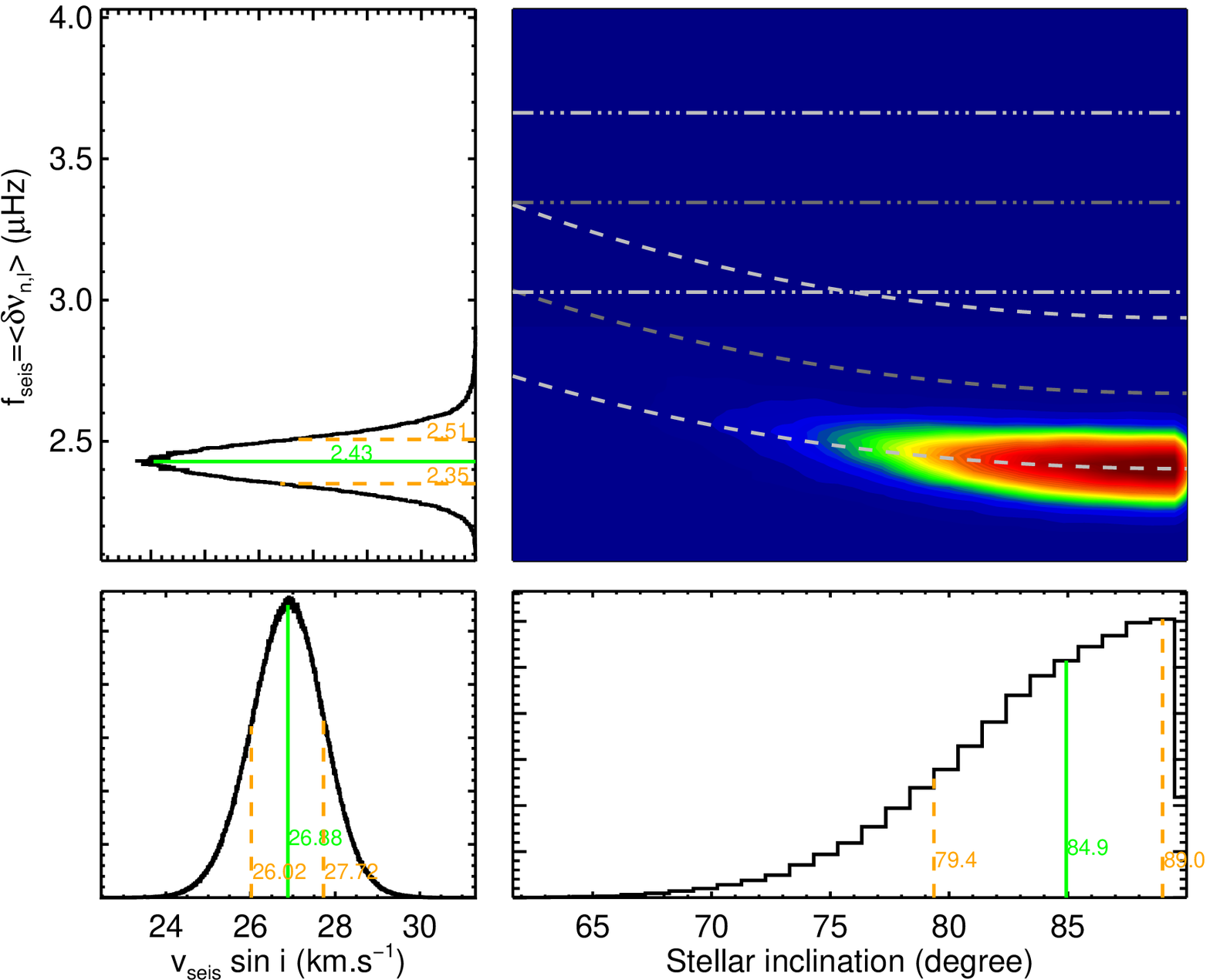}%
		}\hfill%
	\subfigure[KIC 6116048]{%
   \includegraphics[angle=0,width=8cm,height=5.95cm]{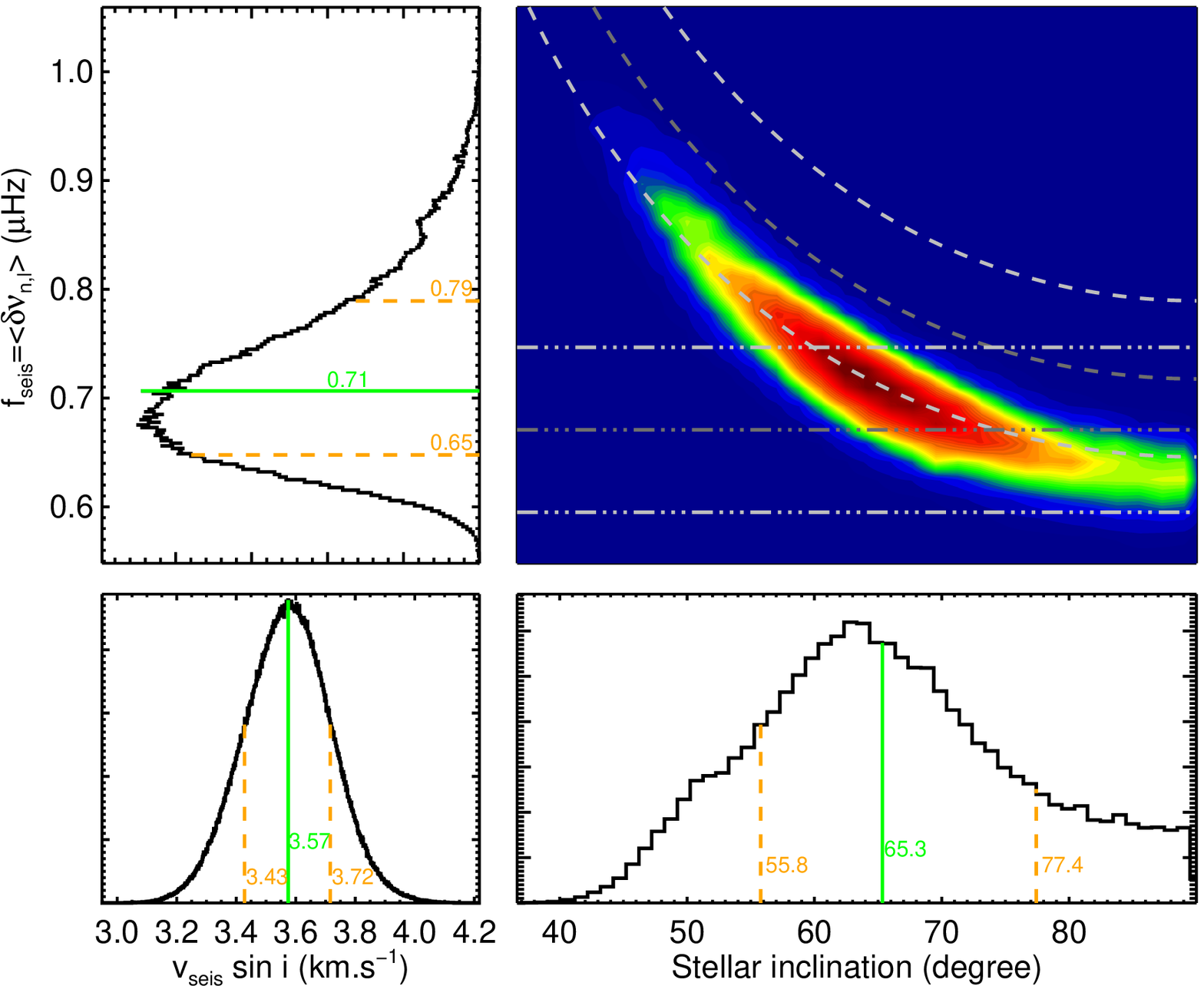}%
		}\hspace*{\fill}%

    \hspace*{\fill}%
    \subfigure[KIC 6508366]{%
   \includegraphics[angle=0,width=8cm,height=5.95cm]{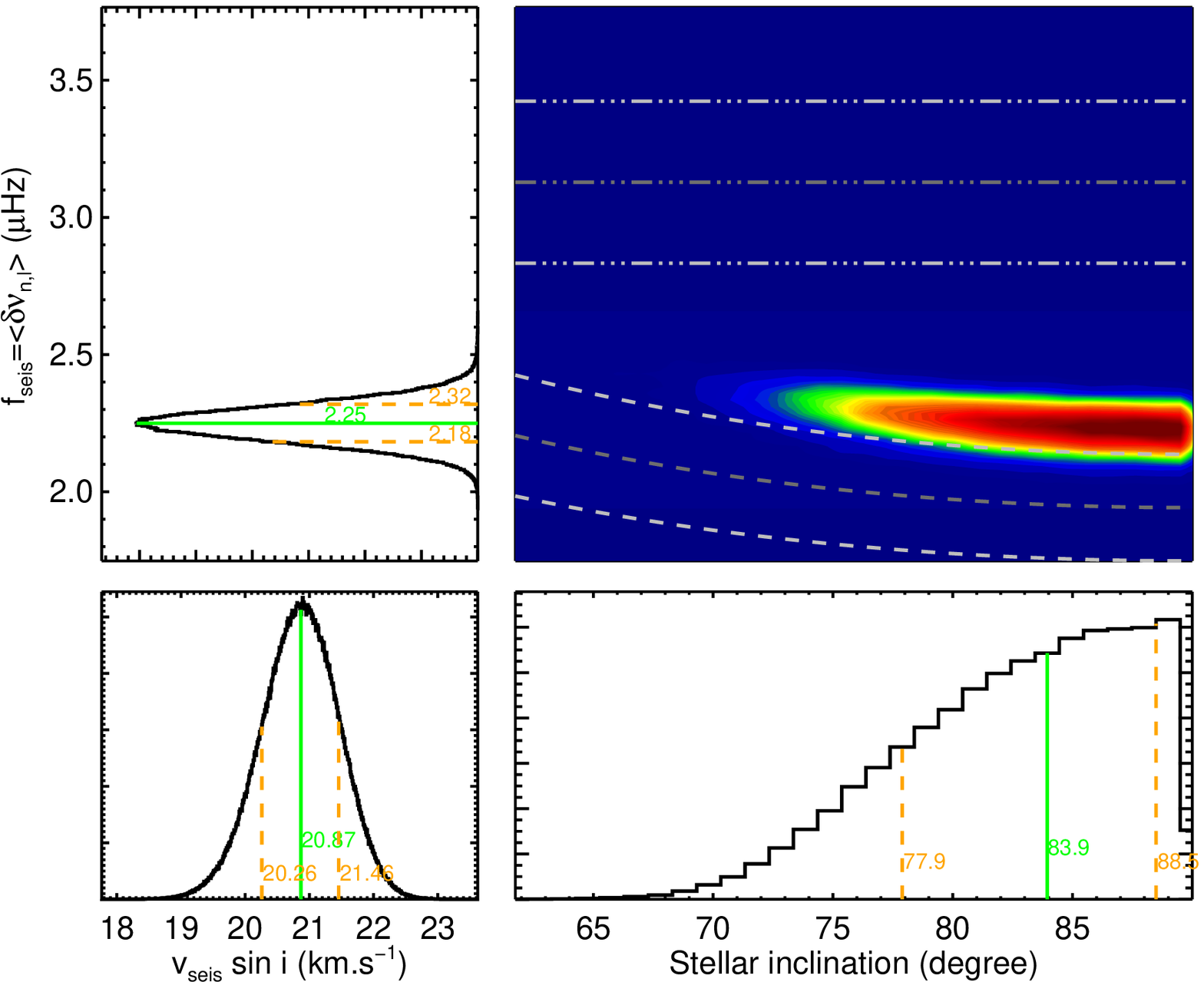}%
		}\hfill%
	\subfigure[KIC 6679371]{%
   \includegraphics[angle=0,width=8cm,height=5.95cm]{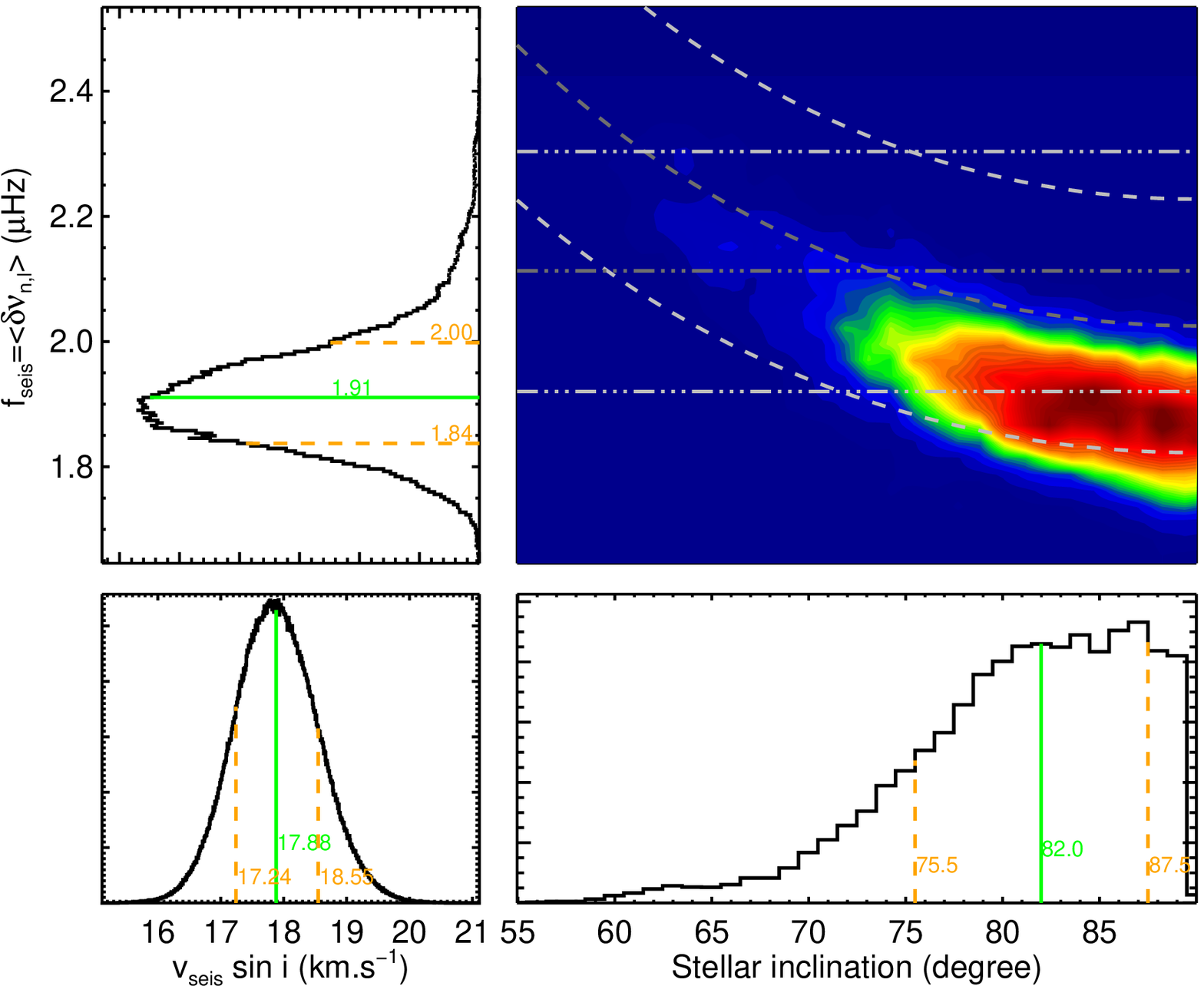}%
    }\hspace*{\fill}%
    
   \end{center}
 \caption{%
 Joint-probability density functions for all of the analysed stars. Horizontal dash-dotted lines indicate the surface rotation rates from lightcurve modulation due to spots (when available), while dashed curves are the rotation rates derived using the spectroscopic $v\sin i$. \red{Unless specified, all spectrum fit include modes of degree $l=0,1,2$.}}  
\label{fig:jointpdf}
\end{figure*} 
\begin{figure*}
  \begin{center}
   \setcounter{subfigure}{6}
   \hspace*{\fill}%
   \subfigure[KIC 7103006]{%
   \includegraphics[angle=0,width=8cm,height=5.95cm]{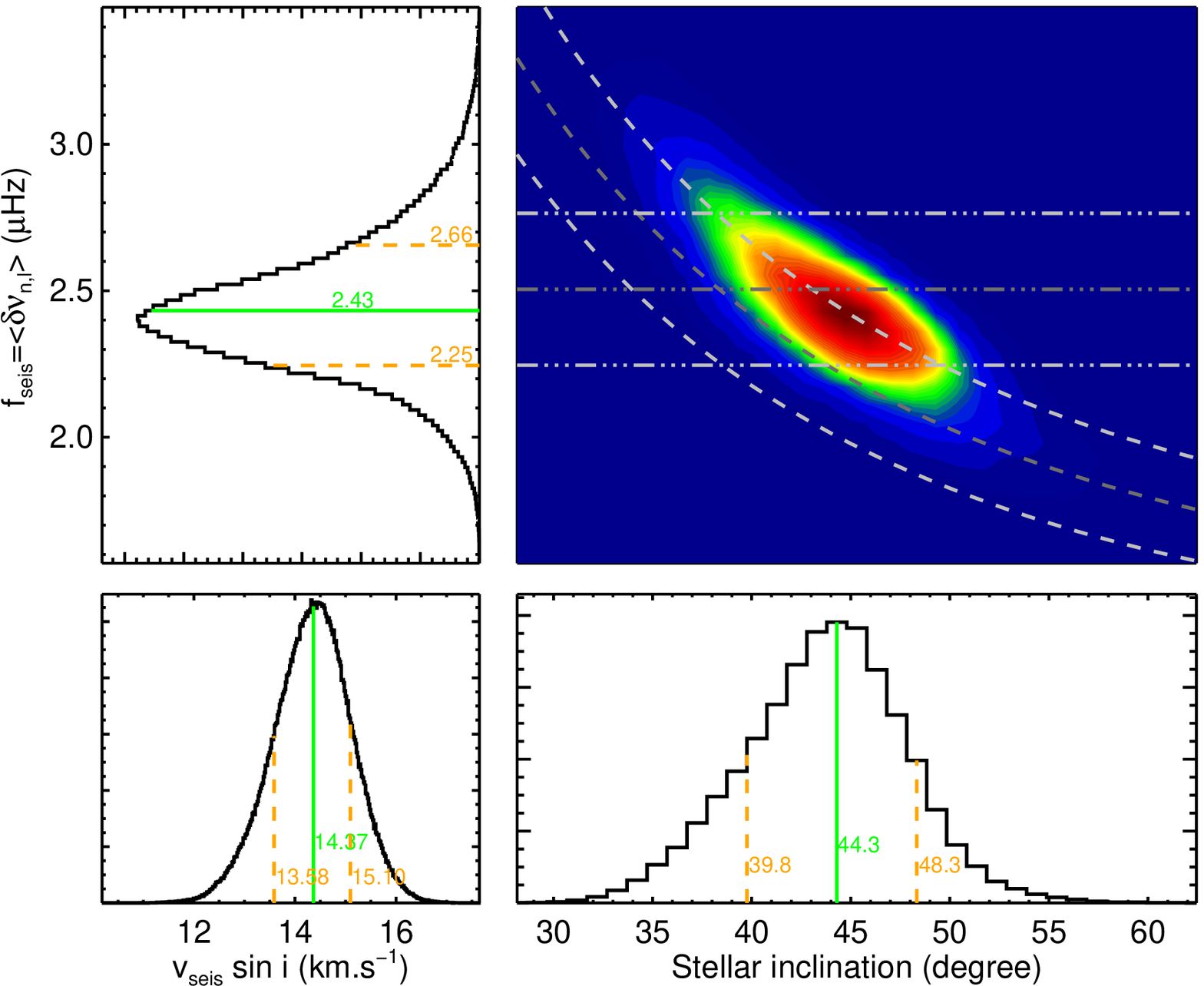}%
		}\hfill%
	\subfigure[KIC 7206837]{%
   \includegraphics[angle=0,width=8cm,height=5.95cm]{Figures/nus-inc-MAP/7206837-nus-inc.eps.eps}%
		}\hspace*{\fill}%

   \hspace*{\fill}%
	\subfigure[KIC 9139151]{%
   \includegraphics[angle=0,width=8cm,height=5.95cm]{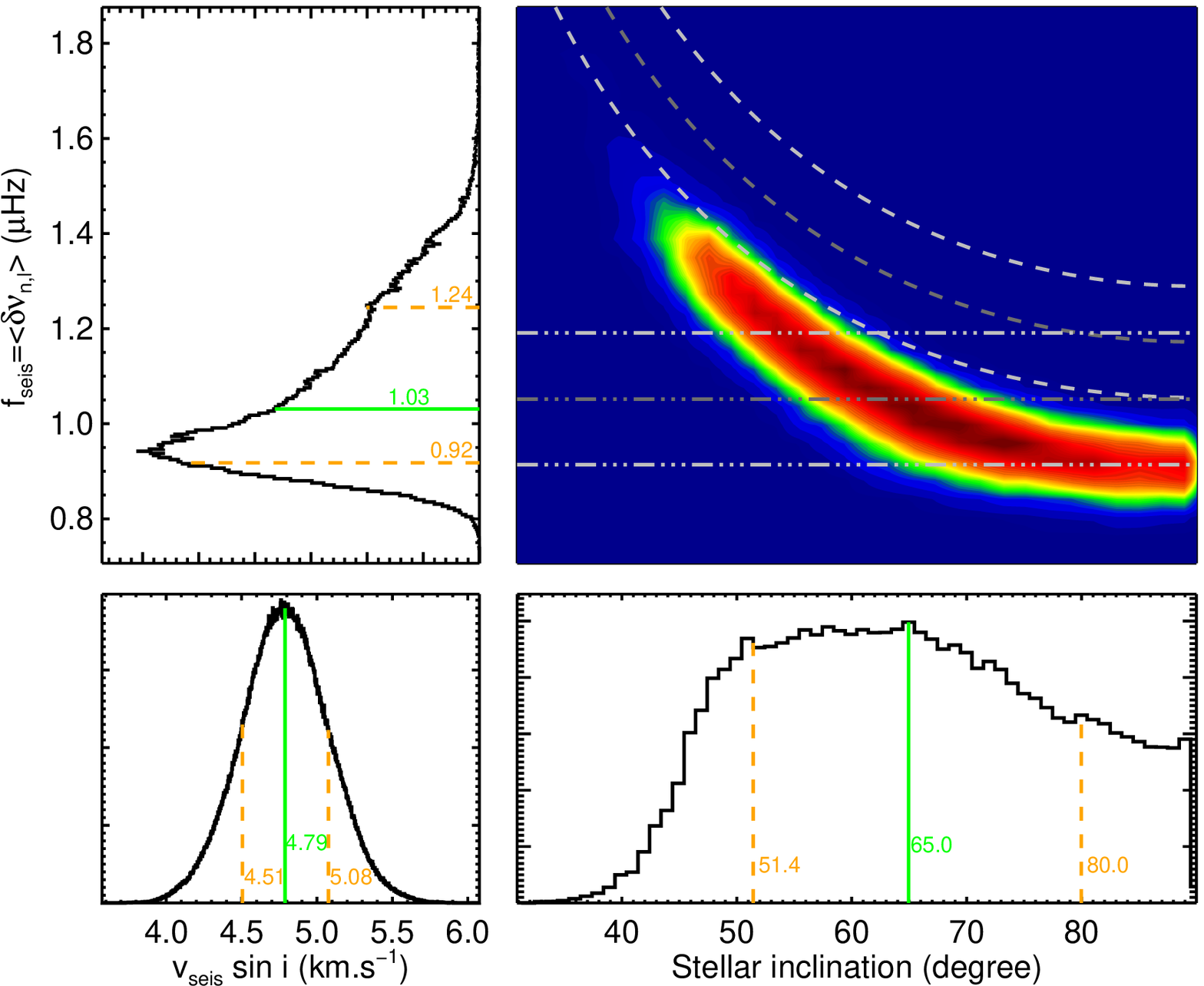}%
		}\hfill%
	\subfigure[KIC 9139163]{%
   \includegraphics[angle=0,width=8cm,height=5.95cm]{Figures/nus-inc-MAP/9139163-nus-inc.eps.eps}%
   }\hspace*{\fill}%
   
   \hspace*{\fill}%
   \subfigure[KIC 9206432 (with $l=0,1,2, 3$ and fixed visibilities)]{%
   \includegraphics[angle=0,width=8cm,height=5.95cm]{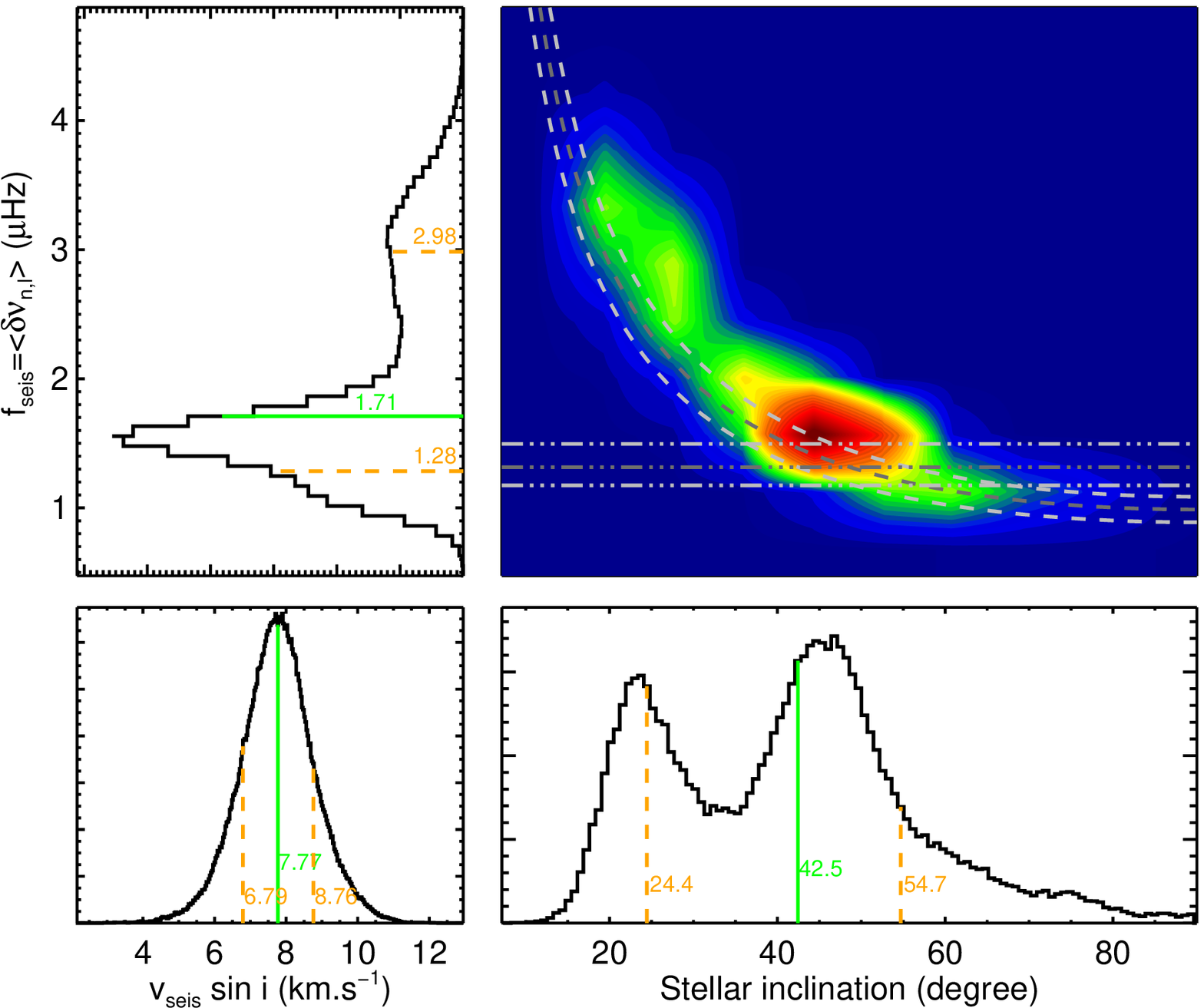}%
		}\hfill%
	\subfigure[KIC 9812850]{%
   \includegraphics[angle=0,width=8cm,height=5.95cm]{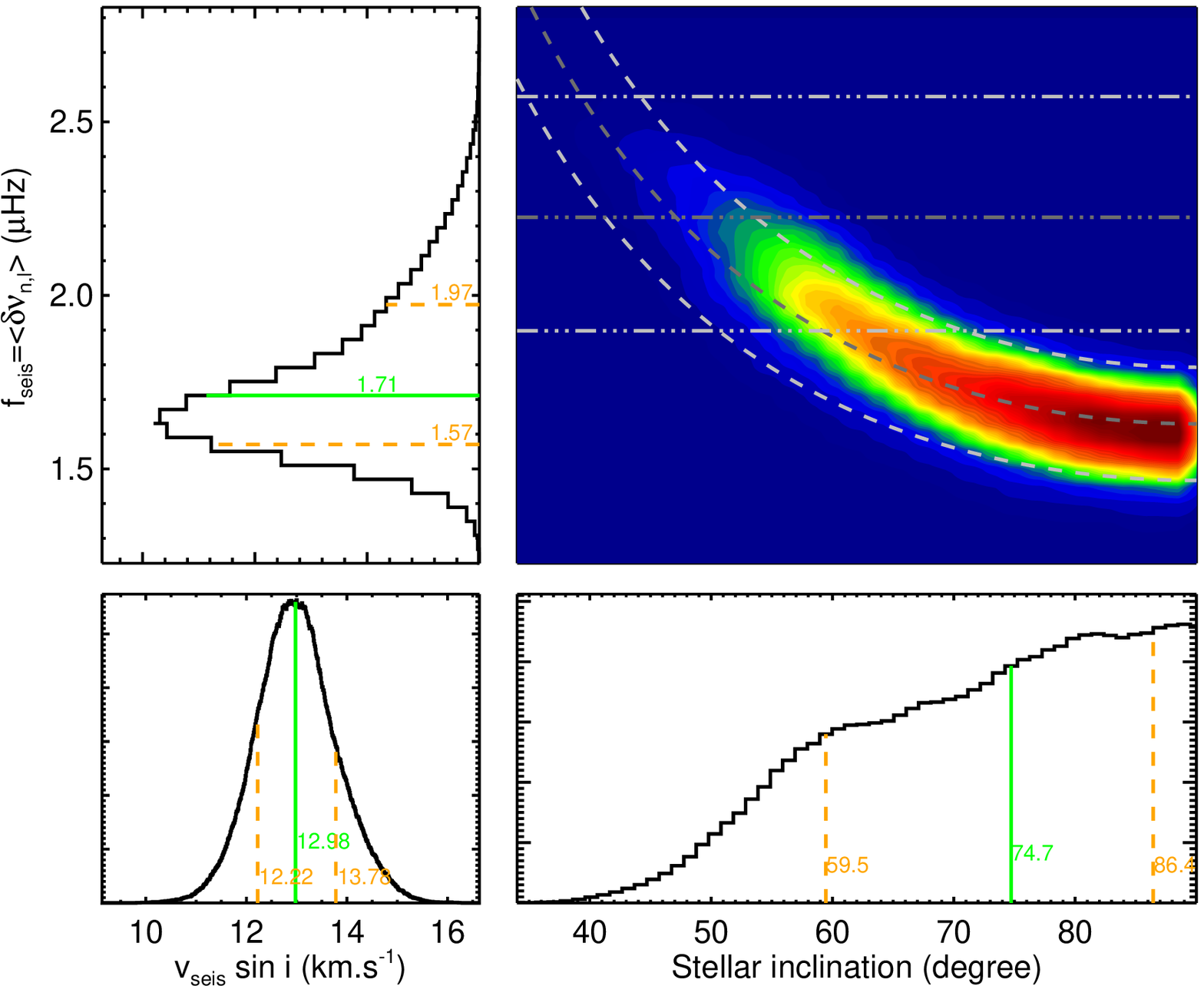}%
   }\hspace*{\fill}%
   
	   \end{center}
\contcaption{}  
\end{figure*} 
\begin{figure*}
  \begin{center}
   \hspace*{\fill}%
   \subfigure[KIC 10162436]{%
   \includegraphics[angle=0,width=8cm,height=5.95cm]{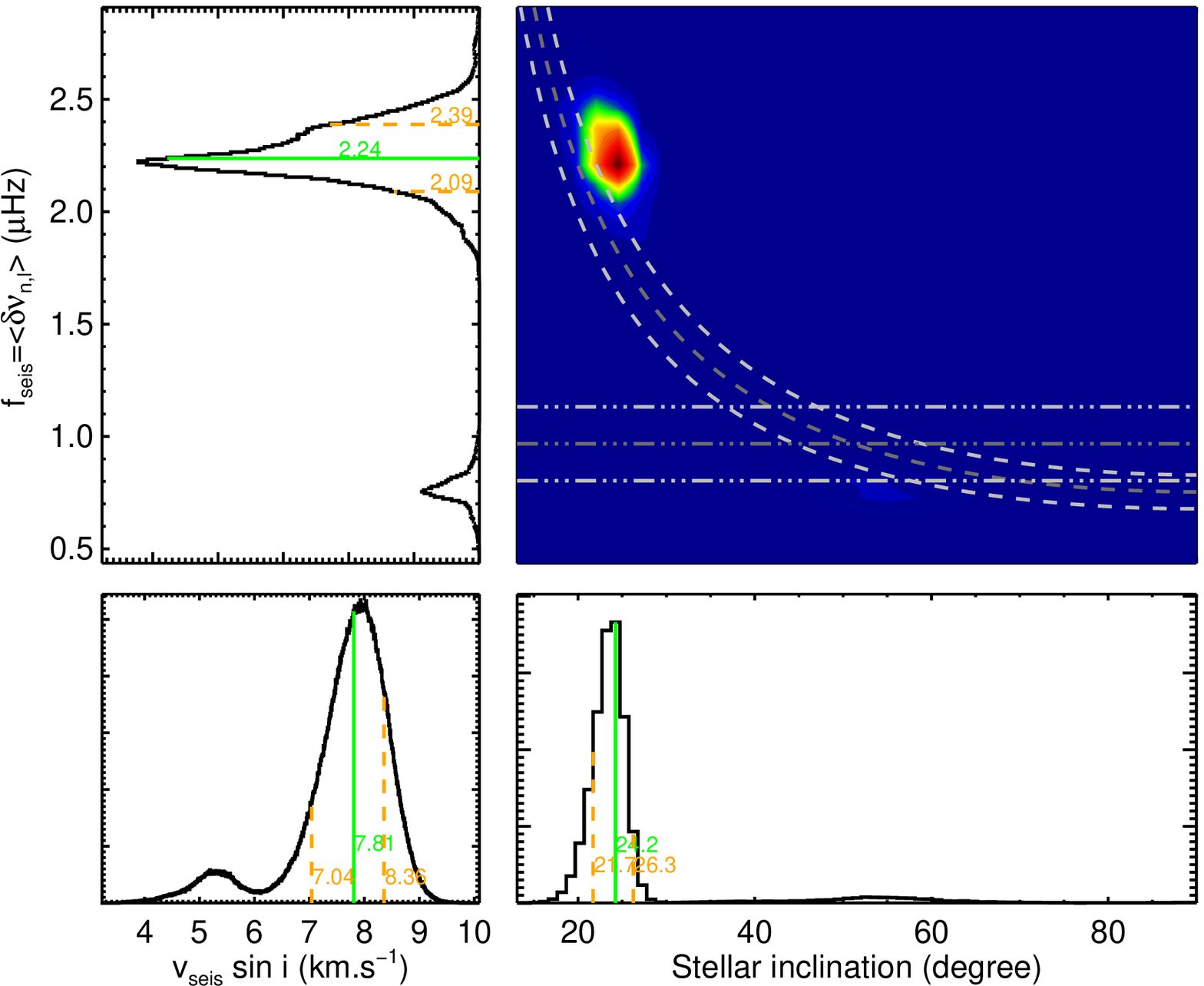}%
		}\hfill%
	\subfigure[KIC 10355856]{%
   \includegraphics[angle=0,width=8cm,height=5.95cm]{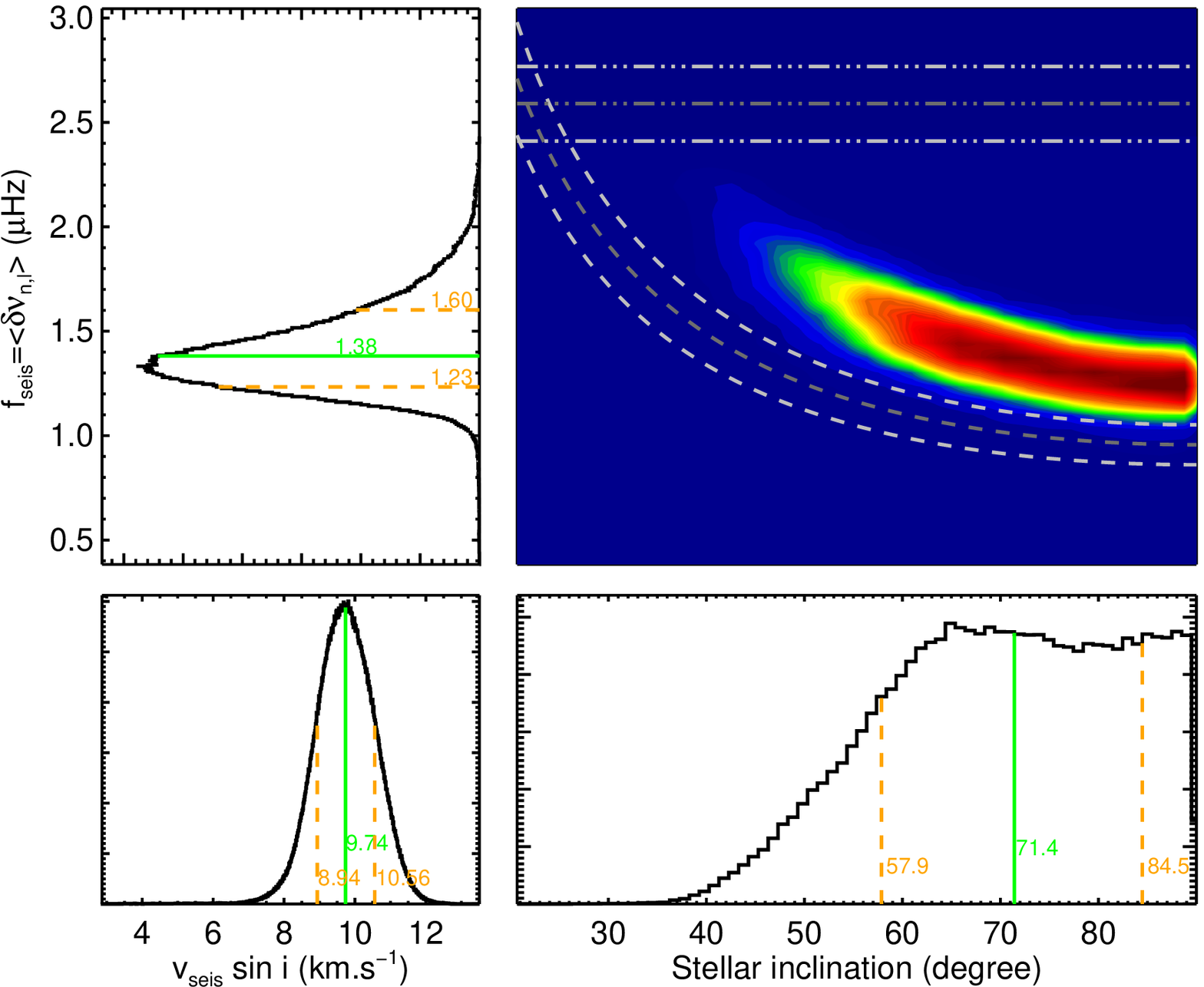}%
   }\hspace*{\fill}%
   
   \hspace*{\fill}%
   \subfigure[KIC 10454113]{%
   \includegraphics[angle=0,width=8cm,height=5.95cm]{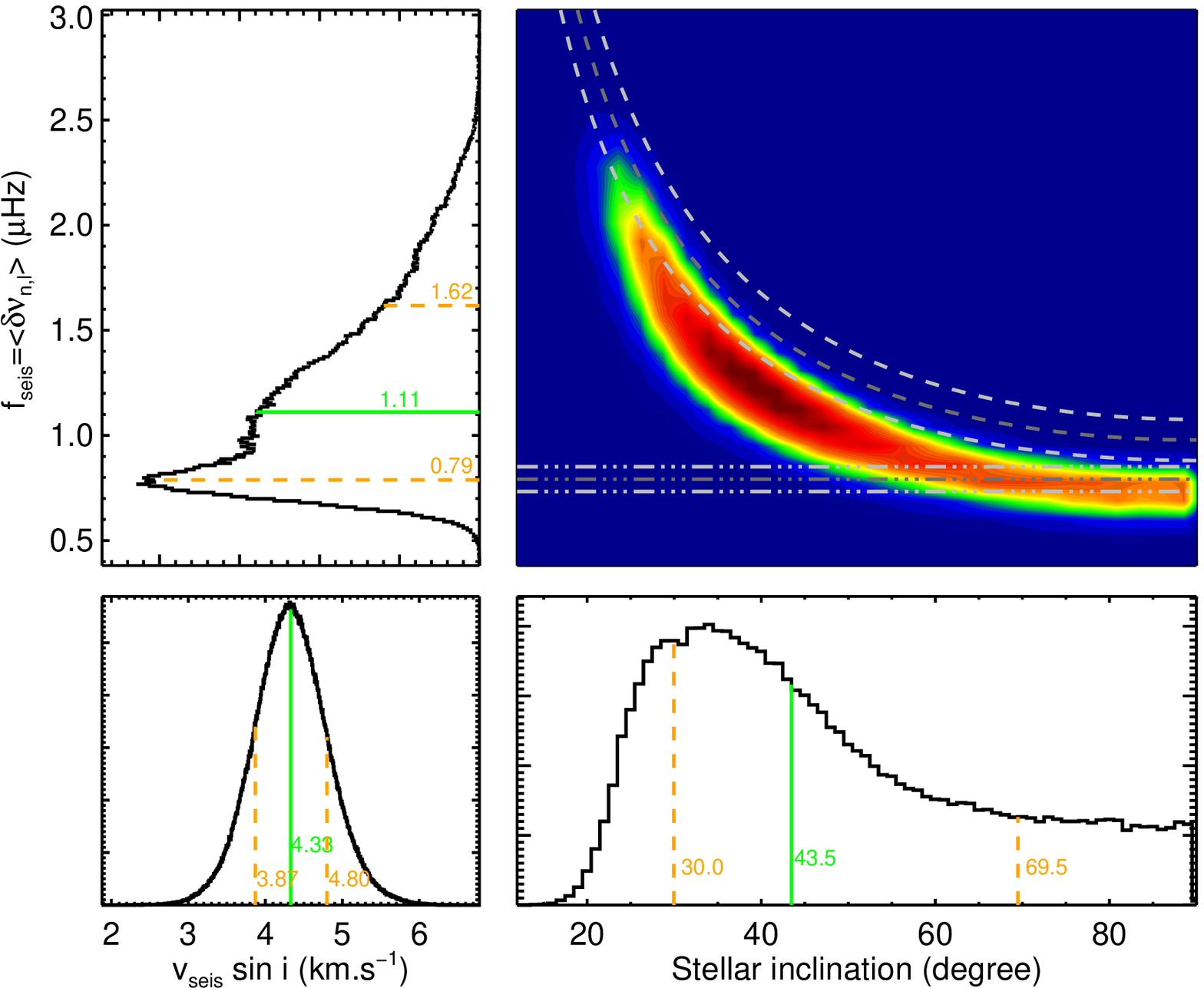}%
		}\hfill%
	\subfigure[KIC 11253226]{%
   \includegraphics[angle=0,width=8cm,height=5.95cm]{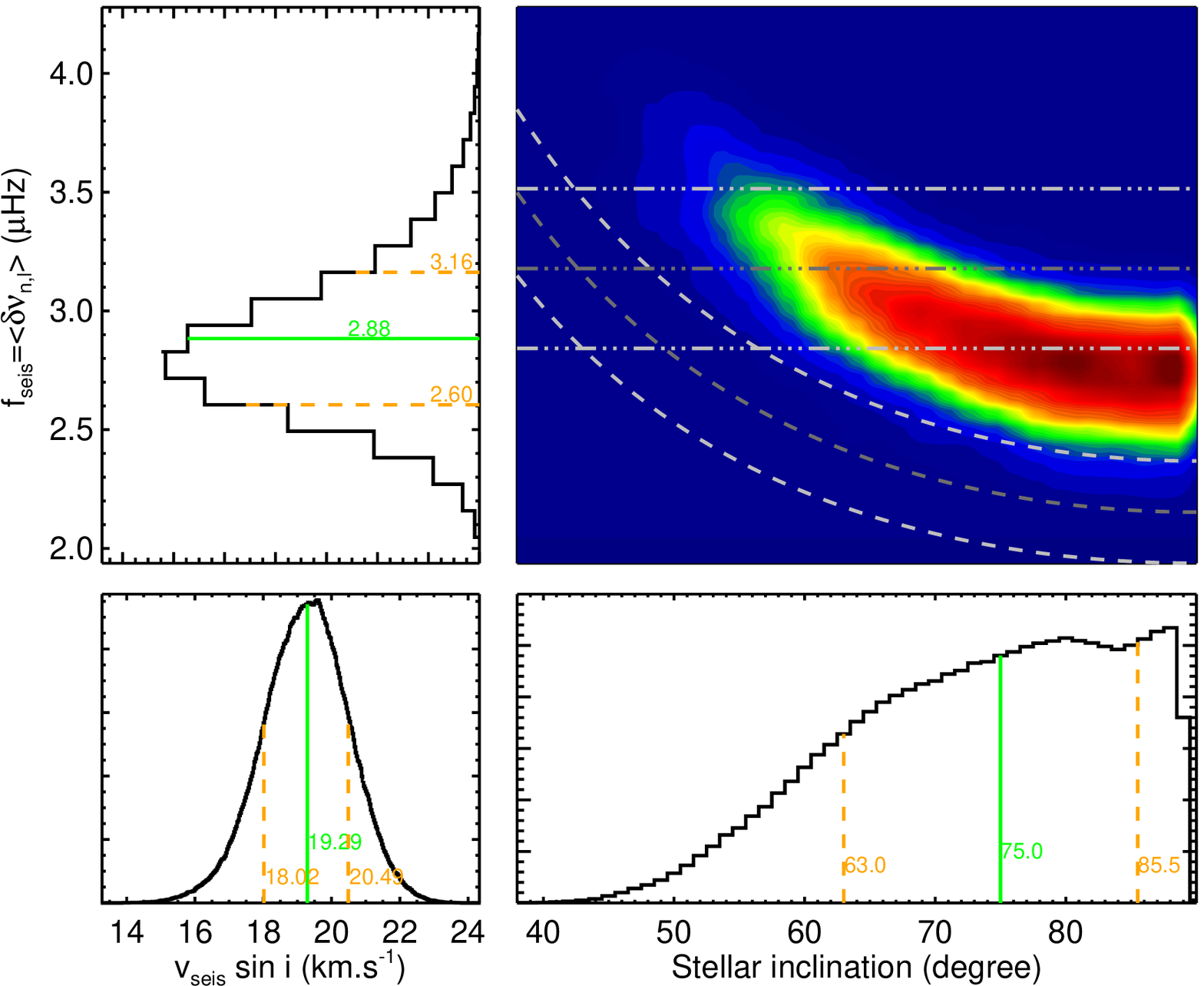}
   }\hspace*{\fill}%
   
   \hspace*{\fill}%
   \subfigure[KIC 12009504]{%
   \includegraphics[angle=0,width=8cm,height=5.95cm]{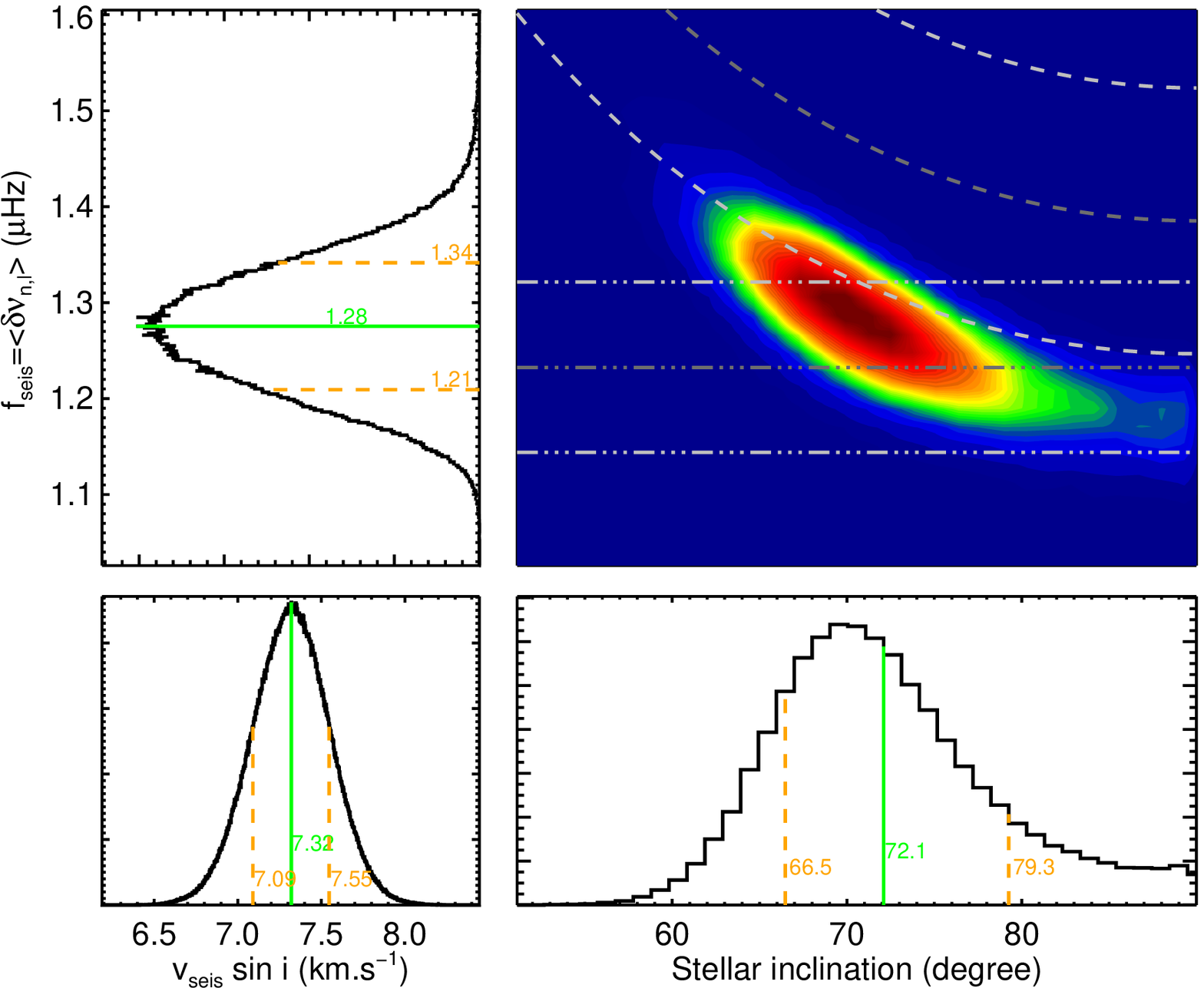}%
		}\hfill%
   \subfigure[KIC 12258514]{%
   \includegraphics[angle=0,width=8cm,height=5.95cm]{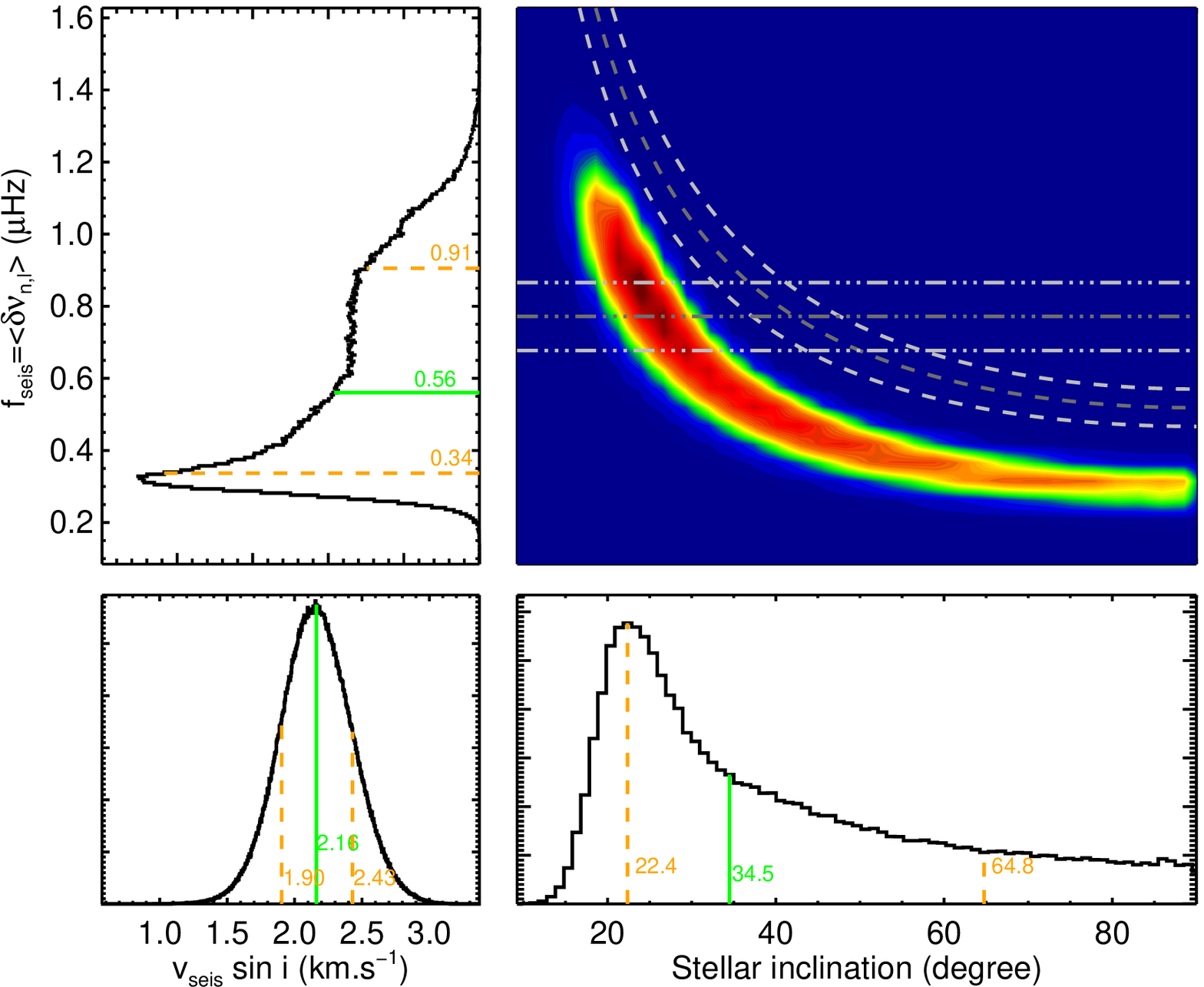}%
		}\hspace*{\fill}%
   
	   \end{center}
\contcaption{}  
\end{figure*} 
\begin{figure*}
  \begin{center}
   \hspace*{\fill}%
	\subfigure[HAT-P-7]{%
   \includegraphics[angle=0,width=8cm,height=5.95cm]{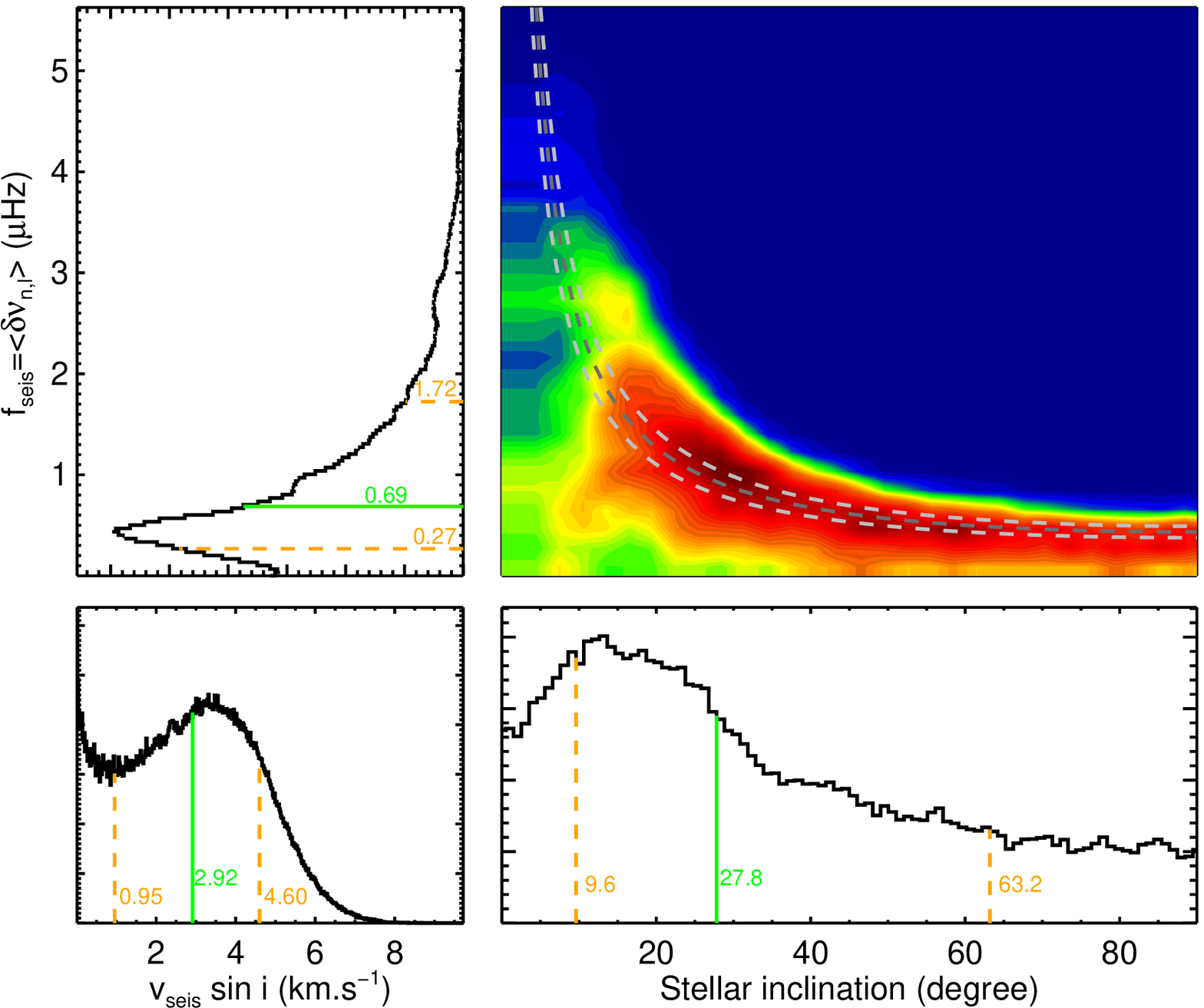}%
		}\hfill%
	\subfigure[{\it Kepler}-25]{%
   \includegraphics[angle=0,width=8cm,height=5.95cm]{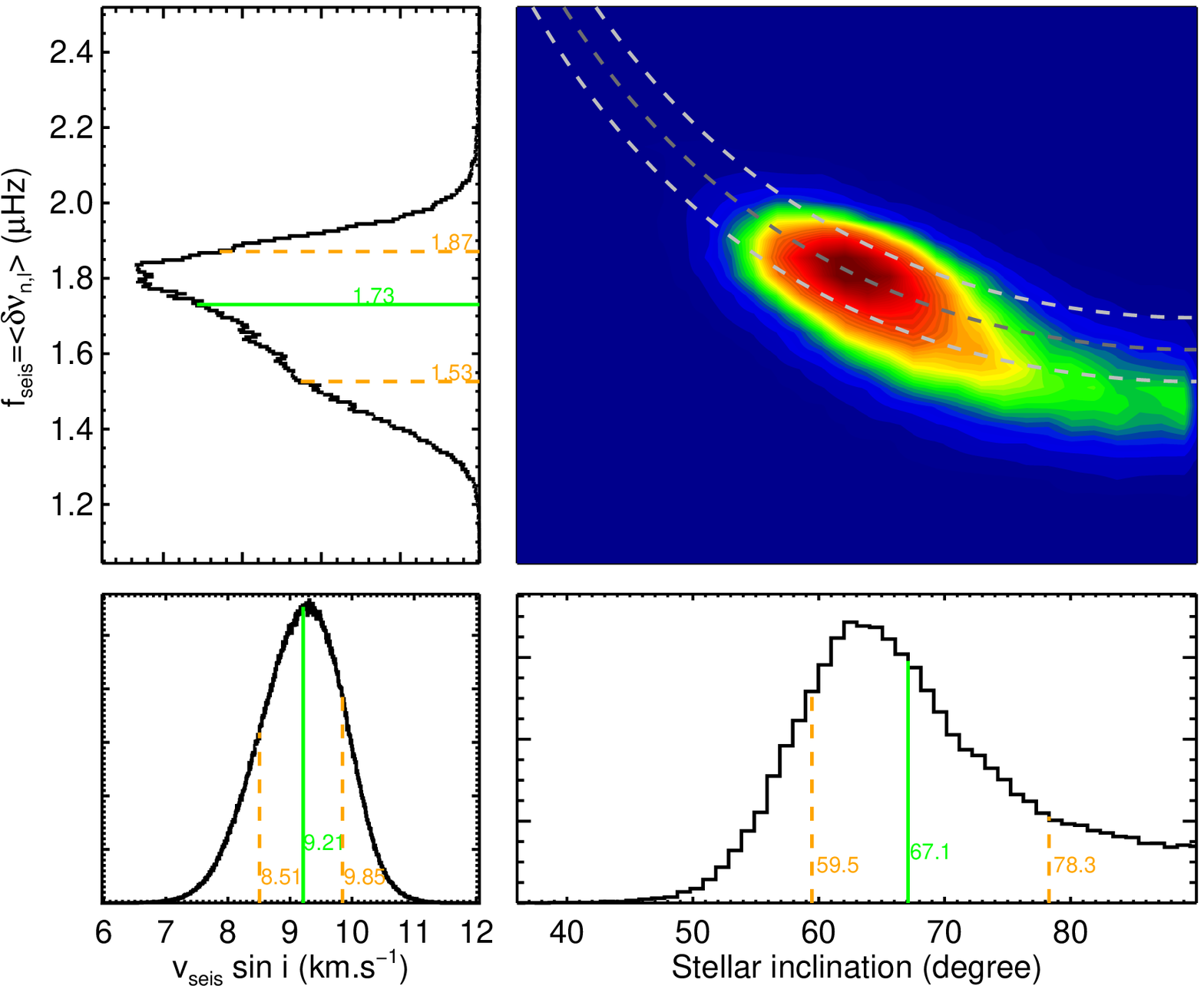}%
		}\hspace*{\fill}%

   \hspace*{\fill}%
	\subfigure[HD 181420]{%
   \includegraphics[angle=0,width=8cm,height=5.95cm]{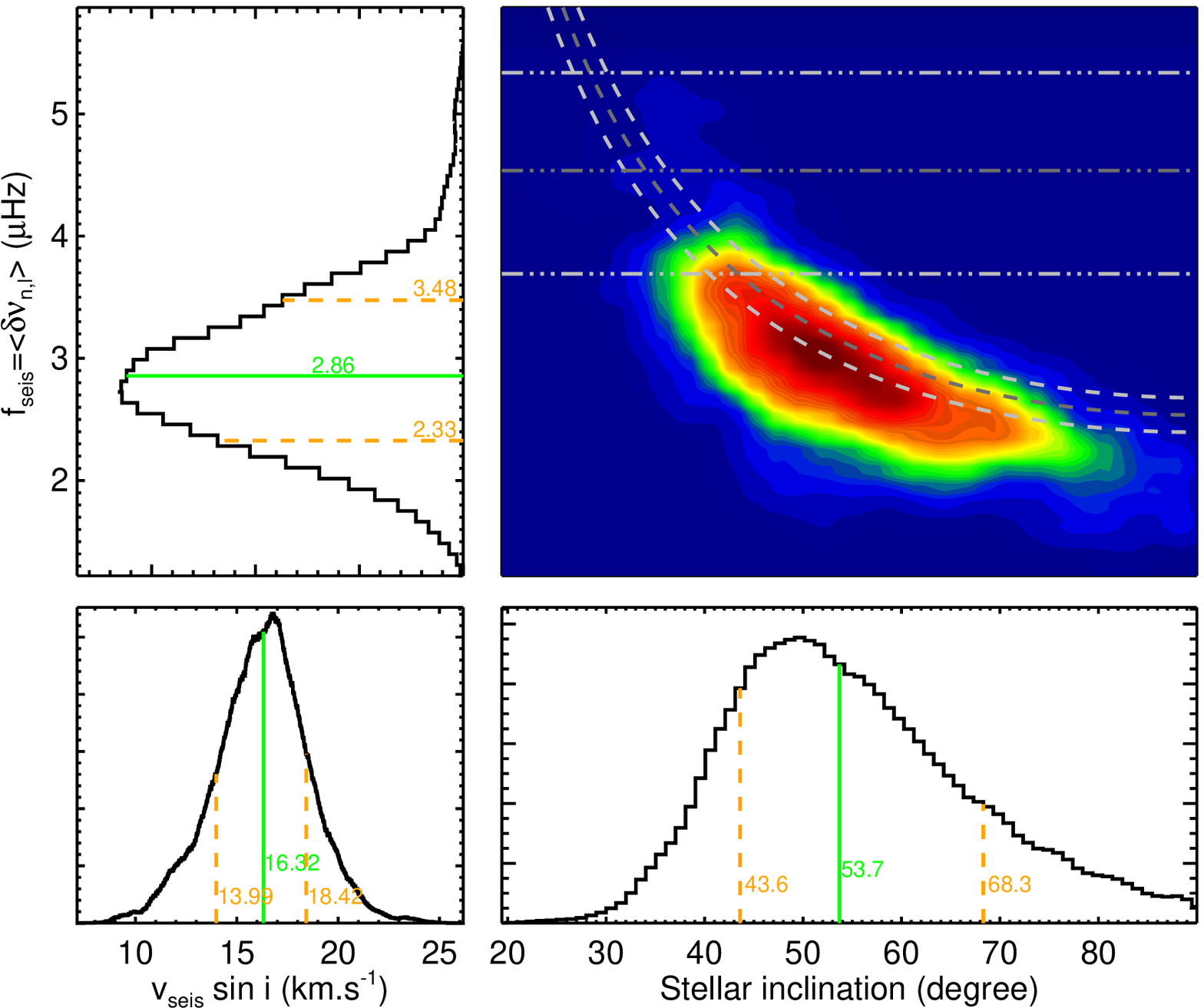}%
		}\hfill%
	\subfigure[HD 49933]{%
   \includegraphics[angle=0,width=8cm,height=5.95cm]{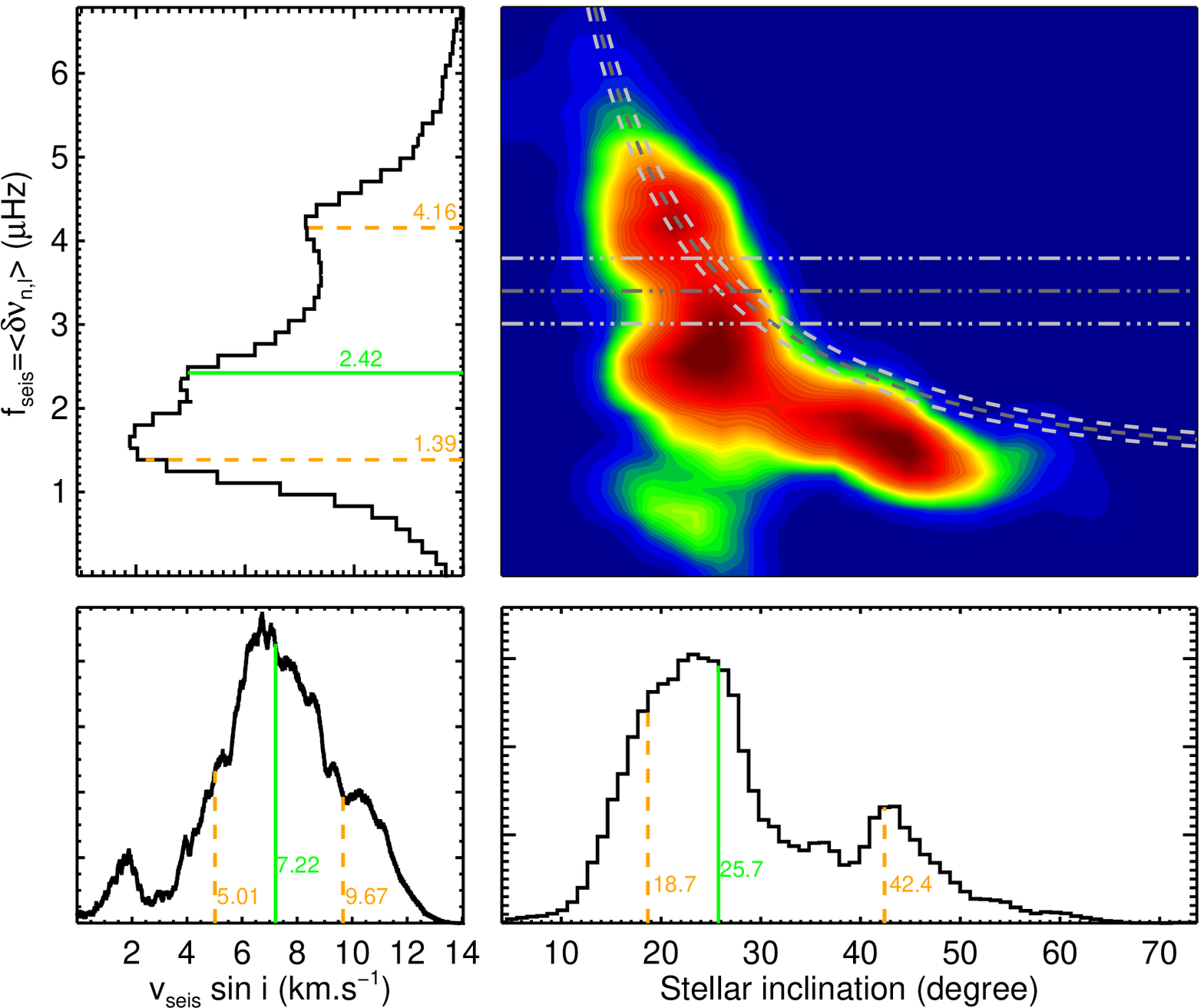}%
		}\hspace*{\fill}%

	   \end{center}
\contcaption{}  
\end{figure*} 

\clearpage
\twocolumn
\section{\red{Supplementary contents for KIC 3424541}}
\label{sec:appendix-342}
\begin{figure*}
  \begin{center}
	\includegraphics[angle=0,width=12cm]{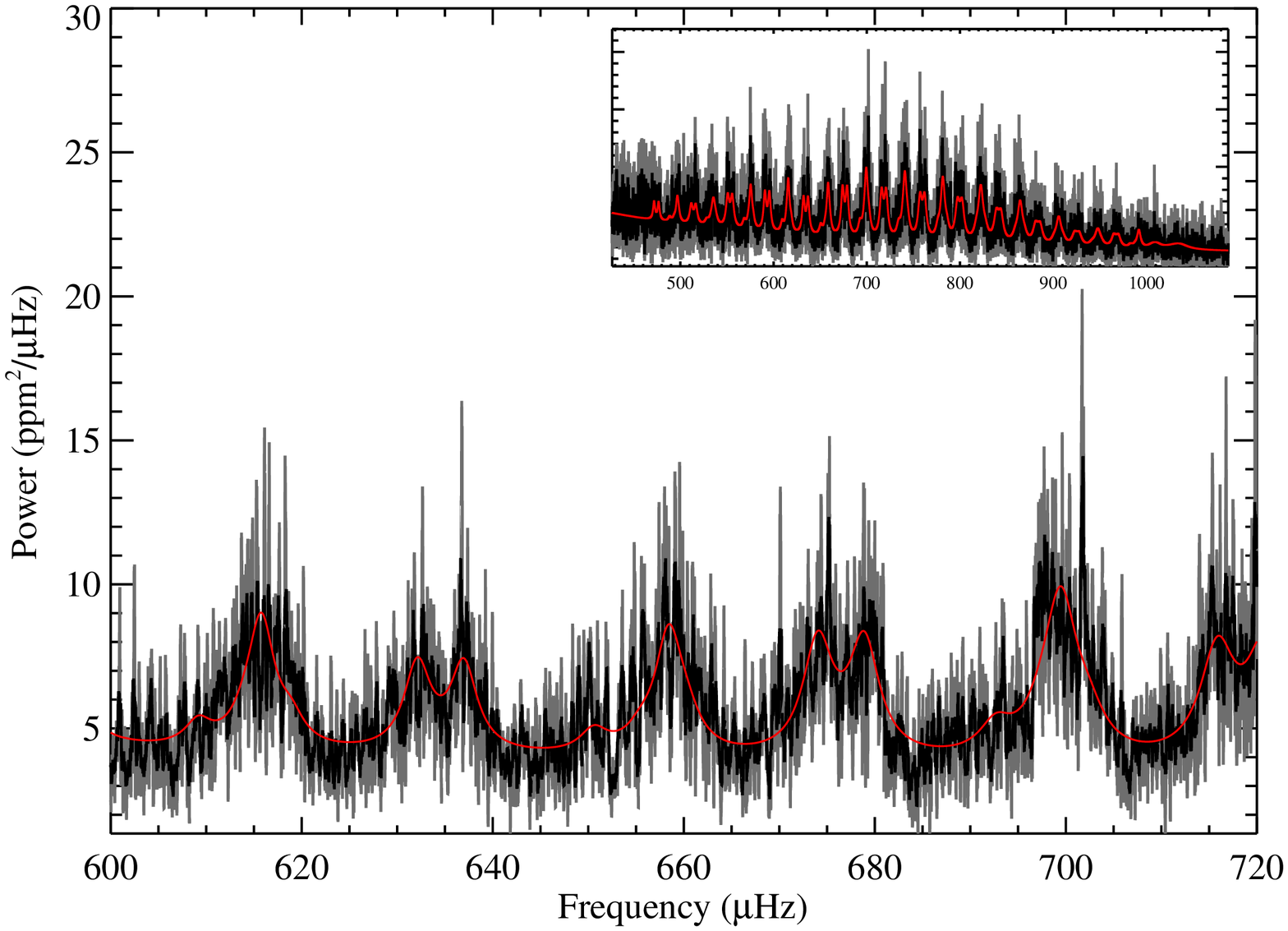}
  \end{center}
  \caption{\red{Fit of the power spectrum of KIC 3424541 with the most likely mode identification (opposite as in \protect\cite{Appourchaux2012}), obtained from the {\it Kepler} data analysis using Quarters 5 to 17. In this identification, the first excess of power on the left of the figure is due to a pair of degree $l=2,0$, and is followed by alternating $l=1$ and $l=2,0$ modes.}}  
\label{fig:KIC3424541:correct}
\end{figure*}

\begin{figure*}
  \begin{center}
 	\includegraphics[angle=0,width=12cm]{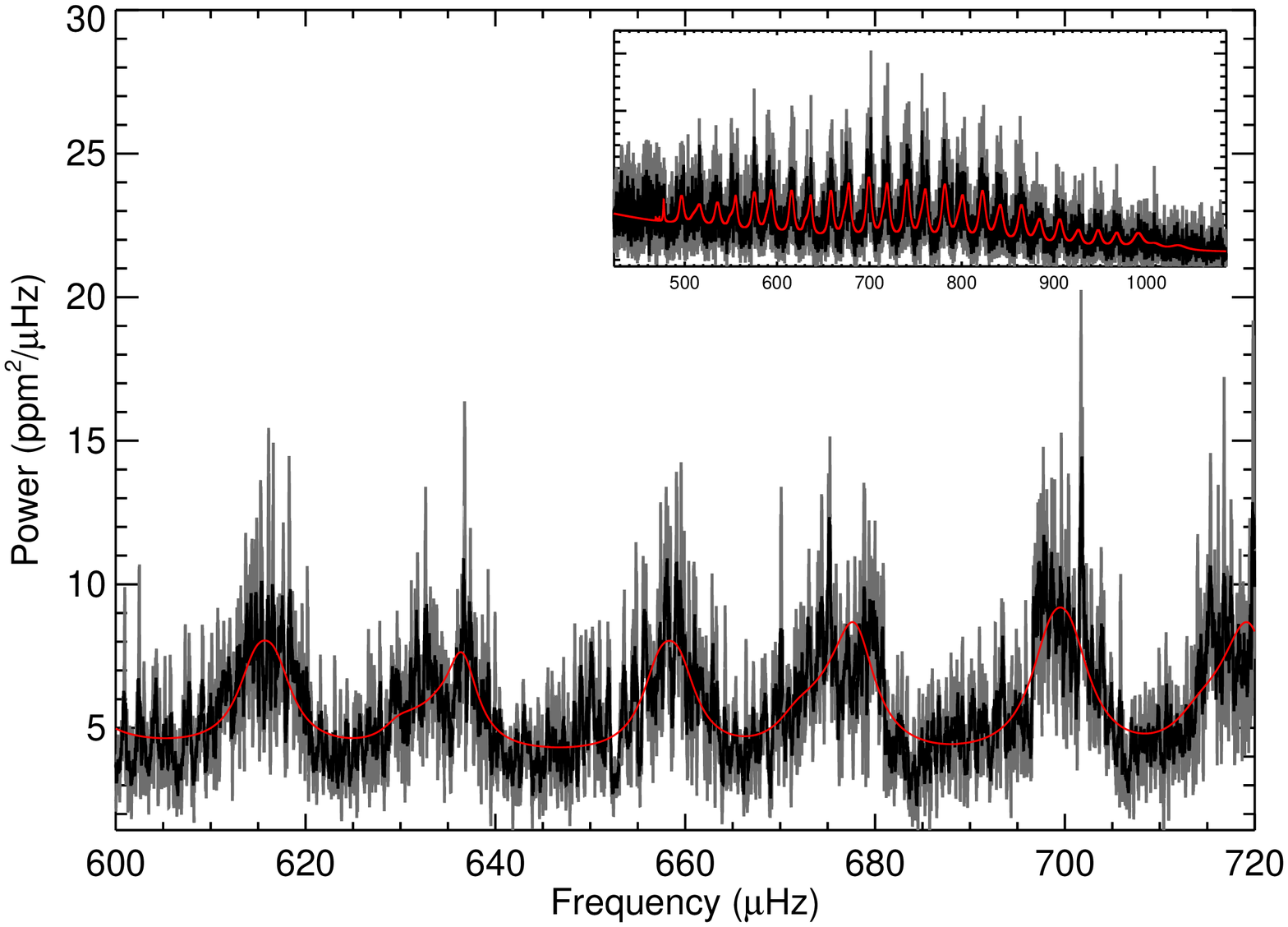}
  \end{center}
\caption{\red{Fit of the power spectrum of KIC 3424541 with the less likely mode identification (same as in \protect\cite{Appourchaux2012}), obtained from the Kepler data analysis using Quarters 5 to 17. In this identification, the first excess of power on the left of the figure corresponds to a degree $l=1$, and is followed by alternating $l=2,0$ and $l=1$ modes.}}  
\label{fig:KIC3424541:wrong}
\end{figure*}

\red{The best fit of the power spectrum for the two possible mode identification of KIC 3424541 is shown in Fig. \ref{fig:KIC3424541:correct} and Fig. \ref{fig:KIC3424541:wrong}. The most probable mode identification was assessed by using the Bayes ratio $O(M_1, M_2)$ between the two possible competing models $M_1$ and $M_2$ \cite[e.g.][]{Benomar2009}. $O(M_1, M_2)$ is the equivalent of the the likelihood ratio used in frequentist approaches. However, contrary to the likelihood ratio, the Bayes ratio can be interpreted in terms of probability. For example a Bayes ratio of $O(M_1, M_2)$=3 in favour of $M_1$ indicates that the model $M_1$ has $75\%$ of probability to be correct. It is often considered that a model is strongly supported when the ratio exceeds 100 \citep{Jeffreys1961}. In the case of KIC 3424541, the mode identification shown in Fig.\ref{fig:KIC3424541:correct} is very strongly supported because the logarithm of the Bayes factor is $7.5\times 10^3$ in favour of this identification.}

\red{The list of measured pulsation frequencies for the most likely mode identification of KIC 3424541 is given in Table \ref{tab:KIC3424541:freq}.}

\begin{table}
 \caption{\red{Mode frequencies for the most likely mode identification of KIC 3424541 presented in this paper.}}
\begin{center}
\begin{tabular}{cc|ccc} 
\hline 
degree  $l$ &  index     & frequency ($\umu$Hz) \\ \hline
\hline 
  0 & 1  &     $496.77  \pm    0.58$   \\  
  0 & 2  &     $535.47  \pm       0.58$ \\ 
  0 & 3  &    $575.46    \pm     0.34$  \\ 
  0 & 4  &    $615.77    \pm     0.35$  \\ 
  0 & 5  &    $658.44    \pm     0.27$  \\  
  0 & 6  &    $699.47    \pm     0.28$ \\  
  0 & 7  &    $740.87   \pm      0.32$  \\ 
  0 & 8  &    $781.62   \pm      0.30$  \\  
  0 & 9  &    $822.33   \pm      0.33$ \\  
  0 & 10 &   $864.54  \pm       0.41$ \\  
  0 & 11 &    $905.98  \pm       0.52$ \\ 
  0 & 12 &   $947.90   \pm      0.57$ \\ 
  0 & 13 &   $991.82  \pm       0.67$ \\  
  0 & 14 &   $1035.99  \pm       1.97$ \\  
1  &  1  &    $474.25   \pm      0.61$ \\  
1 &   2  &    $514.14  \pm       0.38$ \\  
1 &   3  &    $553.20   \pm      0.30$ \\ 
1 &   4  &    $593.10  \pm      0.25$ \\  
1 &   5  &     $634.54   \pm      0.23$ \\ 
1  &  6  &     $676.47   \pm      0.22$ \\ 
1  &  7  &     $718.29   \pm      0.23$ \\  
1  &  8  &     $759.57   \pm      0.24$ \\ 
1  &  9  &     $799.76  \pm       0.30$ \\  
 1 &  10 &      $841.50  \pm       0.36$ \\
 1  & 11 &      $883.52  \pm       0.41$ \\ 
 1  & 12 &     $925.42  \pm       0.56$ \\ 
  1 & 13 &      $967.46  \pm       0.55$ \\ 
 1 &  14 &     $1009.04 \pm       1.16$ \\ 
 2 &  1  &    $493.35   \pm      1.67$  \\   
 2 &  2  &     $533.17  \pm       0.90$ \\  
 2 &  3  &     $573.26   \pm      0.64$ \\  
 2 &  4  &     $614.03  \pm      0.53$ \\  
 2 &  5  &    $655.45  \pm       0.52$  \\  
 2 &  6  &     $697.51  \pm       0.44$ \\  
 2 &  7  &     $738.99  \pm       0.51$ \\ 
 2 &  8  &     $781.14   \pm      0.50$ \\ 
 2 &  9  &     $822.45  \pm       0.55$ \\ 
  2  &10 &      $864.26   \pm      0.63$ \\ 
  2 & 11 &      $905.82  \pm       0.76$ \\  
 2  & 12 &     $946.56  \pm       1.09$ \\  
  2  &13 &    $987.69   \pm      1.46$ \\  
  2 & 14 &     $1028.68   \pm      2.59$ \\ 
\hline
\end{tabular}
\end{center}
\label{tab:KIC3424541:freq}
\end{table}

\section{\red{Supplementary contents for KIC 9206432}}
\label{sec:appendix-920}
\begin{figure*}
  \begin{center}
   \includegraphics[angle=0,width=8cm]{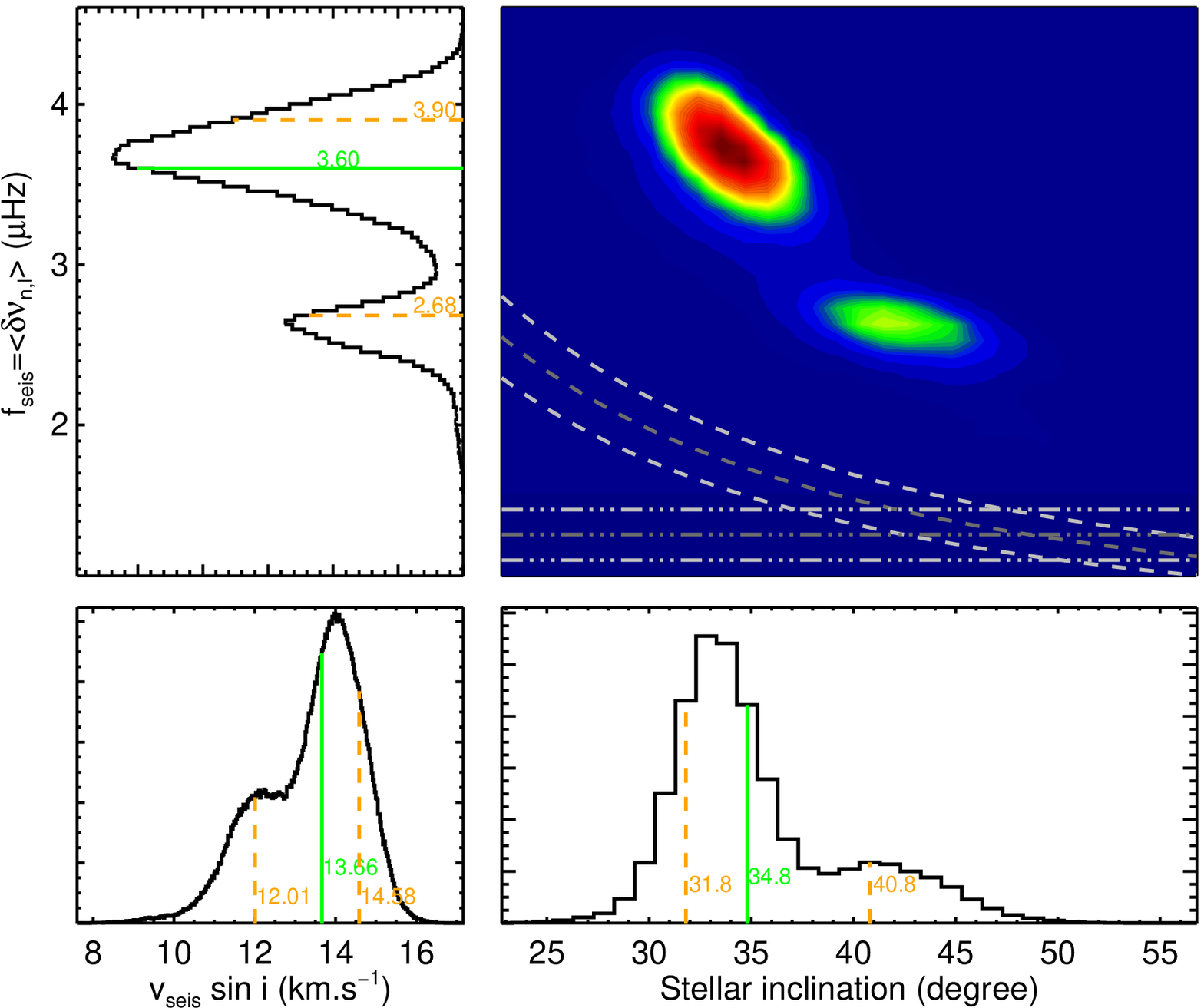} \\
   \includegraphics[angle=0,width=8cm]{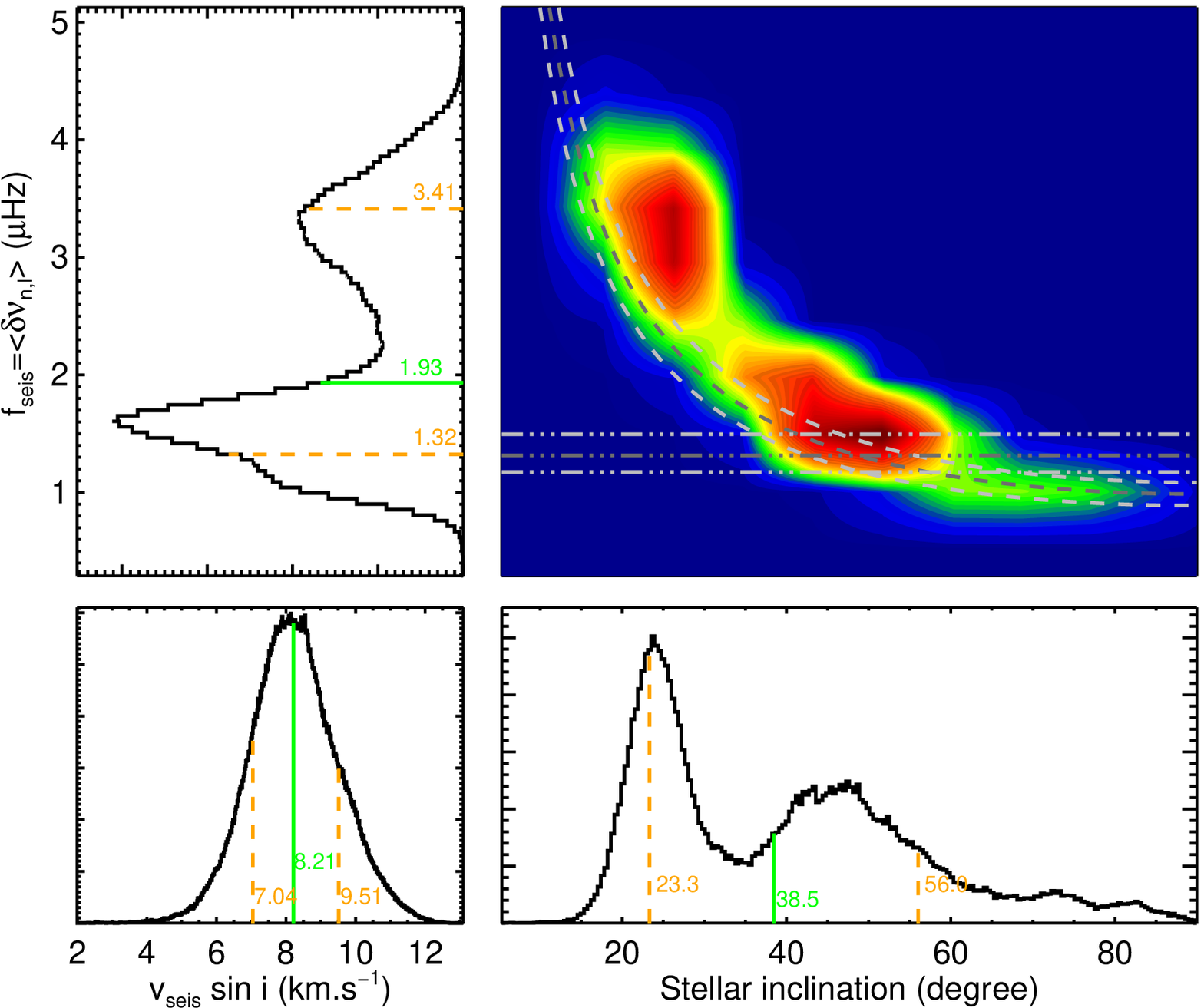} \\
   \includegraphics[angle=0,width=8cm]{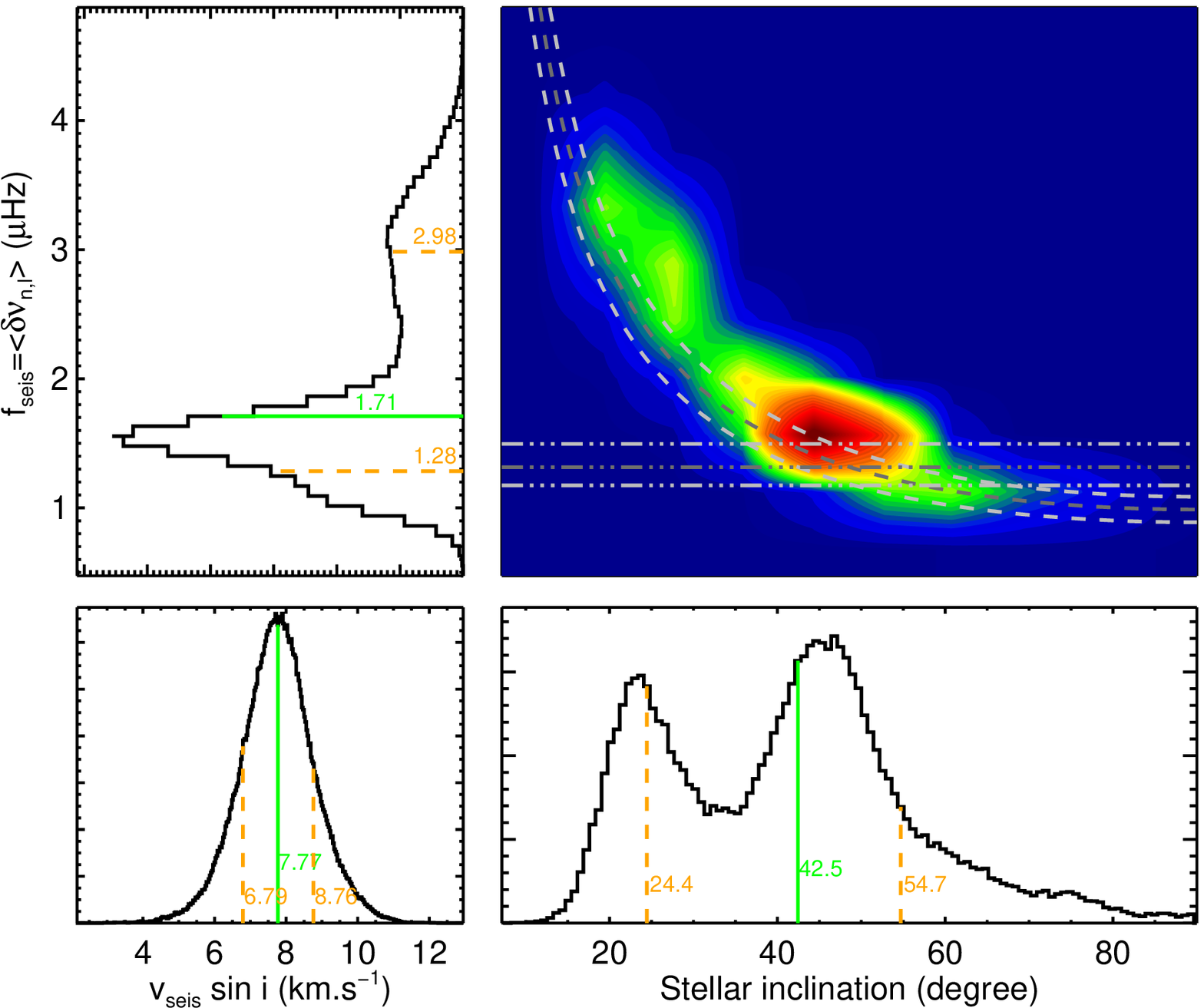} \\
  \end{center}
\caption{\red{\textbf{KIC 9206432.} Comparison of the results obtained for the stellar inclination and rotational splitting by including; (top) modes of degree $l=0,1,2$ mode visibilities as free parameters; (middle) modes of degree $l=0,1,2$ with fixed visibilities; (bottom) modes of degree $l=0,1,2,3$ with fixed visibilities.}}   
\label{fig:KIC9206432-Vis}
\end{figure*}

 	\red{Our thorough analysis of KIC 9206432 enabled us to identify the likely cause for the apparent discrepancy between the surface and the interior rotation. In Fig.\ref{fig:KIC9206432-Vis} (top), we show the results for the stellar inclination and the rotational splitting when the power spectrum is fitted by considering the modes visibilities $V^2_{l=1}$ and $V^2_{l=2}$ as free parameters. In that case, there is an apparent mismatch between surface rotation and rotational splitting.  When the visibilities are fixed to their expected values, the agreement becomes better (middle and bottom).}

\end{document}